\documentclass[aps,preprint,onecolumn,floatfix]{revtex4-2}
\usepackage[english]{babel}
\usepackage{chemformula}
\usepackage{amsmath}
\usepackage{graphicx}
\usepackage{placeins}
\usepackage{subcaption}
\usepackage{adjustbox}

\usepackage{natbib}
\setcitestyle{quare, comma, numbers,sort&compress}
\usepackage{mathptmx}
\usepackage{hyperref}
\usepackage{float}
\usepackage{xcolor}
\usepackage[normalem]{ulem}
\usepackage{mathtools}
\usepackage{amsmath}
\usepackage{SIunits}
\renewcommand{\figurename}{Figure}

\begin{document}

\title{Emergence of non-uniform strain induced exciton species in homo- and heterobilayer transition metal dichalcogenides}
\author{M. Daqiqshirazi and T. Brumme}

\address{Bergstrasse 66c, Theoretical chemistry, Technische Universität Dresden, Dresden, Germany.}

\email{thomas.brumme@tu-dresden.de}

\begin{abstract}
Full control of excitons in 2D materials is an important step to exploit them for applications. Straintronics is one method that can be used to effectively control the movement of excitons.
Unfortunately, the effects of non-uniform strain in 2D materials are not yet well understood theoretically, although these strain fields can be present in experiments in the form of wrinkles, bubbles, and folds, or even explicitly applied to 2D materials through pre-patterned surfaces. The effects of these non-uniform strain fields on multilayers are even less studied due to the sheer size of these systems. In the present investigation, we study wrinkles that form in homo- and heterobilayers of 2D transition metal dichalcogenides using density functional theory. We show that the non-uniform strain leads to the formation of interlayer excitons in homobilayers of \ch{WSe2} and to exciton localization in heterobilayers of \ch{WSe2-MoSe2}.
Our results also reveal that the spin angular momentum is changed due to the mixing of in- and out-of-plane states which can explain the brightening of the formerly dark excitonic states under strain.
Our results will pave the way towards a full understanding of the strain-control of excitons in 2D materials.   
\end{abstract}\maketitle
\thispagestyle{empty} 
\section{Introduction} 
\setcounter{page}{0}

2D materials are an interesting class of materials that can differ significantly in their properties from their bulk crystals \cite{miro2014atlas, novoselov20162d}. One of the important features of these materials are their different or even opposing optoelectronic properties. These properties are indeed very important for the design of new devices. For example, Mo- and W-based transition metal dichalcogenides (TMDCs) undergo a transition from an indirect to a direct band gap when they are thinned down to their monolayers \cite{kuc2011influence, manzeli20172d}. This transition makes them suitable candidates for applications that require a direct band gap semiconductor such as sensors and switches \cite{hassan20232d}.

In addition to the novel intrinsic properties of the 2D materials, these can also be modulated by structural manipulations. Structural variation can be achieved, for example, by placing different 2D materials adjacent to each other, creating heterostructures with unique properties \cite{yuan2023strong} or by using the angular degree of freedom of the layers to form moir\'e (hetero-)structures \cite{du2023moire}.
Furthermore, all these properties can be altered on demand utilizing external fields -- especially strain provides a unique possibility to alter electronic properties of 2D materials \cite{mccreary2022stacking,guo2017tunable}.

Strain is omnipresent in 2D materials. It does not only occur in the synthesis or transfer of 2D materials to substrates, but it can also intentionally be used to tune properties \cite{ahn2017strain, santra2024strain,ghorbani2013strain} which is also known as straintronics. Non-uniform strain is present or can even explicitly be applied in experiments in the form of wrinkles, folds, or when placing 2D materials on pre-patterned surfaces \cite{wang2024locally,kovalchuk2022non,pu2021wrinkle,rosati2021dark}. Wrinkles are particularly interesting as they can unintentionally or intentionally be formed on 2D materials  \cite{chung2011surface, cho2021highly,wood2023curvature,D3NR06261A,lee2022surface}. They can also be used as model system for non-uniformly strained systems such as bubbles or atomic-scale protrusions \cite{krumland2024quantum}.  Although periodically wrinkled structures under compression can be different from delaminated wrinkles formed in TMDC layers which have been transferred to a substrate \cite{wang2021strained}, the strain fields still have similar complex features such that most of the physics will be similar.
Furthermore, in experiments, effects due to the moir\'e potential in heterostructures and due to wrinkling can both lead to similar excitonic effects \cite{alexeev2020emergence, Bai2020, li2021interlayer, hou2024strain} and the investigation of wrinkled structure thus helps to distinguish them.

2D materials also offer a unique platform for fundamental studies of neutral and charged excitons. Among the class of 2D materials, TMDCs are considered as an optimal candidate for optoelectronic investigations and applications due to their unique electronic properties such as the spin-valley optical selection rules \cite{wang2018colloquium} and different excitonic species \cite{kozawa2014photocarrier, lindlau2018role}.
TMDC heterostructures have also attracted a lot of attention due to their type-II band alignment and the resulting formation of interlayer excitons \cite{ge2024unraveling}. Interlayer excitons can be employed more effectively in devices, since they have longer lifetimes due to the reduced overlap of electron and hole wave functions.

In this investigation, we show how the wrinkling of homo- and heterobilayers of 2D TMDCs leads to the localization of the states, the reduction of the local band gap as well as formation of \textbf{inter}layer excitons in \textbf{homo}bilayers.

\begin{figure}
\centering
\begin{subfigure}{0.49\textwidth}
    \includegraphics[width=1.\linewidth]{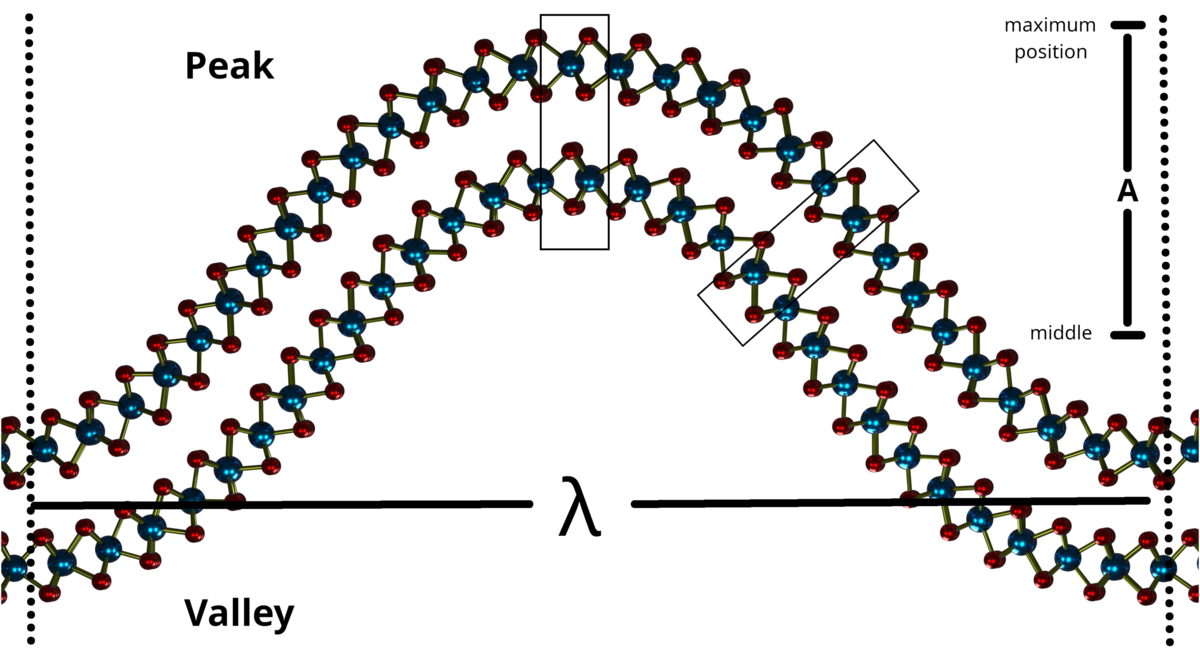}
    \caption{\label{fig_WSe2_bilayer_20percent}}
\end{subfigure}
\begin{subfigure}{0.49\textwidth}
    \includegraphics[width=1.\textwidth]{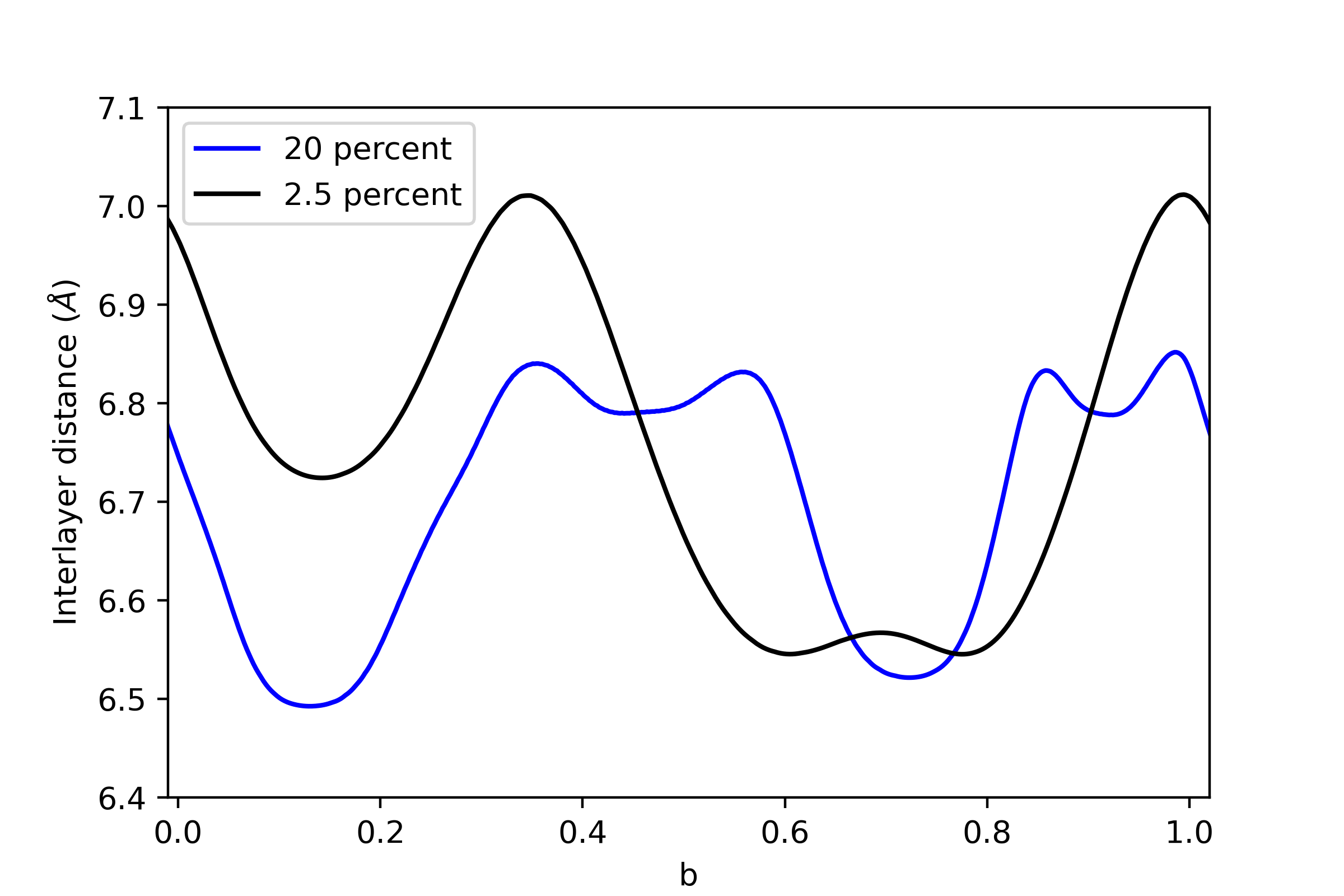}
    \caption{\label{fig_interlayer_homobilayer}}
\end{subfigure} \\
\begin{subfigure}{0.49\textwidth}
     \includegraphics[width=1.\linewidth]{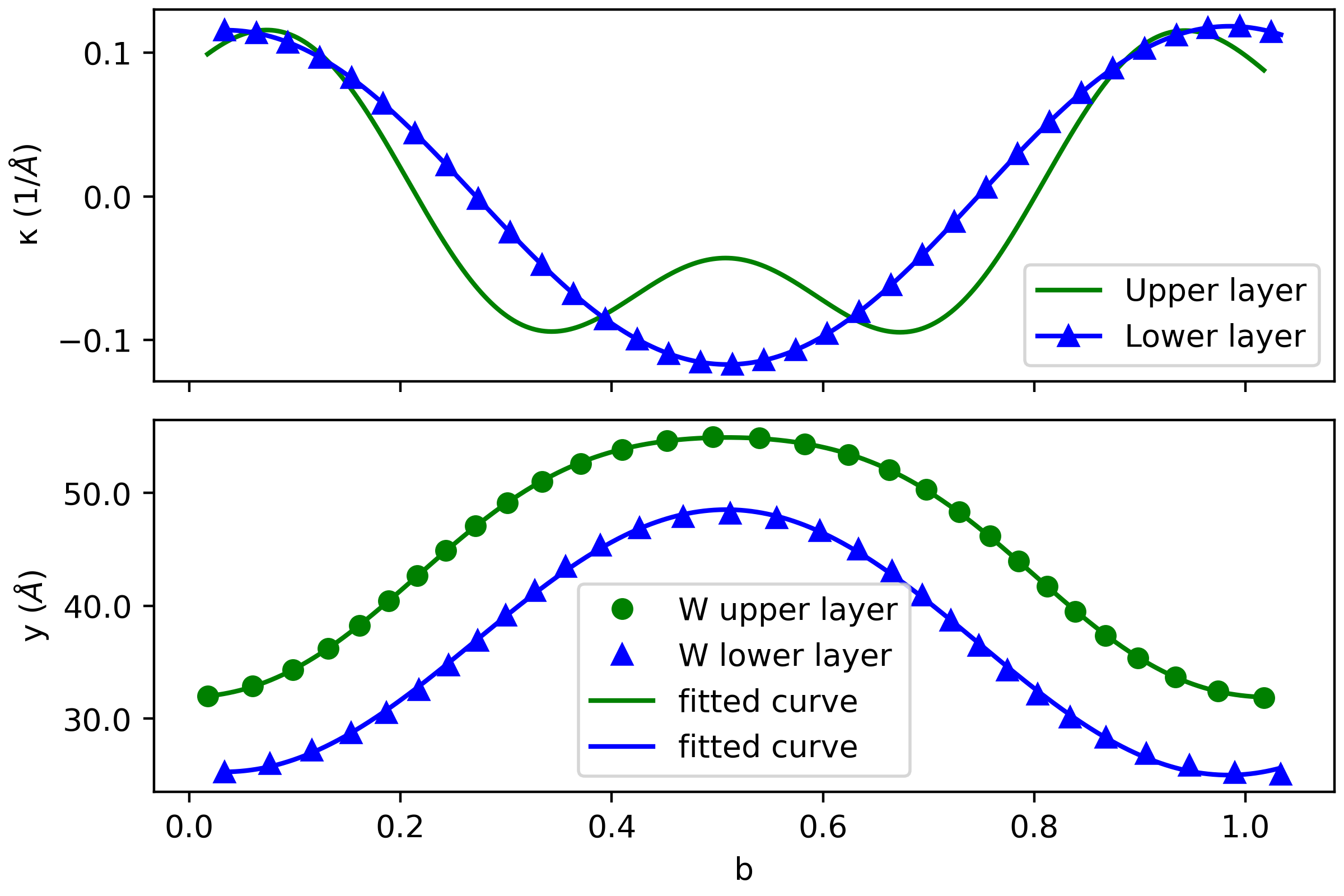}
    \caption{ \label{fig_homobilayer_curvature}}

\end{subfigure}
\caption{Structural changes of the homobilayer \ch{WSe2} under wrinkling. a) The wrinkled homobilayer structure \ch{WSe2} with 20\% compression, two exemplary areas with different stacking are indicated by rectangles. b) Interlayer distance of the \ch{WSe2} homobilayer wrinkle at two different strain values of 2.5\% and 20\%. c) (upper panel) Curvature, $\kappa$, and the atomic positions of the \ch{W} atoms of wrinkled \ch{WSe2} (lower panel) for 20\% compression -- further details on the fit function can be found in supplementary information (SI). }
\end{figure}

\section{Results}
\subsection{\ch{WSe2} bilayer wrinkle}

We create the wrinkled \ch{WSe2} homobilayers by compressing a relaxed flat 2H homobilayer of a $15 \times 1 \times 1$ rectangular \ch{WSe2} unit cell. Due to band folding, we have chosen the direction leading to the armchair termination (please refer to Section \ref{section_method} for more details). The resulting wrinkles cause several variations in the electronic structure of the system. However, to understand the origin of such changes, it is important to first investigate these structural changes. 

\subsubsection{Atomic Structure}

The relaxed lattice parameter of \ch{WSe2} is 3.294 {\AA} which is in a good agreement with literature \cite{brumme2015first,sharma2014strain,emrem2022london}. The interlayer distance of 6.575  {\AA}  is also in an acceptable agreement with the 6.435 {\AA} and 6.50 {\AA} found in Refs.~\cite{brumme2015first} and \cite{emrem2022london}, respectively. After relaxation of the bilayer, it is compressed and the atoms relax in the out-of-plane direction, forming the wrinkled structure. The strain reported in this paper is due to the change of \ch{W-W} bonds even if it is a complex non-uniform one including variations in the \ch{W-Se} bond lengths depending on which side they are. For example, in the relaxed flat homobilayer the \ch{W-W} distance equals 3.294 {\AA}, and at 2.5 \% compression, we have $\max(\ch{W-W})=3.313$ {\AA} and $\min(\ch{W-W})=3.286$ {\AA} which correspond to 0.5 \% and -0.2 \% strain in the system (please refer to table \ref{table_strain_homobilayer} for more detail). 

The non-uniform strain effects on monolayer TMDCs in different spatial position has been discussed in several theoretical and experimental studies \cite{lee2020switchable,jiang2022analysis,daqiqshirazi2023funneling}. However, its effects on bilayers are not investigated thoroughly. Figure \ref{fig_WSe2_bilayer_20percent} shows the relaxed structure of homobilayer \ch{WSe2} after 20 \% compression. It obtains higher strain at the peaks and valleys of the structure and retains a lower strain in the connecting area. Also, at the peak (valley), the upper layer has lower (higher) strain in comparison to the lower layer. Moreover, the bilayers have two extra degrees of freedom. They do not only relax in the out-of-plane direction, but also their interlayer distances as well as their stacking are affected. These two points are not considered in previous studies to the best of our knowledge. Figure \ref{fig_interlayer_homobilayer} depicts the minimum distance between the splines passing through the positions of the metal atoms on different layers. It is evident that the distance fluctuates between two values. At a lower compression of 2.5\%, the two layers have larger flat areas and the curved regions are smaller. On the other hand, at 20\% compression, the flat areas are smaller and curved regions are larger. Moreover, peak and valley positions are flatter (Figure \ref{fig_homobilayer_curvature}). As it can be seen in Figure \ref{fig_WSe2_bilayer_20percent}, the bilayer stacking is changed and does not correspond to the 2H stacking any longer ($H_h^h$ in the nomenclature of Ref.~\citenum{yu2018}). There are sections in the structure where the chalcogen and metal atoms are not above each other and the stacking is closer to e.g. $H_h^W$. The layers additionally have different curvature along the wrinkles. Hence, each layers' strain state is different at the highest curvature sections (valley and peaks). These  structural variations and the localization of curvature cause the electronic structure alteration that we discuss in the following section.


%

\subsubsection{Electronic structure} 

The flat homobilayer \ch{WSe2} is an indirect band gap semiconductor with its valance band maximum (VBM) at  K and its conduction band minimum (CBM) at Q points of the reciprocal space (see Figure \ref{fig_band_structure_flat_hetero}).  In our calculation, its indirect band gap is 1.159~\unit{eV} (without SOC 1.382~\unit{eV}) and the difference between indirect and direct band gaps is about 174~\unit{meV}. These values are in good agreement with previous investigations \cite{sharma2014strain,zhao2013origin, lu2014mos}.

Figure \ref{fig_band_structure_homobilayer} shows the band structures of the wrinkled \ch{WSe2} homobilayer at different compressions along the $\Gamma$--X line to which the bands of the hexagonal unit cell are folded.

Due to the small difference between direct and indirect band gap of the unstrained \ch{WSe2} homobilayer, we find that the wrinkled structure undergoes an indirect to direct transition even for small strain values of 0.6 \% which occur for 10\% compression and above (Table \ref{table_strain_homobilayer}). This will lead to an increased photoluminescence as shown in several experimental investigations \cite{desai2014strain,uddin2022engineering}. With increasing strain the CBM at the backfolded K point moves down and the Q point moves up to higher energy, therefore, leading to an indirect-to-direct transition \cite{uddin2022engineering} (see also Figure \ref{fig_band_structure_flat_hetero}). Moreover, as seen in Figure \ref{fig_band_structure_homobilayer_noshift}, both valance and conduction bands shift in the same direction with respect to the vacuum level. Yet, the conduction band shifts stronger, leading to the observed reduction of the band gap.

\begin{figure}
\begin{subfigure}{0.99\textwidth}
        \includegraphics[width=1.0\textwidth]{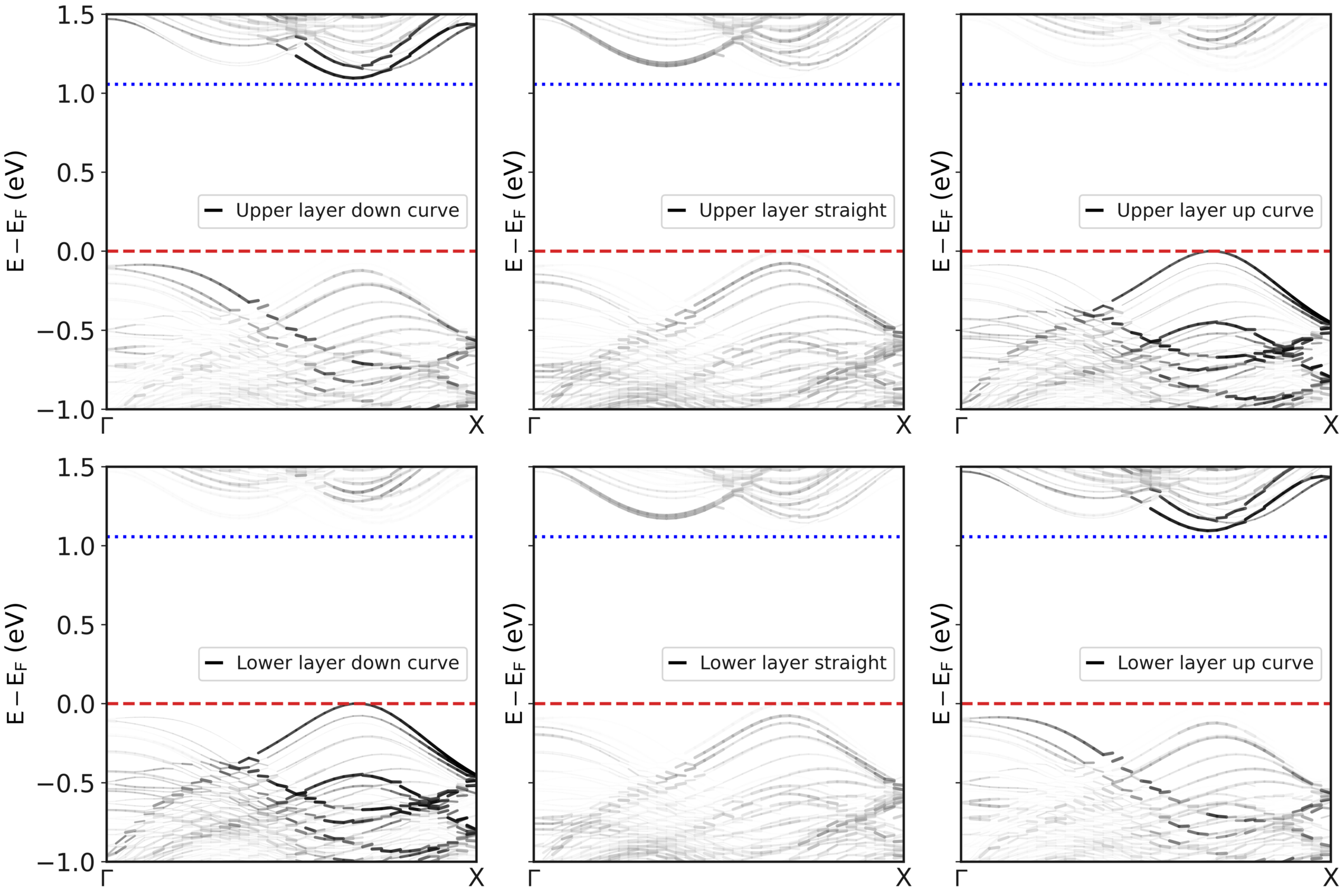}
        \caption{}
\end{subfigure}
\\
\begin{subfigure}{0.5\textwidth}
        \includegraphics[width=1.0\textwidth]{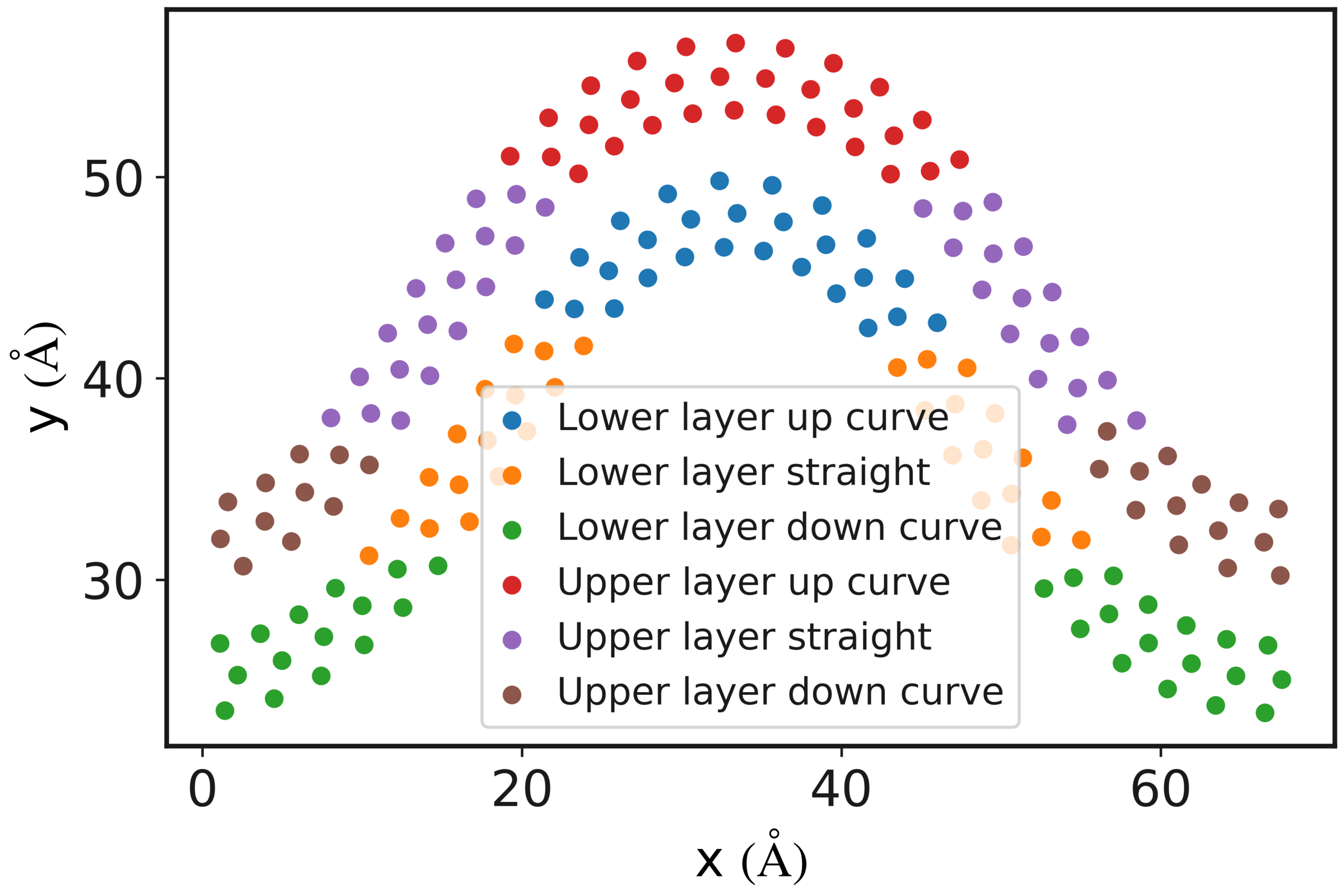}
        \caption{}
\end{subfigure}
    \caption{Localization of states on different regions and layers. a) Band structures of \ch{WSe2} homobilayer at 20\% compression projected to different sections of the wrinkle as sketched in b).}
    \label{fig_band_bilayer_WSe2_different_layer_all_different_sections}
\end{figure}

Figure \ref{fig_band_bilayer_WSe2_different_layer_all_different_sections} shows the contribution of different sections of the wrinkles to the band structure of the 20\%-compressed bilayer \ch{WSe2}.
Due to the (approximate) inversion center at the inflection point of the wrinkle, CBM and VBM are both formed  by two energy-degenerate bands which are localized in either of the two curved regions. Focusing on the curved area at $\approx0.5\mathrm{b}$ (see Figure \ref{fig_homobilayer_curvature}), the CBM is mostly contributed from the states located on the lower layer, and the VBM is more localized on the upper layer. Table \ref{table_contribution_homobilayer_sections} shows the quantitative values of the contribution to the band edges for different sections of the 20\% compressed wrinkle (the sections are indicated in Figure \ref{fig_band_bilayer_WSe2_different_layer_all_different_sections}). For this case, almost 80\% of the valence band (VBM-1) is localized on the upper layer and more than 90\% of the conduction band (CBM) is localized on the lower layer.  

The non-uniform strain thus effectively leads not only to the localization of the states and hence a funneling of the excitons to the strained regions but also to the formation of \textbf{inter}layer excitons in the \ch{WSe2} homobilayer. This separation of the electron and hole wave functions on different layers is significant, as it will lead to a longer lifetime of the excitons. Yet, in contrast to the case of heterobilayers, the energy of the interlayer excitons will be similar to those of the intralayer excitons -- as can be seen in Figure \ref{fig_band_bilayer_WSe2_different_layer_all_different_sections} also for the intralayer exciton the electron and hole are localized in different regions of the layer however further away from each other. 
Table \ref{table_band_gap_homo_sections} additionally summarizes the local band gaps in different sections of the homobilayer at different compressions.  

\begin{figure}
    \centering
    \includegraphics[width=0.32\textwidth]{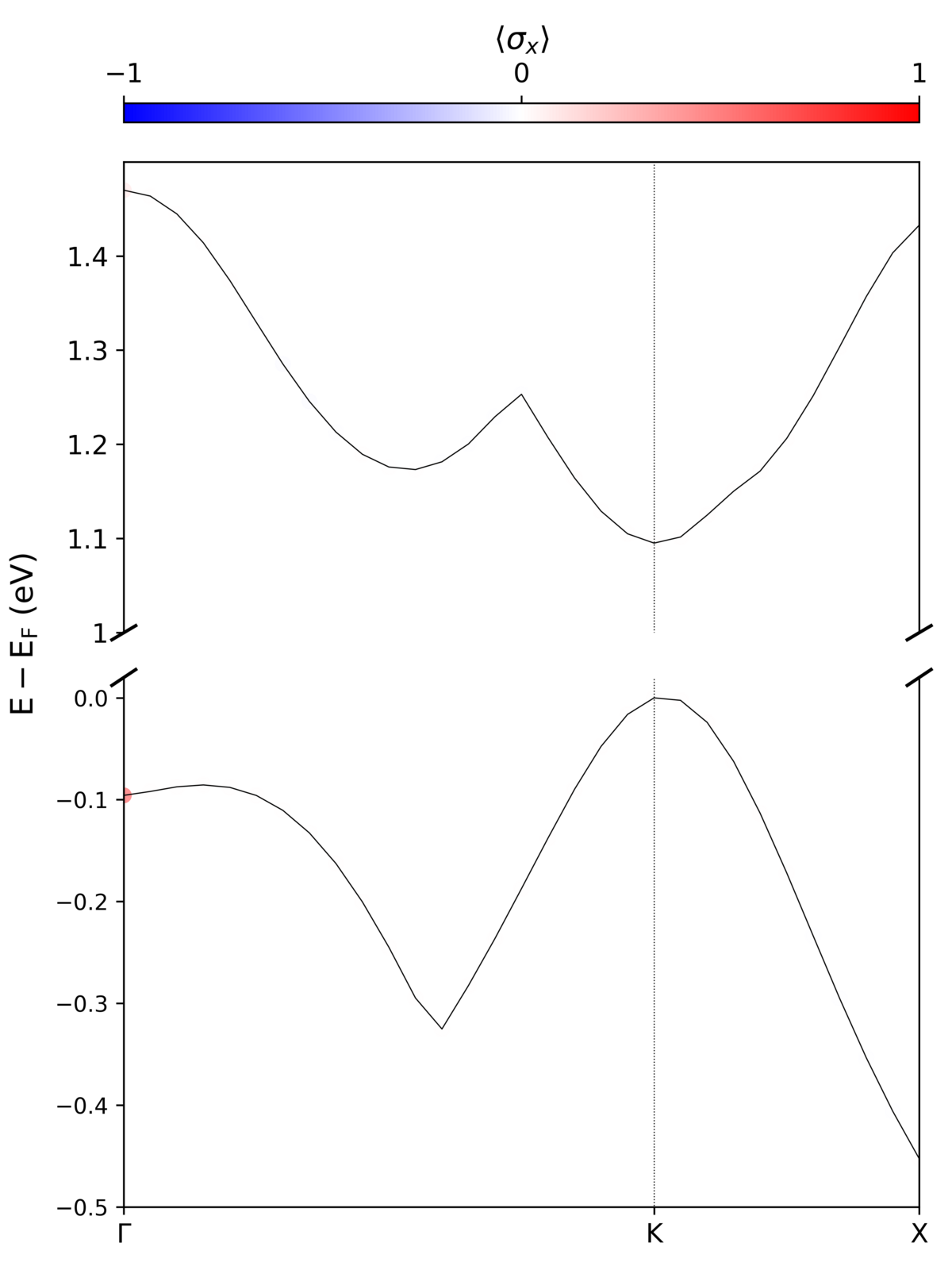}
    \includegraphics[width=0.32\textwidth]{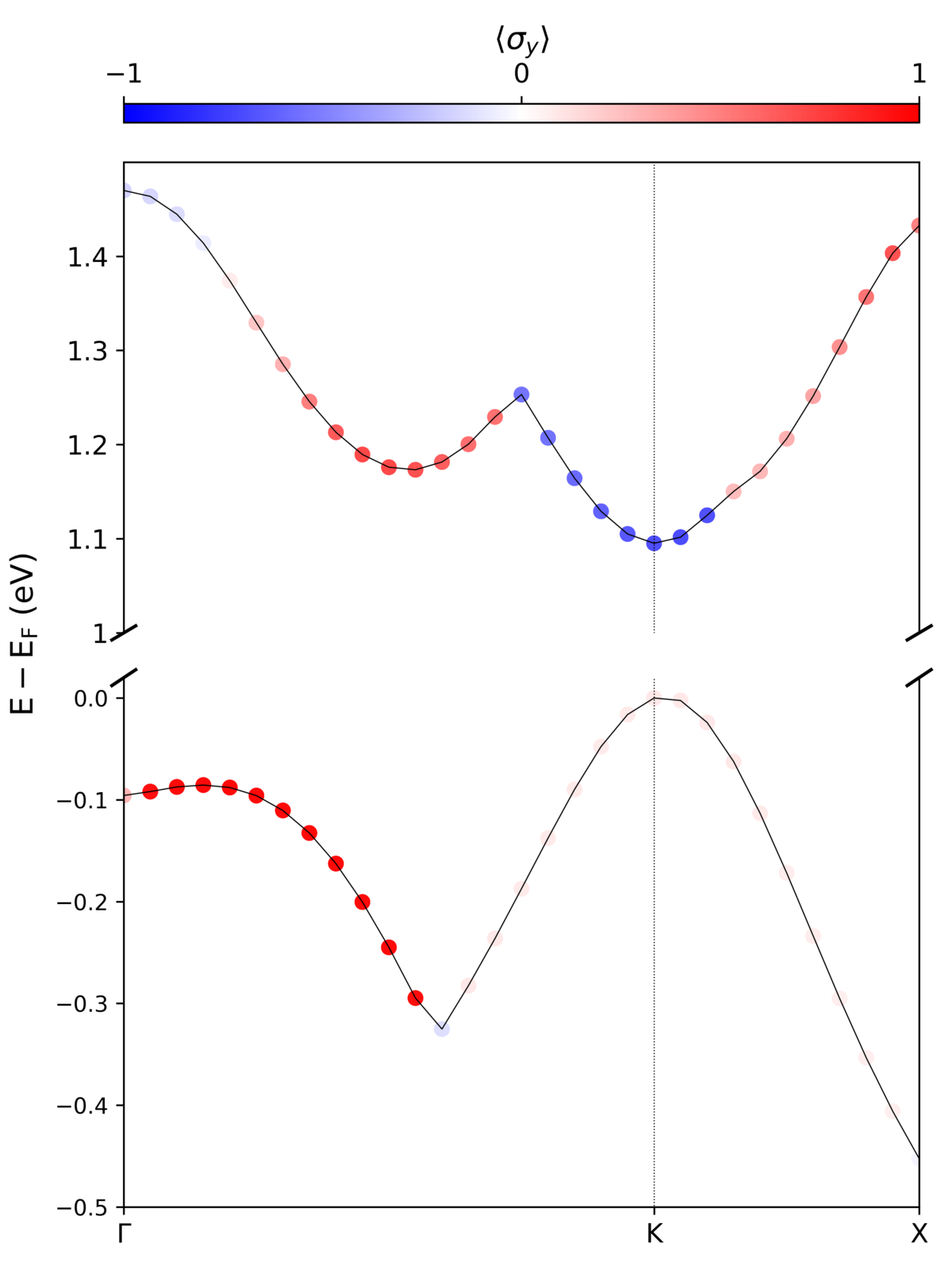}
    \includegraphics[width=0.32\textwidth]{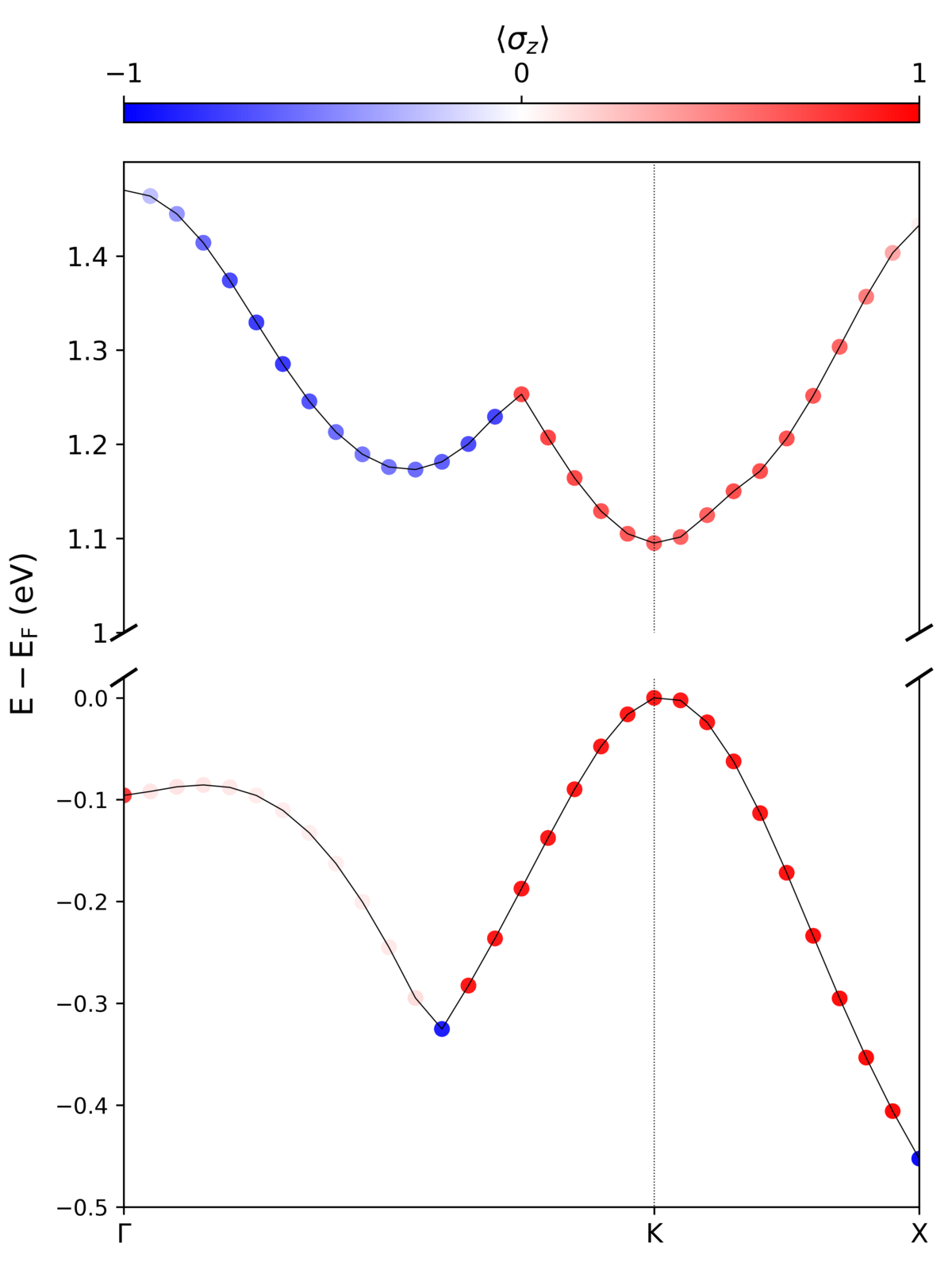}
      \includegraphics[width=0.32\textwidth]{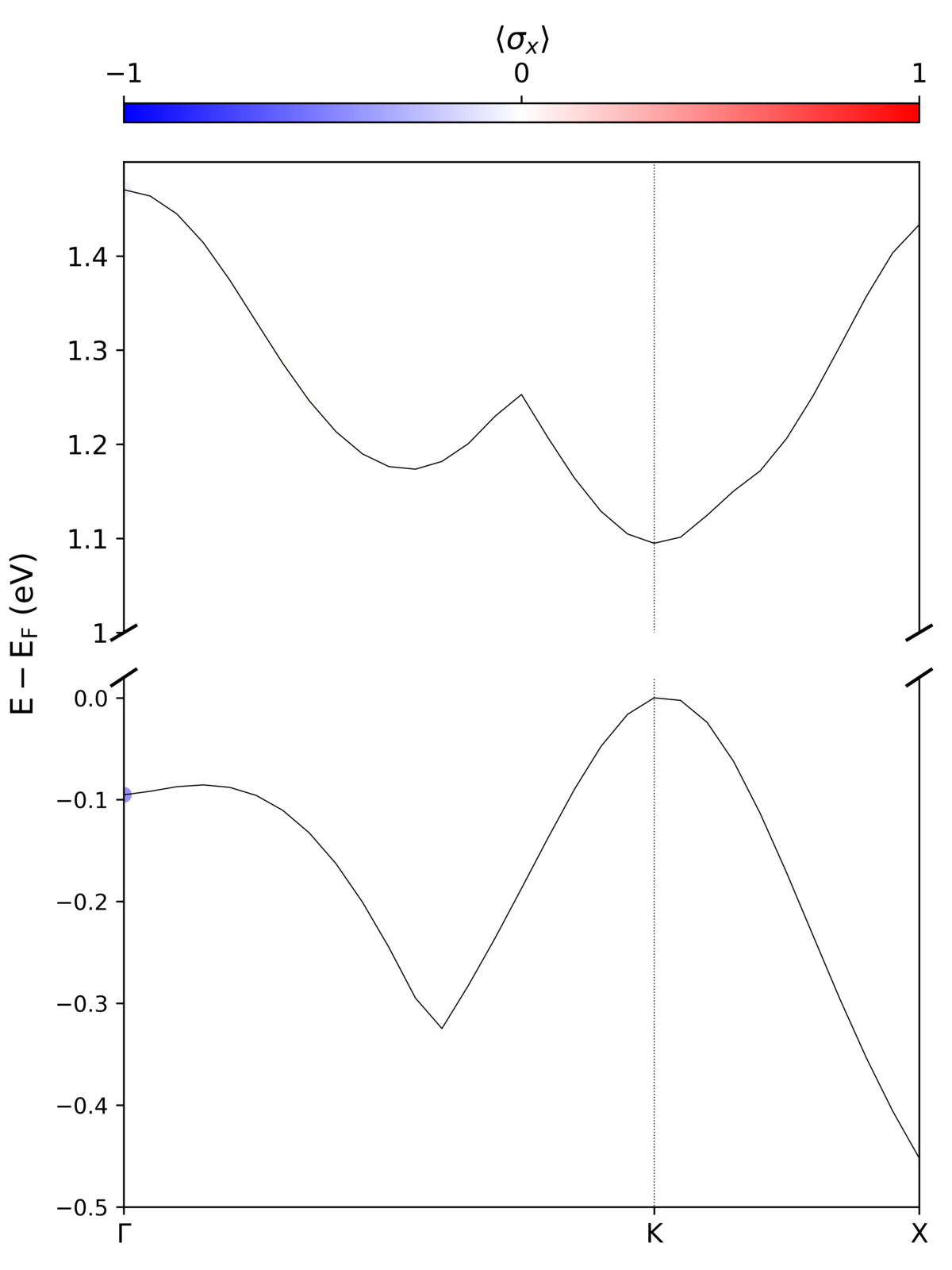}
    \includegraphics[width=0.32\textwidth]{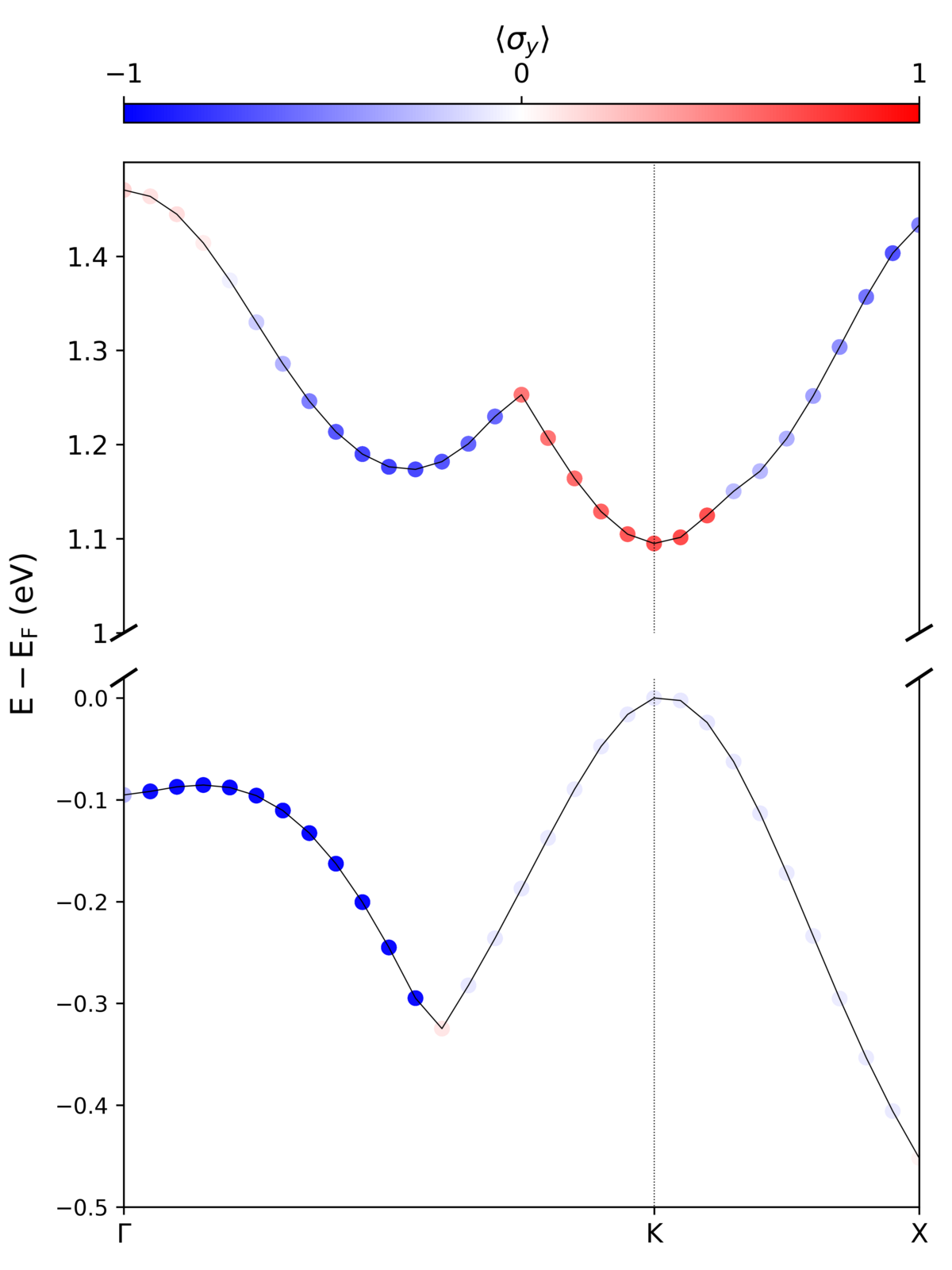}
    \includegraphics[width=0.32\textwidth]{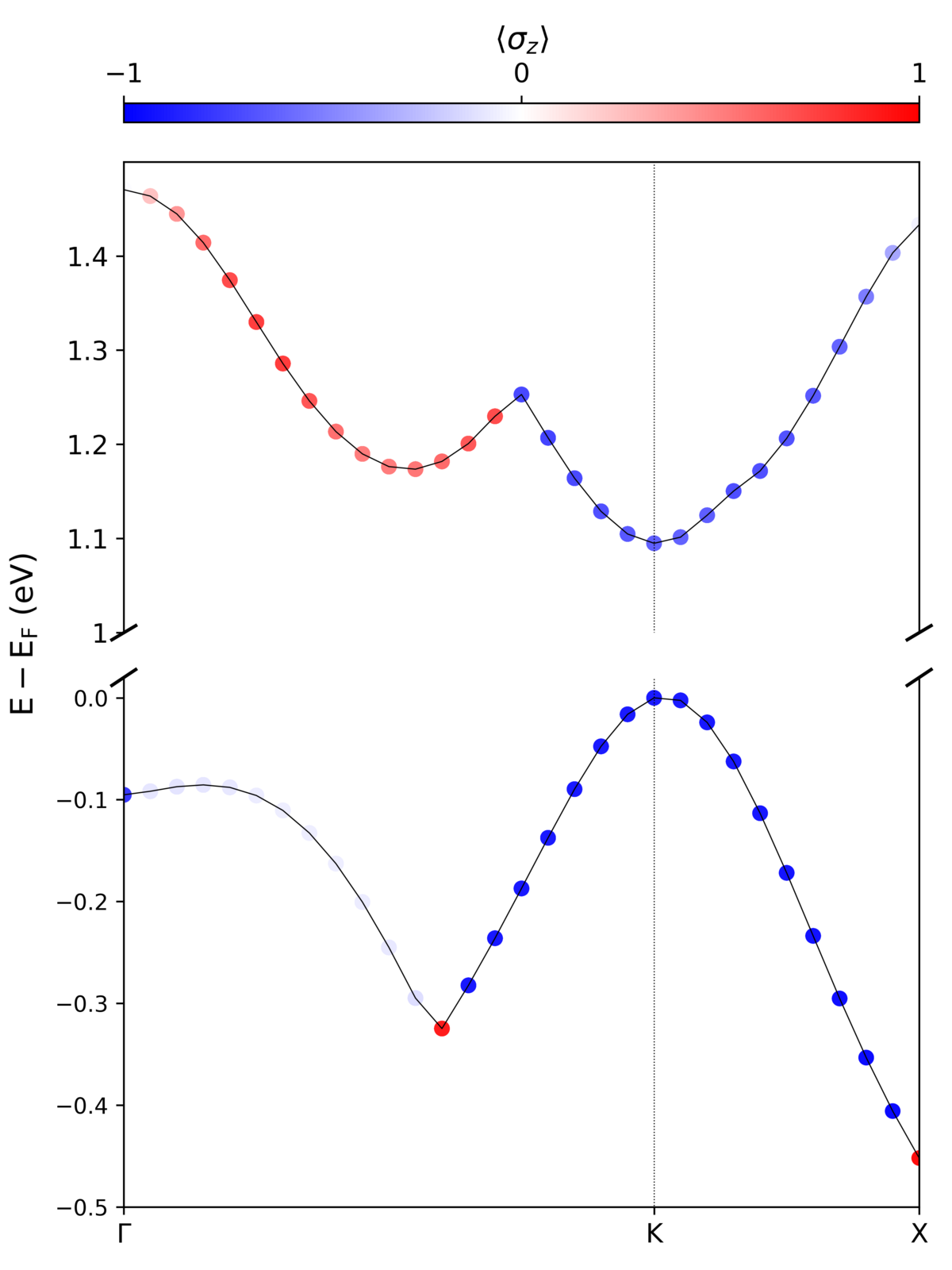}
    \caption{Valence and conduction bands of the homobilayer \ch{WSe2} are made of two energy-degenerate bands with opposite spin directions, each localized on a different layer \cite{spin_layer_locking}. The first row depicts the expectation values $\langle\sigma_i\rangle$, i.e. the spin texture, of VB and CB+1 states localized in the same curved region but on different layers. The second row shows their energy degenerate partners (VB-1 and CB) localized on the other curved region. Plots are for the homobilayer wrinkle at 20\% compression.}
    \label{fig_spin_texture_homo_20_individual_bands}
\end{figure}

For small strain, the backfolded bands of the Q point (which is still lower in energy) hide the CBM at K. Additionally, the CBM and VBM are degenerate states, however plotting the spin textures of the VB-1, VB+1, CB, CB+1 individually (see Figure \ref{fig_spin_texture_homo_20_individual_bands}) unravels that the smallest direct gap at K which is dark in monolayer \ch{WSe2} (spin forbidden) becomes brighter after compression.
It should be noted that this brightening is different from the brightening of the excitons due to the mixing with defect states\cite{hernandez2022strain} and that it is directly linked to the formation of the interlayer exciton -- the spin-layer locking \cite{spin_layer_locking} leads to a spin-allowed transition if the electron and hole are localized on different layers instead of being on the same. Figures \ref{fig_spin_texture_homo_2.5} and \ref{fig_spin_texture_homo_20} show the expectation value of the Pauli matrices of all states close to CBM and VBM for 2.5\% and 20\% compression.

Furthermore, the VBs near the $\Gamma$ point show a Rashba-like splitting due to the curvature-induced electric field (see Figure \ref{fig_band_structure_homobilayer_20_zoom} for a magnified view) which is, however, smaller than the splitting in the wrinkled monolayer \cite{daqiqshirazi2023funneling}. This reduction of the splitting is due to the reduction of the out-of-plane dipole moments and the curvature (Table \ref{table_structural_homobilayer}) at the outer surface.  


The change of the metal atoms in one of the layers of a bilayer can change the properties of the system, not only by breaking the inversion symmetry -- TMDC heterobilayers are particularly interesting because they can host interlayer excitons due to their type-II band alignment. Therefore, in the next section we consider the wrinkling of \ch{WSe2/MoSe2} heterobilayers and investigate the alterations of the electronic structure due to non-uniform strain in the system.


\subsection{\ch{WSe2/MoSe2} heterobilayer wrinkle}

The structure of the heterobilayer is created by replacing \ch{W} in the lower layer with \ch{Mo} atoms. The heterobilayer wrinkles then are created by the same procedure as the homobilayers. In the following section, we first discuss the geometrical variation in the structures due to the wrinkling and in the subsequent section we explain how this changes the electronic structure.

\begin{figure}
    \begin{subfigure}{0.49\textwidth}
        \includegraphics[width=0.99\linewidth]{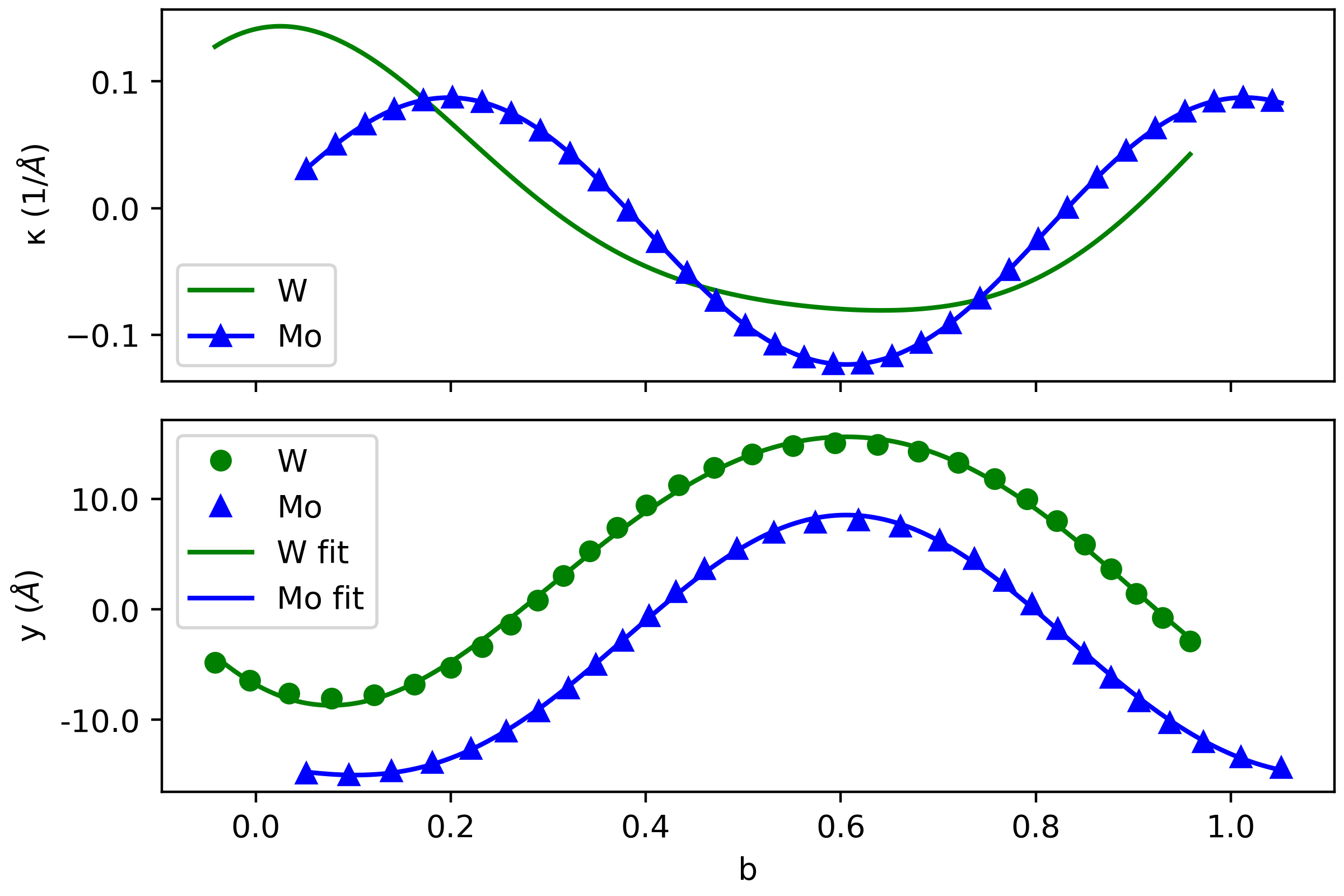}
        \caption{  \label{fig_heterobilayer_curvature}}
    \end{subfigure}
    \begin{subfigure}{0.49\textwidth}
        \includegraphics[width=1\textwidth]{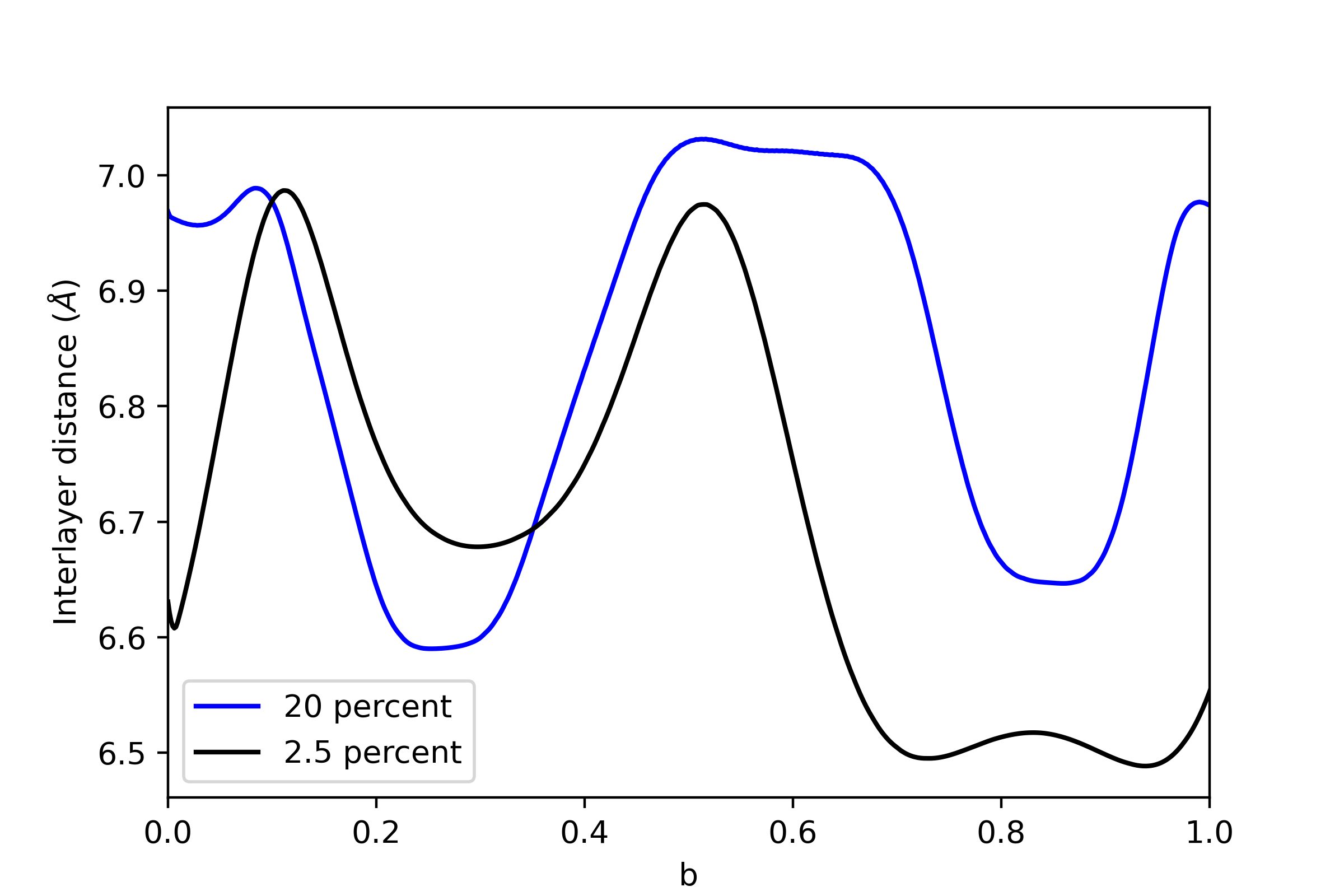}
\caption{ }
    \end{subfigure}
     \caption{Structural changes due to wrinkling of the heterobilayer \ch{WSe2$/$MoSe2}. a) (upper panel) Curvature, $\kappa$, of the fit function of the metal atoms position of wrinkled \ch{WSe2$/$MoSe2} (lower panel) with a compression of 20\% b) Interlayer distance of the \ch{WSe2$/$MoSe2} heterobilayer wrinkle at two different compression values of 2.5\% and 20\% -- please note that the wrinkles have a different wave length and the distance is shown w.r.t. the fixed cell axis.\label{fig_interlayer_heterobilayer}}
  
\end{figure}

\subsubsection{Atomic structure}

The heterobilayer relaxation is quite similar to the homobilayer \ch{WSe2}. Table \ref{tabel_strain_heterobilayer} shows the maximum and minimum of the strain in each layer of the heterobilayer. The strain in these structures is of a complex nature; the layers adjust the interlayer distance and the stacking. Figure \ref{fig_interlayer_heterobilayer} demonstrates the variation of the distance of the metal atoms of the heterobilayer \ch{WSe2$/$MoSe2} in the two cases of 2.5\% and 20\% compressions. It is obvious that the distance between the layer is fluctuating along the wrinkle. Interestingly, at lower strain (i.e. lower compression), the variance of the interlayer distance is larger than in the cases with higher compression. Moreover, a similar stacking variation as in the homobilayer occurs in the heterobilayer.

\subsubsection{Electronic structure}

The flat heterobilayer \ch{WSe2$/$MoSe2} is also an indirect band gap semiconductor (band gap equals 1.005~\unit{eV} and 1.229~\unit{eV} with and without SOC, respectively, \textit{cf.} Fig.~\ref{fig_band_structure_flat_hetero}) with the VBM and CBM localized at the K and Q point, respectively (experimental indirect band gap 1.31~eV\cite{khalil2022hybridization}). The difference between direct and indirect band gap (76 \unit{meV}) is smaller than in the case of the homobilayer (174 \unit{meV}). Having an indirect band gap close to the bright exciton energy can lead to non-radiative losses in devices based on this material \cite{uddin2022engineering}. Additionally, in \ch{WSe2$/$MoSe2} heterobilayers, the CB and VB are localized on different layers due to the type-II band alignment. This leads to an increased lifetime of the corresponding \textbf{inter}layer excitons in comparison to the \textbf{intra}layer excitons. Moreover, heterostructures have an intrinsic electric dipole that changes their band structures \cite{lu2014mos} -- it can for example lift the degeneracy of states as well as couple with the intrinsic spin momentum of the system.
Figure \ref{fig_band_structure_heterostructure} shows the band structure of the heterobilayer of \ch{WSe2$/$MoSe2} for different compressions. At 2.5\% compression the band structure is still very similar to the flat system. At this compression, the maximum strain in the structure is 0.6\%. (cf. Table \ref{tabel_strain_heterobilayer}). VBM and CBM occur at the backfolded K point, however, the difference between the direct band gap and indirect band gap (at the backfolded Q point) is only 13 \unit{meV}. By the increase of the strain in the system, it goes through an indirect to direct transition which will lead to a stronger photoluminescence of the system. Moreover, the direct band gap reduces from 1.082 \unit{eV} to  0.931 \unit{eV}.

 Although the compressive strain drives the system to transit to a direct band gap \cite{sharma2014strain}, the out-of-plane relaxation reduces the strain present in the system. For example, in Figure \ref{fig_band_structure_5percent_w_wo_wrinkle} of the SI the 5\% compressed system with and without wrinkle formation is shown. The wrinkled system still has an indirect band gap in contrast to the direct band gap of the uniaxial strained system.

\begin{figure}
    \includegraphics[width=1.0\textwidth]{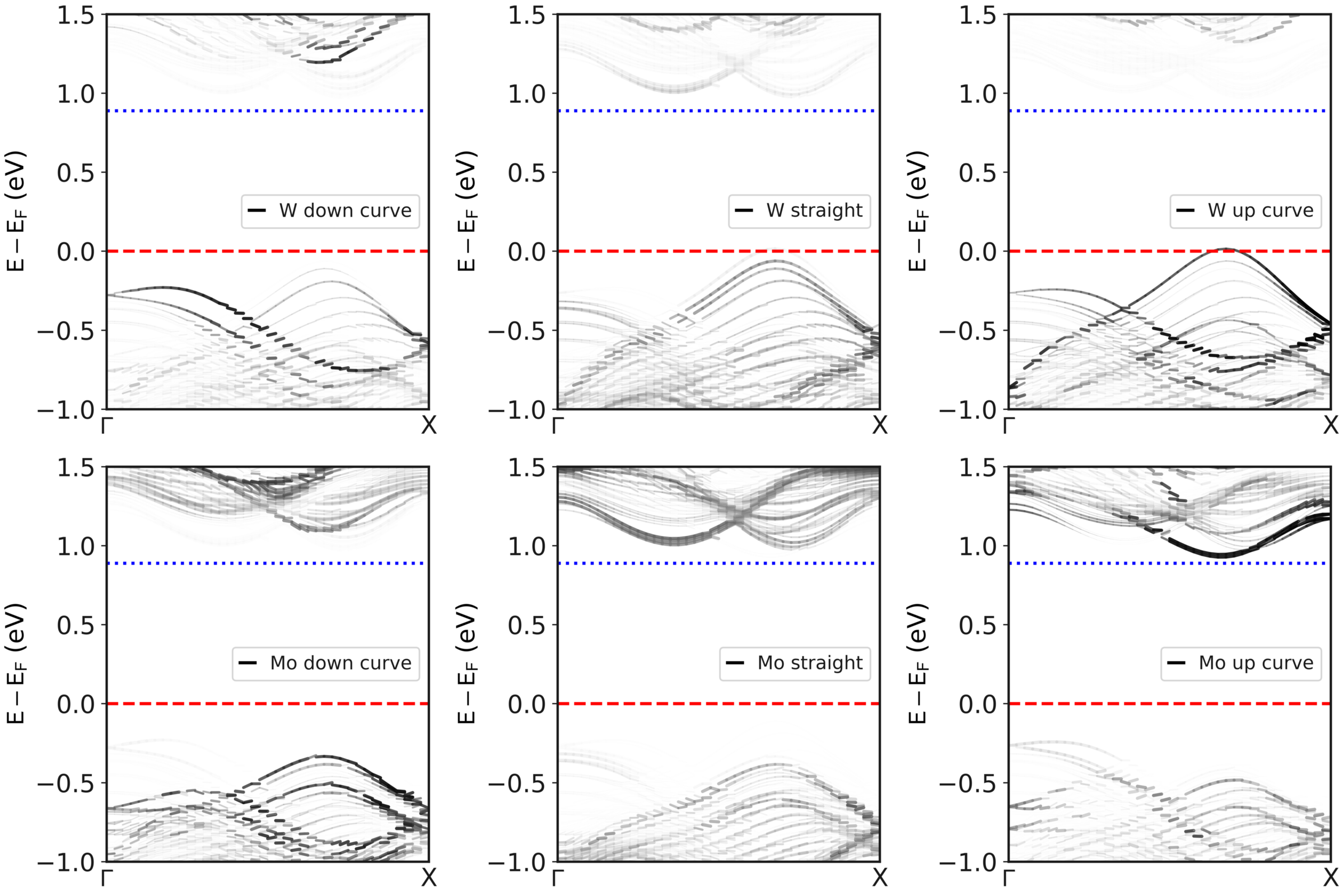}
    \includegraphics[width=0.5\textwidth]{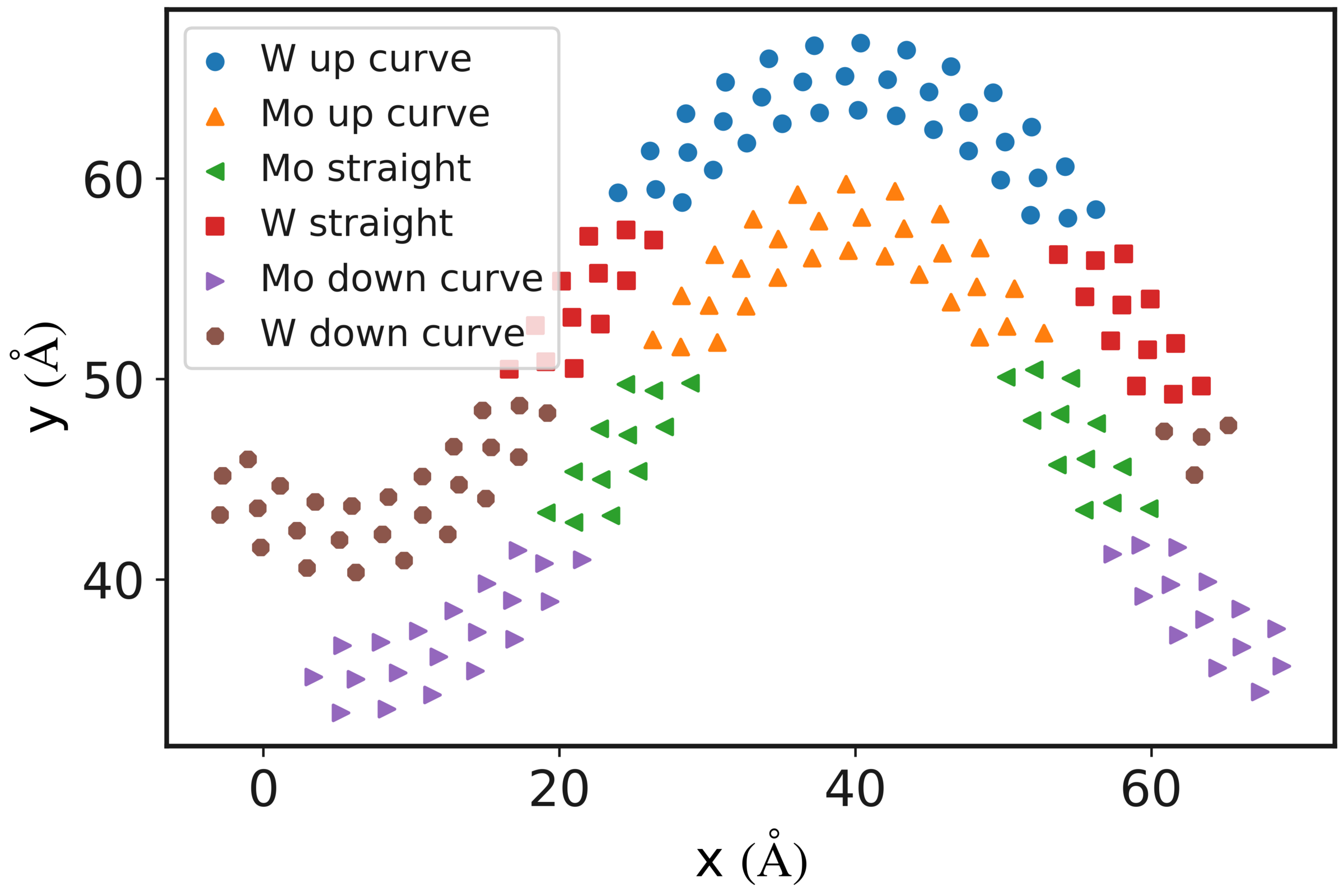}
    \caption{Band structure of \ch{WSe2$/$MoSe2} heterobilayer at 20\% compression projected to metal atom in different sections of the wrinkle, the location of VBM (red) and CBM (blue) are also indicated to help the eye, additionally (below) the position of contributing atoms to the band projection is given.}
    \label{fig_band_heterobilayer_WSe2MoSe2_different_layer_different_sections}
\end{figure}

Figure \ref{fig_band_heterobilayer_WSe2MoSe2_different_layer_different_sections} shows the band structure projected on different sections of the wrinkled \ch{WSe2$/$MoSe2} at 20 \% compression. It can be inferred that the VBM is mostly made of \ch{W} states at the upper curved area, whereas the CBM originates from the \ch{Mo} states in this section. This is particularly important as the spatial separation of electron and holes can enhance the lifetime of the excitons which will be relevant for the effective employment of these structures in devices. For reference, Figure~\ref{fig_band_heterobilayer_WSe2MoSe2_different_layer_different_sections_2.5} shows the projected bands on different sections at 2.5\% compression. Strain dependence of the localization is also apparent from the Table \ref{table_band_gap_hetero_section} which summarizes the band gap at different section of the heterobilayers for different compressions.

 Figure \ref{fig_spin_texture_2.5} and \ref{fig_spin_texture_20} show the expectation values of the Pauli matrices for the cases with 2.5\% and 20\% of compression. At the backfolded K point, the CB and CB+1 state 
 become spin mixed. Due to the conduction band and valence band possessing different orbital magnetic quantum numbers, $m_l$, the allowed excitations in TMDC monolayers are those where the states have the same spin state \cite{wang2018colloquium}. The spin mixing we find in the wrinkled heterobilayers can be the reason why some researchers, in experiment, observed the dark exciton states \cite{hernandez2022strain}.
Furthermore, the curvature will lead to a mixing of in-plane with out-of-plane states similar to the case of carbon nanotubes\cite{klinovaja2011}. This mixing of states with different $m_l$ will also influence the optical selection rules and thus change the photoluminescence at the strained regions at the peaks.
For a general overview of the $z$ component of the expectation value of Pauli matrices at various strain states, please refer to the Figures \ref{fig_spin_texture_cb} and \ref{fig_spin_texture_vb} of the SI. 

Heterobilayers have an intrinsic dipole moment that breaks the symmetry of the system and leads to momentum direction splitting (i.e. Rashba-like splitting \cite{bihlmayer2022rashba}). 
As it can be inferred from Figure \ref{fig_band_structure_heterostructure}, the splitting increases with strain. Unfortunately, unlike for wrinkled TMDC monolayers \cite{daqiqshirazi2023funneling}, we have not observed any case with the splitting in the heterostructure to occur above the VBM energy at K. In this way, it is difficult to observe this splitting in the experiment. Nevertheless, the presence of such splitting is significant, as it can be further modified by external electric fields \cite{guo2023switching} or proximity effects \cite{szalowski2023spin}.

In order to see the environment effect (i.e. the second layer) on Rashba-like splitting of the monolayer TMDC, we also considered the separated wrinkled layers by keeping their shape profiles. Looking at Figure \ref{fig_layer_sperated_heterobilayer12.5} in the SI, the Rashba-like splitting of each layer is larger than in the combined structure. Hence, we conclude that the interaction of the two layers reduces the Rashba-like splitting in the TMDC heterostructures which may be due to screening effects.


\begin{figure}
   \centering
   \includegraphics[width=0.99\linewidth]{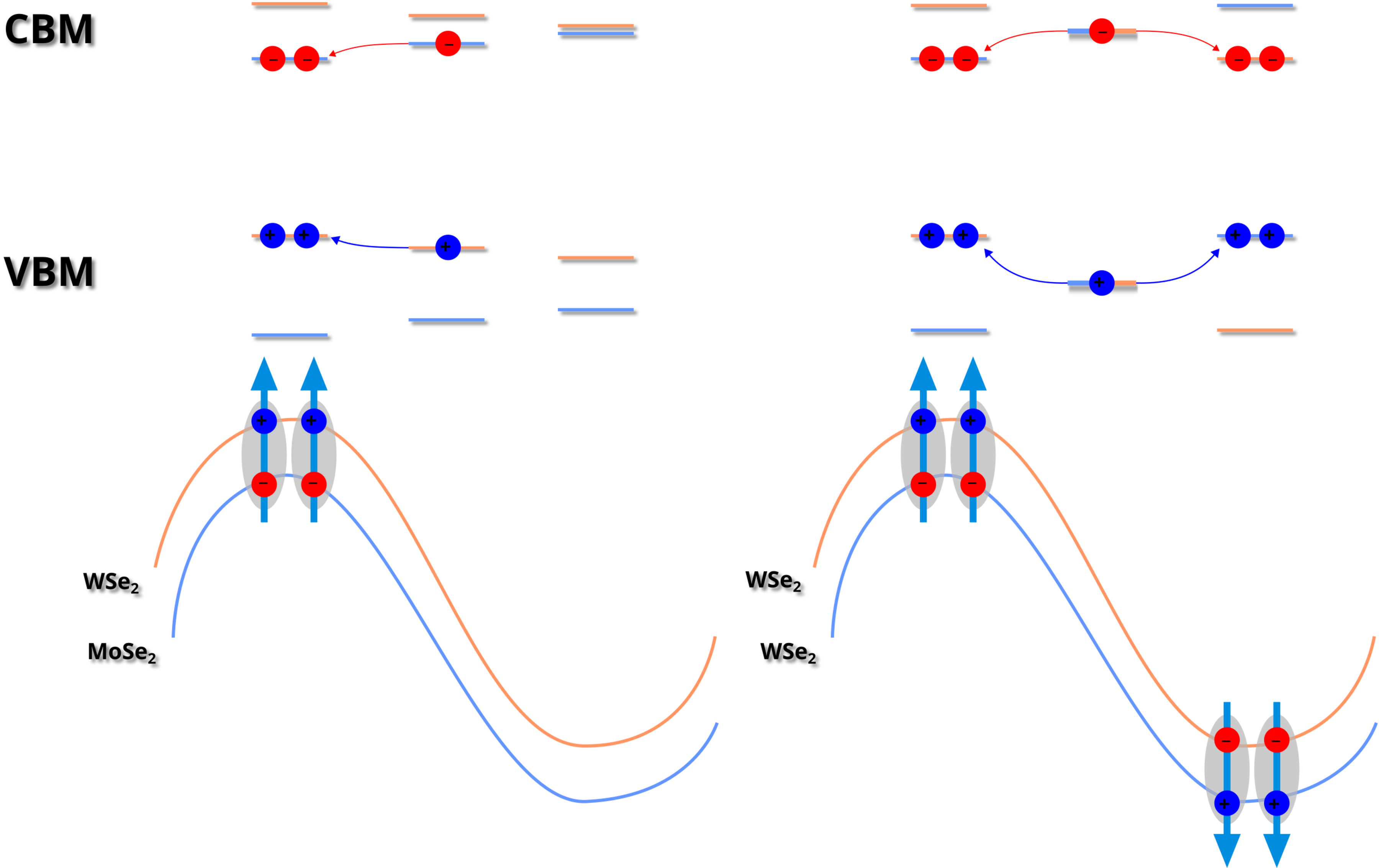}
    \caption{Electron and holes are localized in different spatial positions of the bilayer system due to non-uniform strain, giving rise to the formation of interlayer excitons in homobilayer \ch{WSe2} and localized excitons in heterobilayer \ch{MoS2-WSe2}. Left) heterobilayer \ch{MoS2-WSe2} and right) homobilayer \ch{WSe2}, the VBM and CBM edges are depicted above the spatial positions of the wrinkle shown below it. The excitons on the peaks of the wrinkles with arrows showing their dipoles orientation are also sketched.}
    \label{fig_state_summary}
\end{figure}

\section{Discussion}
Although uniform strain is known as a tool to modify the exciton dynamics in TMDCs, the role of non-uniform strain on excitons is more elusive. We have shown here that non-uniform strain can lead to unique localization of the band-edge states in bilayer TMDCs. This will result in rather unexpected properties of the excitons in the homo- and heterobilayers which are shown in Figure~\ref{fig_state_summary} and can be summarized as follows:
\begin{itemize}
    \item localization of the states due to non-uniform strain leads to formation of localized bright interlayer excitons in homobilayers TMDCs 
    \item interlayer excitons in wrinkled/inhomogeneously-strained homobilayers localize in high-strain regions
    \item intralayer excitons in wrinkled/inhomogeneously-strained homobilayers will have very low oscillator strength due to the larger electron-hole separation
    \item interlayer excitons in wrinkled/inhomogeneously-strained heterobilayers localize in the area with the largest strain in the \ch{MoSe2} layer
    \item intralayer excitons will also have lower oscillator strength due to the larger electron-hole separation
\end{itemize}
This will influence the exciton diffusion in both systems. In homobilayers strain will lead to an increased photoluminescence due to the brightening of the formerly dark states (spin forbidden transitions) and due to the indirect-direct band-gap transition. Furthermore, the interlayer excitons which will form have an out-of-plane dipole in contrast to the intralayer excitons in monolayers. Thus, they will repel each other, which will lead to an increased diffusion. However, also the intralayer excitons (if they form even if the Coulomb interaction will be smaller due to larger separation) have now a preferred direction for the dipole moment which is from one curved region to the other. This will again lead to a repulsion. For heterobilayers the preferred formation of interlayer excitons in the high-strain areas of \ch{MoSe2} will lead to a higher exciton density in this region and accordingly to a larger diffusion. Intralayer excitons will again have a preferred orientation for the dipole moment. Yet, this will be opposite for \ch{MoSe2} and \ch{WSe2} -- where we find the electrons in \ch{WSe2}, there are the holes in \ch{MoSe2} and vice versa. This will again lead to different behaviors for the intralayer excitons which in turn will also influence the interlayer ones. 
Even though the exciton behavior of the wrinkled TMDC bilayers was elucidated in this research, the strain fields in various systems such as bubbles, pre--strained surfaces and even folds due to pre--strained substrate should have similar behaviors. 

This work paves the way to the effective strain-induced manipulation of excitons in interesting applications such as exciton transistors or single quantum emitters.

\section{Acknowledgments}

We thank Agnieszka Kuc and Alexey Chernikov for fruitful discussions. The authors gratefully acknowledge the computing time made available to them on the high-performance computer at the NHR Center of TU Dresden and on the high-performance computers Noctua 2 at the NHR Center PC2. These are funded by the German Federal Ministry of Education and Research and the state governments participating on the basis of the resolutions of the GWK for the national high-performance computing at universities (www.nhr-verein.de/unsere-partner). We furthermore acknowledge the Gauss Centre for Supercomputing e.V. (GCS -- www.gauss-centre.eu) for funding this project by providing computing time through the John von Neumann Institute for Computing on the GCS Supercomputer JUWELS \cite{alvarez2021juwels} at J\"ulich Supercomputing Centre. We would like to thank the German Science Foundation for supporting this work via the SFB 1415, Project ID No. 417590517.

\section{Methods \label{section_method}}
The wrinkles of homobilayer \ch{WSe2} and heterobilayer \ch{WSe2$/$MoSe2} have been investigated by means of density functional theory as implemented in the all-electron code FHI-aims \cite{blum2009ab}.  The Perdew-Burke-Ernzerhof exchange correlation functional \cite{perdew1996generalized}, Tkatchenko-Scheffler dispersion correction \cite{tkatchenko2009accurate} and a non-self-consistent SOC \cite{huhn2017one} have been utilized to calculate the electronic structure of the systems. The initial structures were created with a 15$\times$1$\times$1 supercell of the relaxed rectangular unit cell of bulk-like stacked 2H homo- or heterobilayers with the longer in-plane supercell vector being along the armchair direction. For the heterobilayer we placed \ch{MoSe2} in the \ch{WSe2} lattice which results in a minor mismatch of 0.15\%. Subsequently, the flat structures have been fully relaxed to forces and pressures below 0.001 \unit{eV/\text{\AA}} and 0.1 bar, respectively. Then the supercell was compressed along the armchair direction and the atomic positions have been relaxed again keeping the reduced lattice direction constant to retain the strain. The systems usually deform in the out-of-plane direction and wrinkles are formed. However, it is worthwhile to mention that even with perturbing the positions of some atoms, few systems remained in the local minimum of a flat structure. Therefore, for these systems, we started from systems with larger compression which show the wrinkle formation and increased again the lattice parameter. Tables \ref{table_structural_homobilayer} and \ref{table_structural_heterobilayer} summarize the structural parameters of wrinkled homo- and heterobilayer systems in this investigation. Mulliken band structures and spin textures were obtained from the final wrinkled structures. We chose the armchair direction as the longer supercell axis due to the band folding -- compression and wrinkle formation along the zigzag direction leads to many bands being folded back to the $\Gamma$ point which obscures the identification of the valance band maximum. The interested reader might refer to the supporting material of Ref.~\cite{daqiqshirazi2023funneling} for further discussion.


\newpage
\section*{References}
\bibliography{Reference.bib}
\newpage
\appendix
\setcounter{figure}{0}
\setcounter{table}{0}
\renewcommand{\figurename}{Figure}
\renewcommand{\thefigure}{S\arabic{figure}}
\renewcommand{\thetable}{S\arabic{table}}
\newpage
\section{Supporting Information}
\subsection{Fitting Procedure}
The location of metal atoms is fitted by a trigonometric function as follows:
\begin{equation}
    f(x,a_0,a_1,a_2,a_3,a_4,a_5)=a_0+a_1 \times sin(a_2x)+a_3 \times sin(a_4x+a_5)
\end{equation}
Then, the first and second derivatives are calculated analytically, and the curvature is calculated by,
\begin{equation}
    \kappa=\frac{y''}{(1+y'^2)^\frac{3}{2}}
\end{equation}
\newpage
\FloatBarrier
\section*{\ch{WSe2} bilayer wrinkle}

\begin{figure}
    \includegraphics[width=0.49\textwidth]{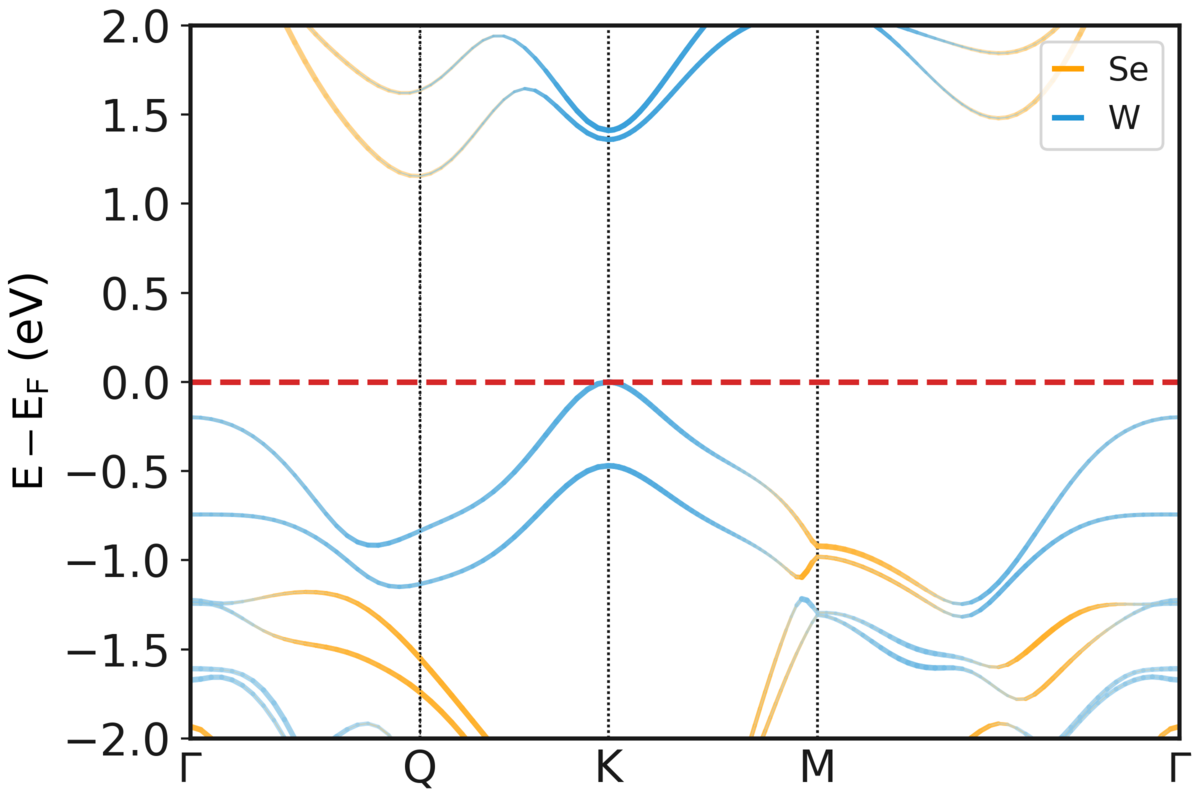}
    \includegraphics[width=0.49\textwidth]{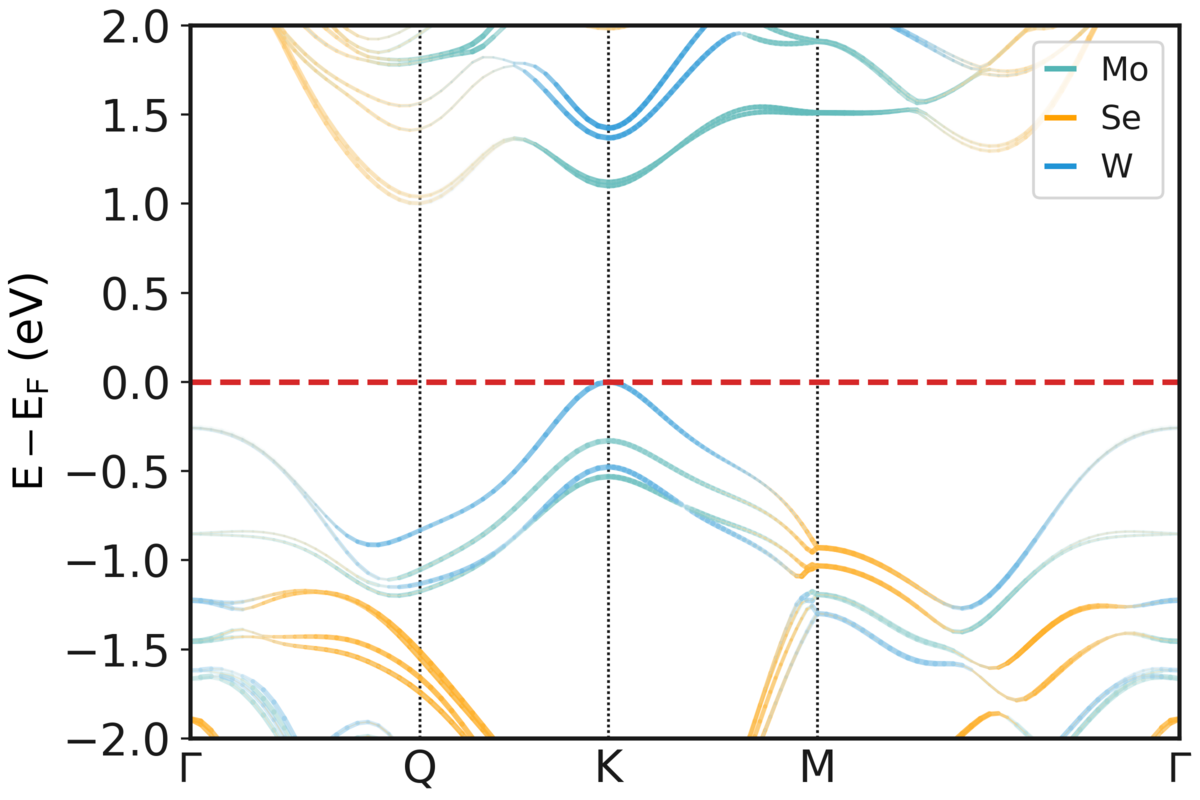}
    \caption{The  Mullikan projected band structure of flat unstrained homobilayer \ch{WSe2} and \ch{WSe2$/$MoSe2} heterobilayer in hexagonal unit cell. Both structures are indirect band gap with heterobilayer having smaller difference between direct and indirect band gaps}
    \label{fig_band_structure_flat_hetero}
\end{figure}

\begin{table}
\caption{Maximum and minimum strain at each layer of the wrinkled homobilayer \ch{WSe2} structure, defined as $\epsilon=\frac{d-d_\mathrm{flat}}{d_\mathrm{flat}}$, where $d$ and $d_\mathrm{flat}$ are the \ch{W-W} distance in the strained and flat, unstrained system, respectively. \label{table_strain_homobilayer}}
\begin{adjustbox}{max width=\textwidth}
\begin{tabular}{ c|c|c|c|c} 
  \textbf{Compression}     & \textbf{Maximum strain upper layer}  & \textbf{Minimum strain upper layer} & \textbf{Maximum strain lower layer} & \textbf{Minimum strain lower layer} \\ \hline
2.5          & 0.0034   & -0.0032    &  0.0033    & -0.0032\\
5            & 0.0048  &-0.0022  & 0.0050  &  -0.0022 \\
6            & 0.0037  & -0.0037 & 0.0035  & -0.0038 \\
7.5          & 0.0034  & -0.0049 & 0.0033  & -0.0049   \\
10           & 0.0058  & -0.0044 &  0.0059 & -0.0044 \\
12.5         & 0.0077  & -0.0063 &  0.0082 & -0.0062 \\
15           & 0.0135  & -0.0018 & 0.0131  & -0.0018  \\
17.5         &0.0157   & -0.0018 &  0.0162 & -0.0017 \\
20           & 0.0187  & -0.0023 & 0.0187  & -0.0023 \\
\end{tabular}
\end{adjustbox}
\end{table}

\begin{table}
\caption{Contribution of the atoms of different sections of the 20\% compressed \ch{WSe2} homobilayer wrinkle to different bands at the reciprocal lattice K. \label{table_contribution_homobilayer_sections} and please refer to the Figure \ref{fig_band_bilayer_WSe2_different_layer_all_different_sections} for the sections location.}
\begin{adjustbox}{max width=\textwidth}
\begin{tabular}{ c|c|c|c|c|c|c} 
   \textbf{Band index}&  \textbf{Upper layer straight} &  \textbf{Lower layer straight} &  \textbf{Upper layer up curve} & \textbf{Upper layer down curve} & \textbf{Lower layer up curve} & \textbf{Lower layer down curve} \\ \hline
  \textbf{CBM+1}          &  0.087             &     0.001          &  0.002           & 0.906               &    0.000              & 0.003              \\
  \textbf{CBM}            &  0.001             &     0.08           &  0.003           & 0.                  &    0.913              & 0.003              \\
  \textbf{VBM}            &  0.001             &    0.15            &  0.000              & 0.003               &    0.015              & 0.831               \\ 
  \textbf{VBM-1}          &  0.182             &    0.001           &  0.799           & 0.015               &    0.003              & 0.                  \\  

\end{tabular}
\end{adjustbox}
\end{table}


\begin{table}
    \centering
    \caption{Different local band gaps of homobolayer \ch{WSe2} wrinkles at the reciprocal space K point for different compression. a) Interlayer band gap, b) the smallest apparent (interlayer) band gap at VBM. Intralayer bands gaps: c) upper layer down curve, d) upper layer straight, e) upper layer up curve, f) lower layer down curve, g) lower layer straight, h) lower layer up curve (at least 30\% localization in the respective region). \label{table_band_gap_homo_sections}} 
    \begin{tabular}{c|c|c|c|c|c|c|c|c}
     \textbf{Compression}  & \textbf{a} & \textbf{b} & \textbf{c} & \textbf{d} & \textbf{e} & \textbf{f} & \textbf{g} & \textbf{h}  \\

\hline
        2.5   & 1.272& 1.204& 2.386& 1.252& 1.35 & 1.35 & 1.204& 1.794  \\
        5    & 1.245& 1.196& 1.325& 1.196& 1.342& 1.342& 1.196& 1.325   \\
        6    & 1.198& 1.198& 1.336& 1.198& 1.397& 1.397& 1.198& 1.277   \\
        7.5  & 1.197& 1.197& 1.279& 1.197& 1.334& 1.334& 1.198& 1.279   \\
         10   & 1.193& 1.193& 1.316& 1.226& 1.318& 1.318& 1.226& 1.283  \\ 
        12.5  &1.192 & 1.192& 1.287& 1.289& 1.306& 1.305& 1.289& 1.287   \\
         15   & 1.128& 1.128& 1.246& 1.281& 1.307& 1.307& 1.281& 1.246  \\
        17.5  &  1.11 & 1.11 & 1.233& 1.245& 1.295& 1.294& 1.245& 1.232 \\
         20   & 1.094& 1.094& 1.22 & 1.28 & 1.281& 1.281& 1.28 & 1.219  \\
    \end{tabular}

    \label{tabel_inter-intralayer_bandgap_homobilayer}
\end{table}

\begin{table}
\caption{Structural parameters of the homobilayer \ch{WSe2} in $\mathrm{A} $ for different compressions- A and radius of curvature, R, values are extracted using the fitted curve explained in the SI \label{table_structural_homobilayer}}
\begin{tabular}{ c|c|c|c|c|c|c} 
     \textbf{Compression}   &  \textbf{$\rm \lambda$} &\textbf{$ \rm  A_{upper layer}$}& \textbf{$\rm A_{lower layer}$} &\textbf{$\rm A_{both}$} & \textbf{$\rm R_{{min}_{upper layer}} $} & \textbf{$\rm R_{{min}_{lower layer}} $} \\ \hline
     2.5   & 83.137 &  4.171&  4.177&  7.669 & 43.54  & 32.949     \\
     5     & 81.005 &  5.994&  5.982&  9.462 & 28.345 & 32.929    \\
     6     & 80.152 &   6.538&  6.524& 9.997 & 25.846 & 32.614     \\
     7.5   & 78.874 &   7.297&  7.280& 10.746& 22.673 & 20.623    \\
     10    & 76.741 &   8.435&  8.415& 11.875& 18.752 & 16.265  \\
     12.5  & 74.610 & 9.403 &  9.383& 12.812 &16.517& 19.125     \\
     15    & 72.479 &  10.075& 10.091& 13.494& 14.53 & 11.307    \\     
     17.5  & 70.347 & 10.857& 10.872& 14.275& 12.863&  9.662     \\
     20    & 68.215 & 11.562 & 11.582 & 14.973 & 8.637 &  8.444   \\   
\end{tabular}
\end{table}

\begin{figure}
\begin{subfigure}{0.32\textwidth}
        \includegraphics[width=1\textwidth]{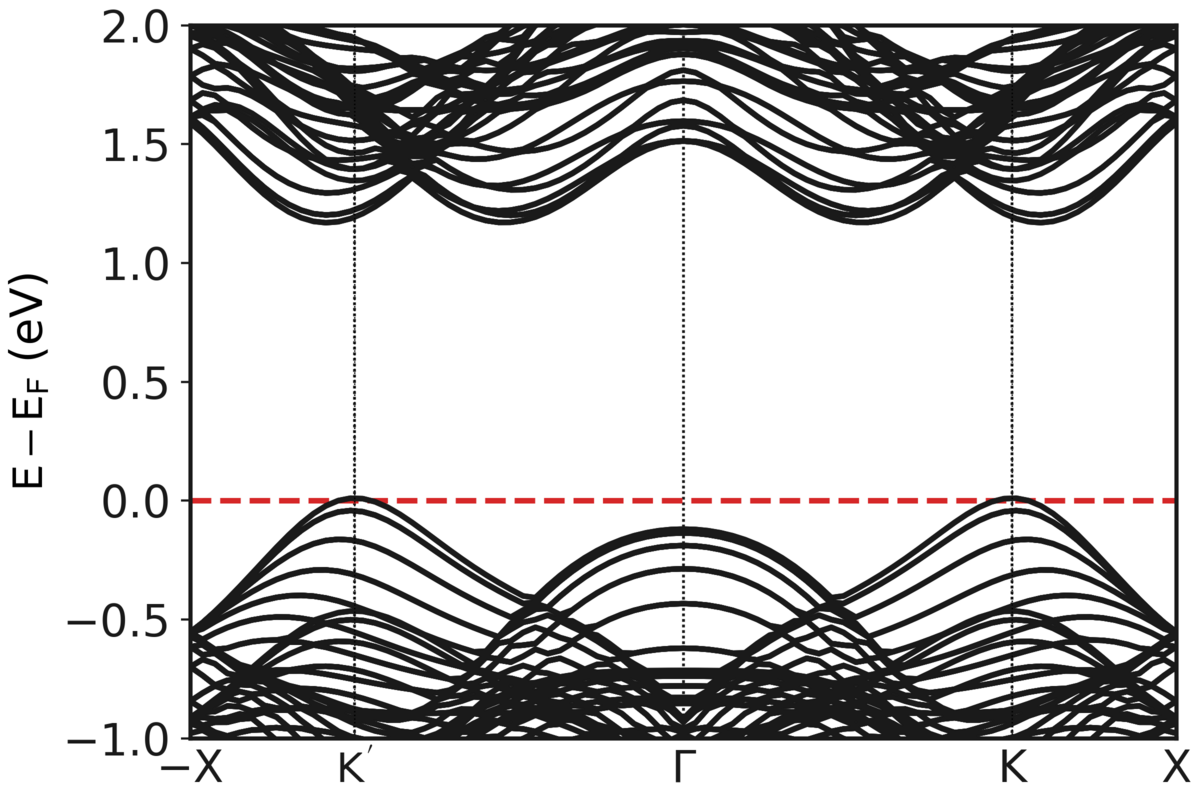}
        \caption{}
\end{subfigure}
\begin{subfigure}{0.32\textwidth}
        \includegraphics[width=1\textwidth]{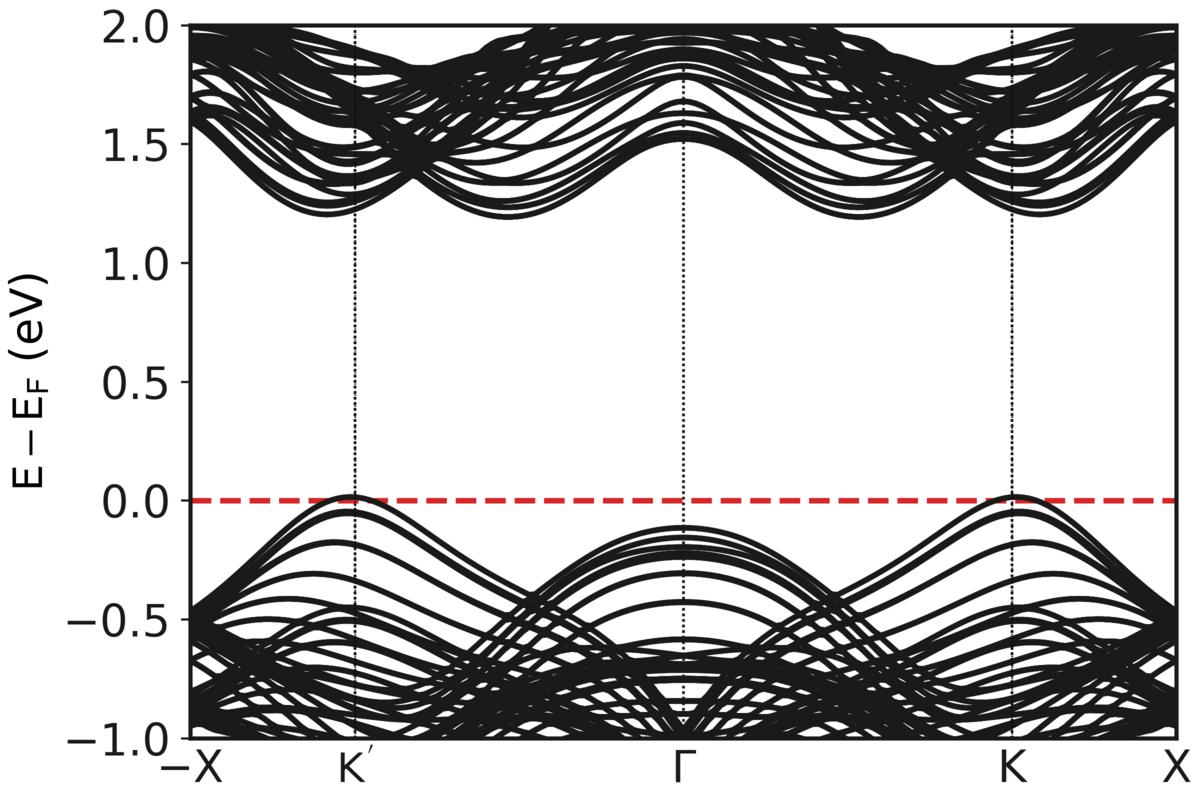}
        \caption{}
\end{subfigure}
\begin{subfigure}{0.32\textwidth}
        \includegraphics[width=1\textwidth]{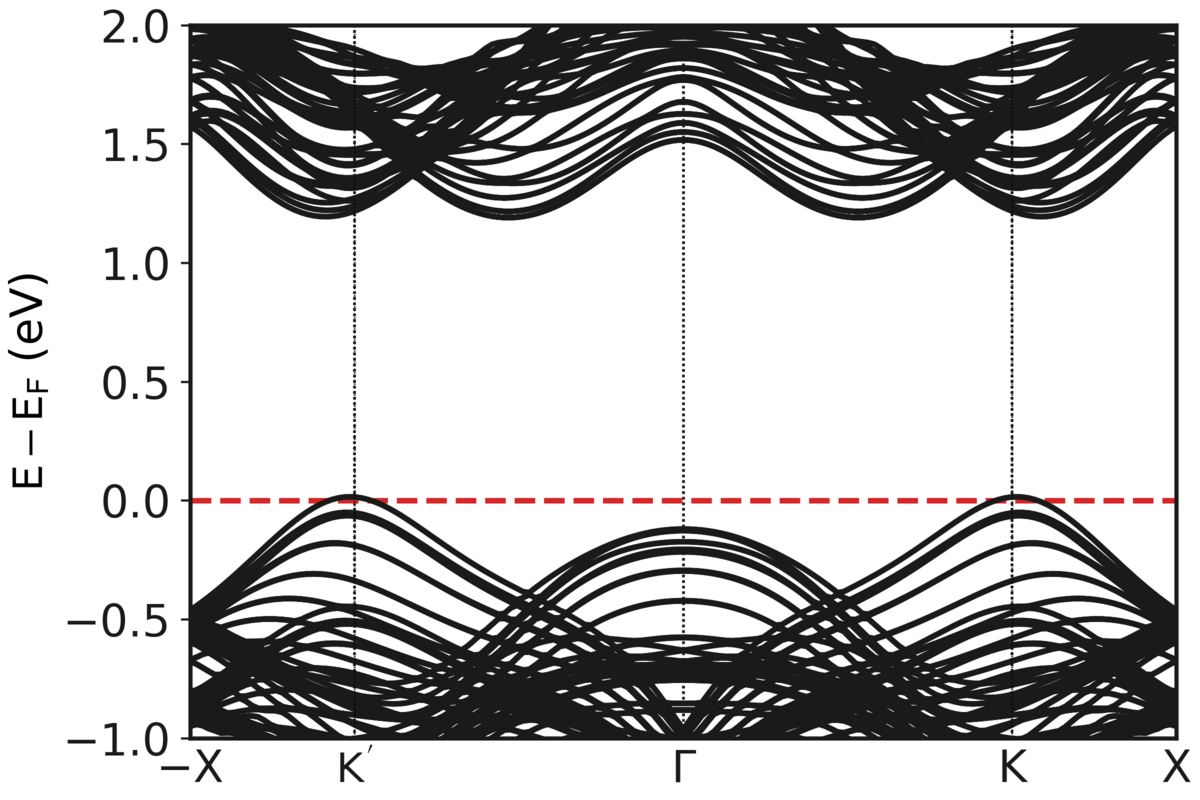}
        \caption{}
\end{subfigure}
\begin{subfigure}{0.32\textwidth}
        \includegraphics[width=1\textwidth]{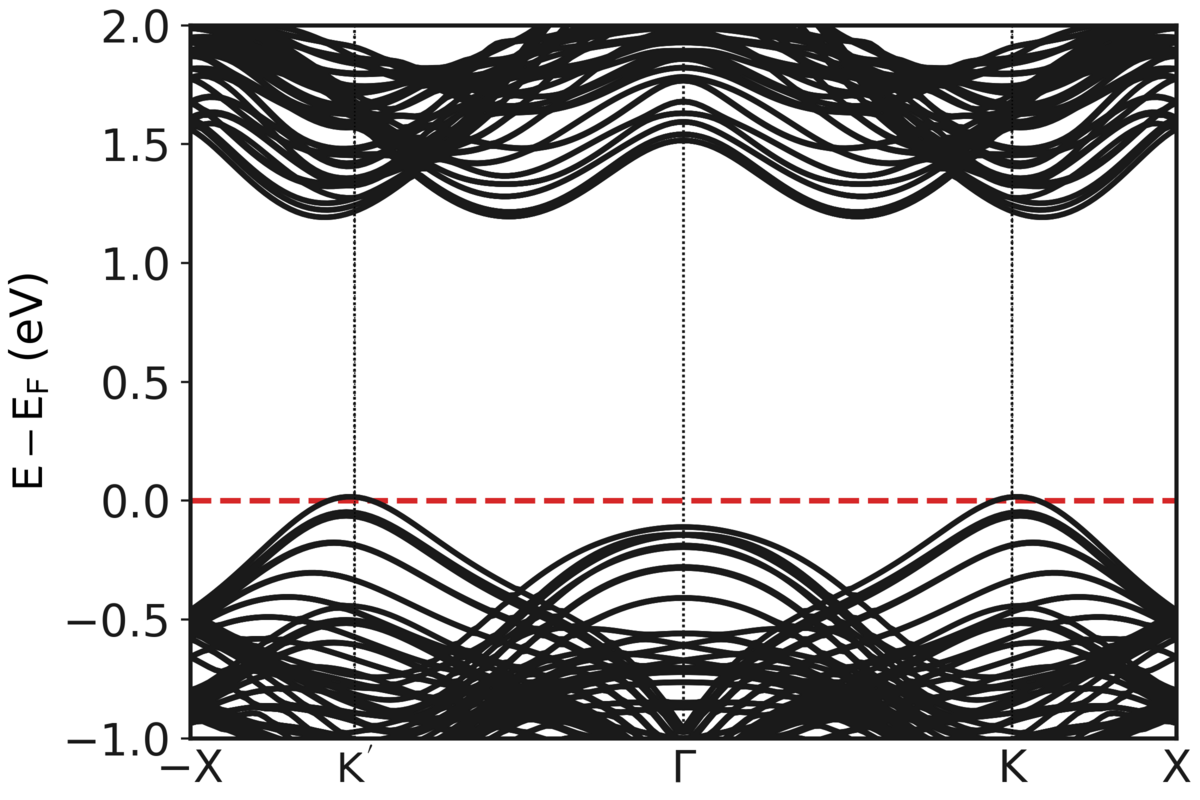}
        \caption{}
\end{subfigure}
    \begin{subfigure}{0.32\textwidth}
         \includegraphics[width=1\textwidth]{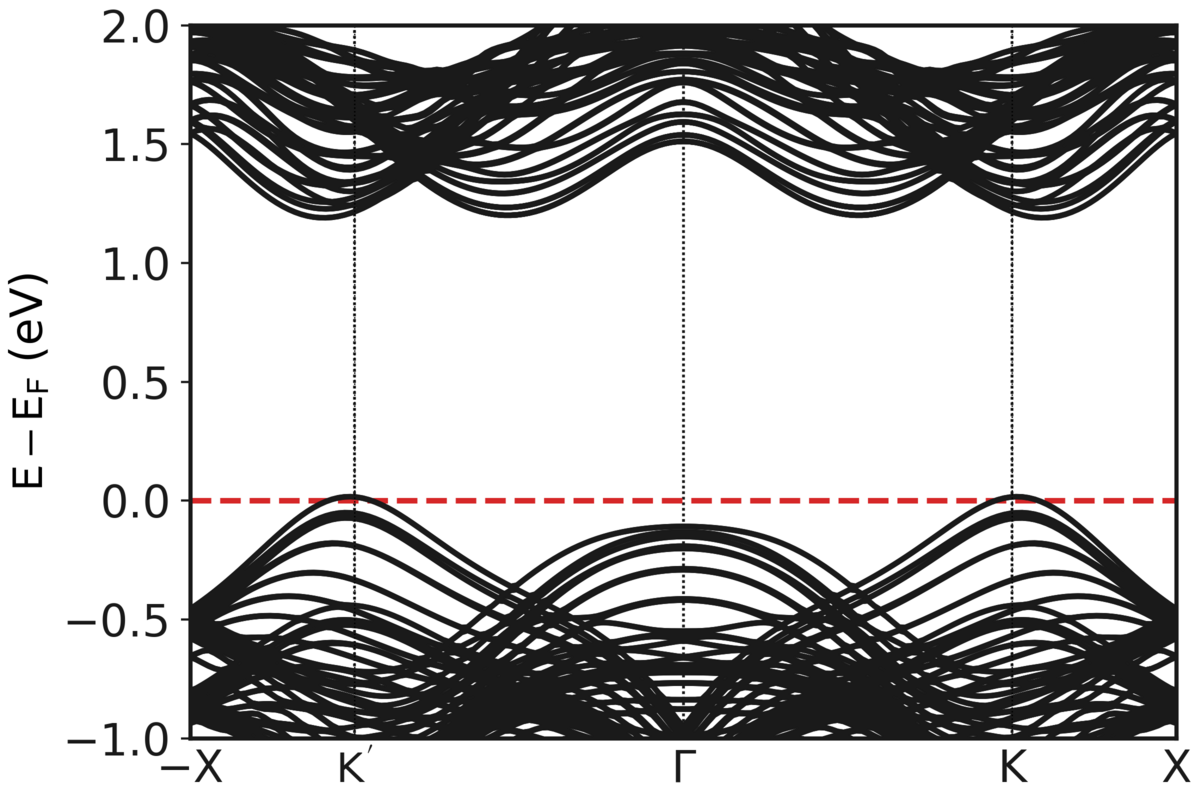}
         \caption{}
    \end{subfigure}
    \begin{subfigure}{0.32\textwidth}
         \includegraphics[width=1\textwidth]{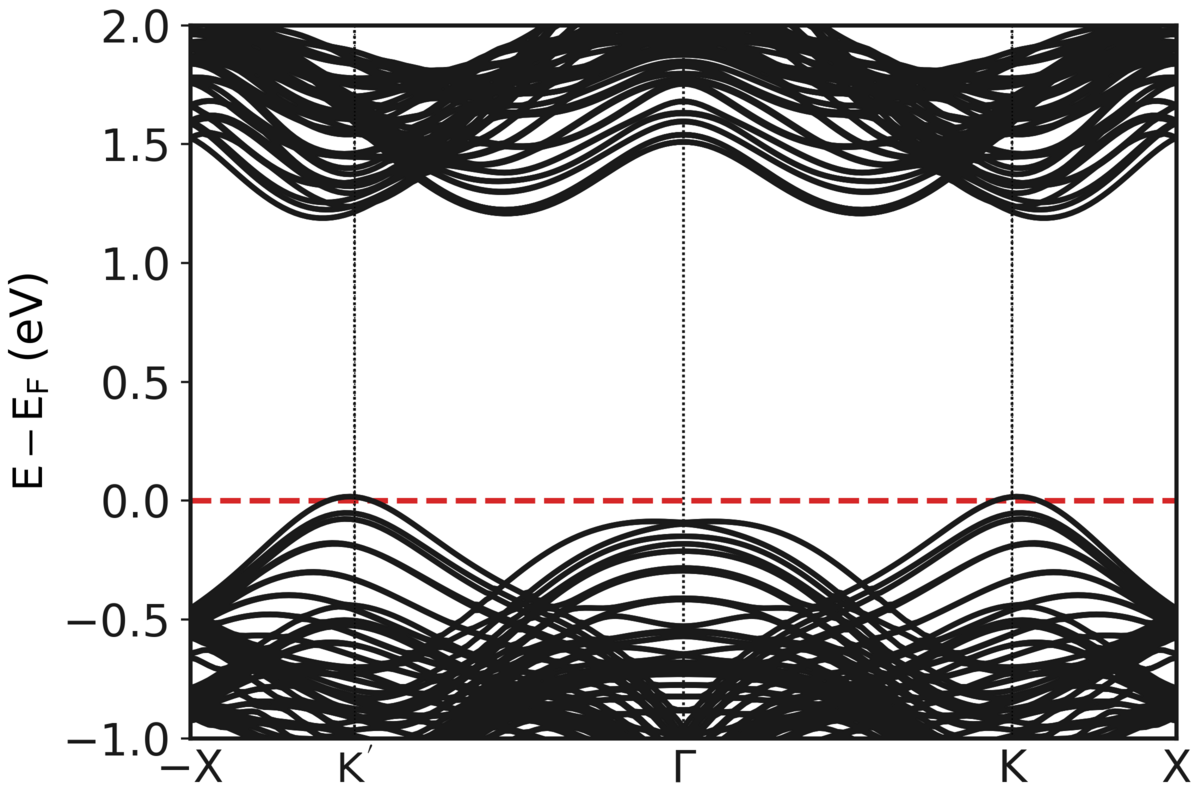}
         \caption{}
    \end{subfigure}
    \begin{subfigure}{0.32\textwidth}
         \includegraphics[width=1\textwidth]{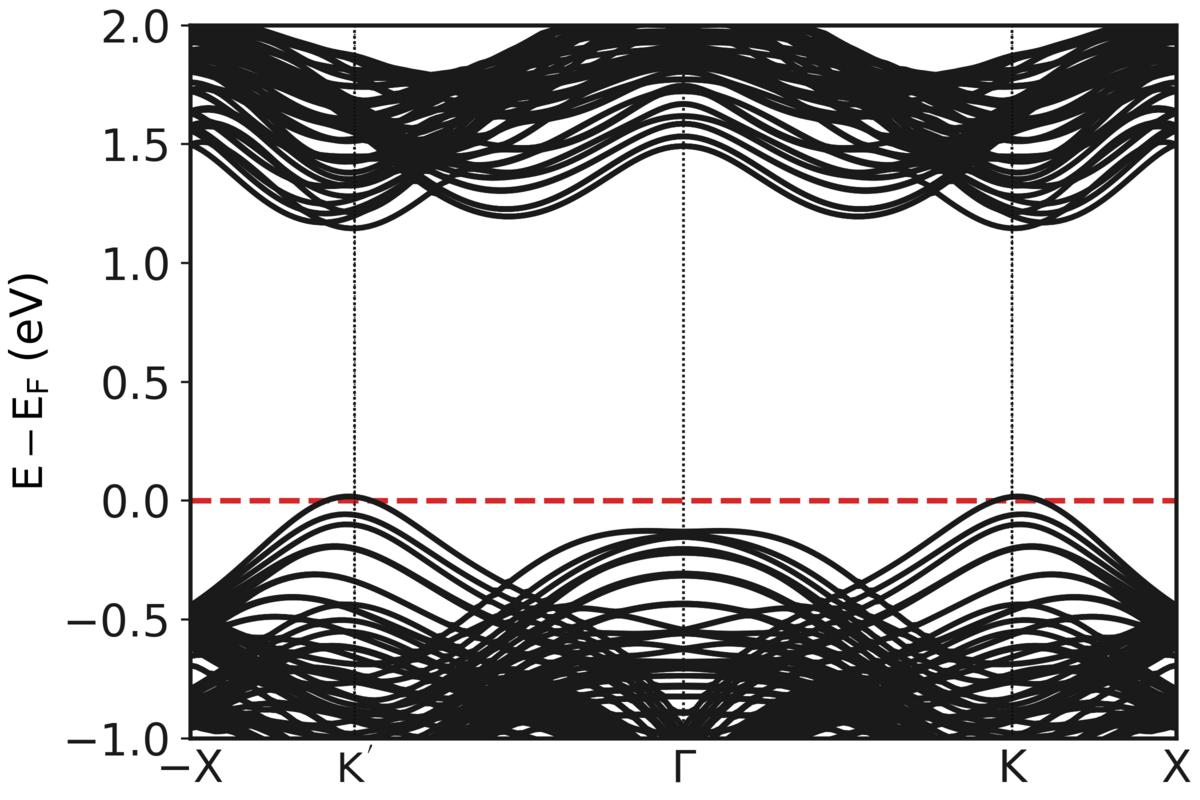}
         \caption{}
    \end{subfigure}
    \begin{subfigure}{0.32\textwidth}
         \includegraphics[width=1\textwidth]{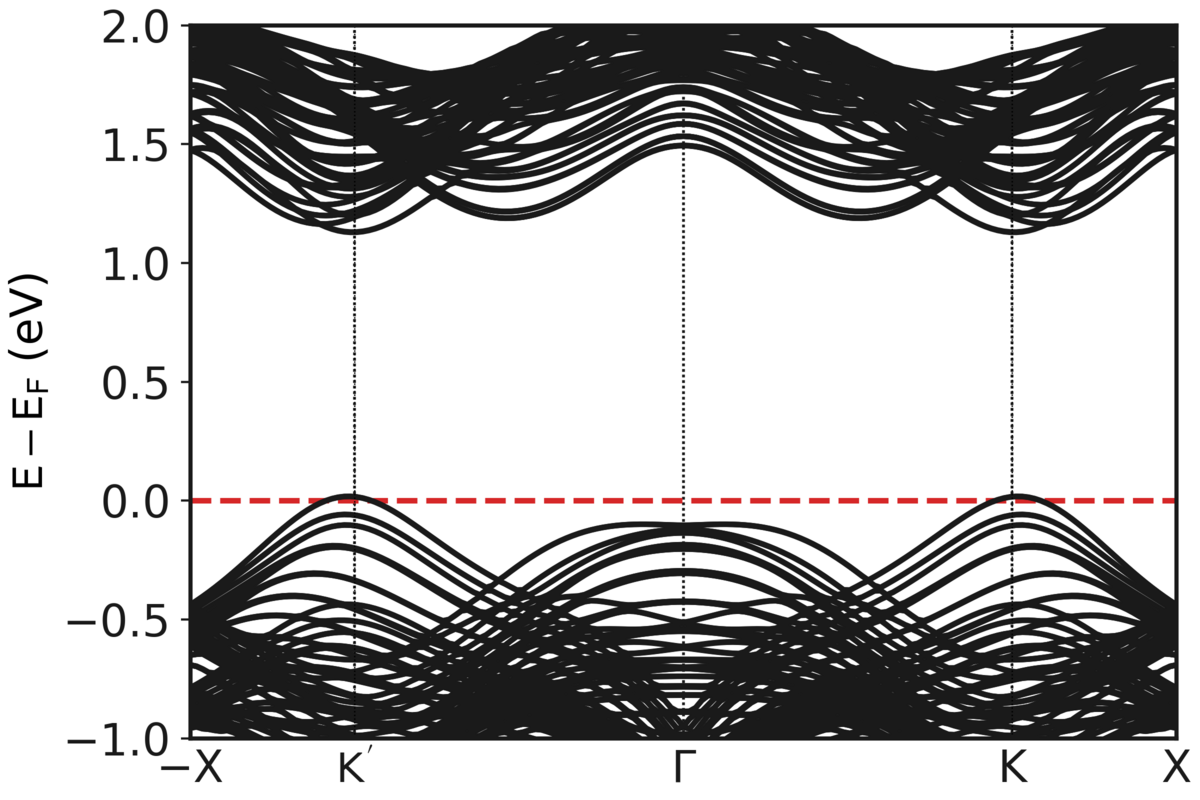}
         \caption{}
    \end{subfigure}
    \begin{subfigure}{0.32\textwidth}
          \includegraphics[width=1\textwidth]{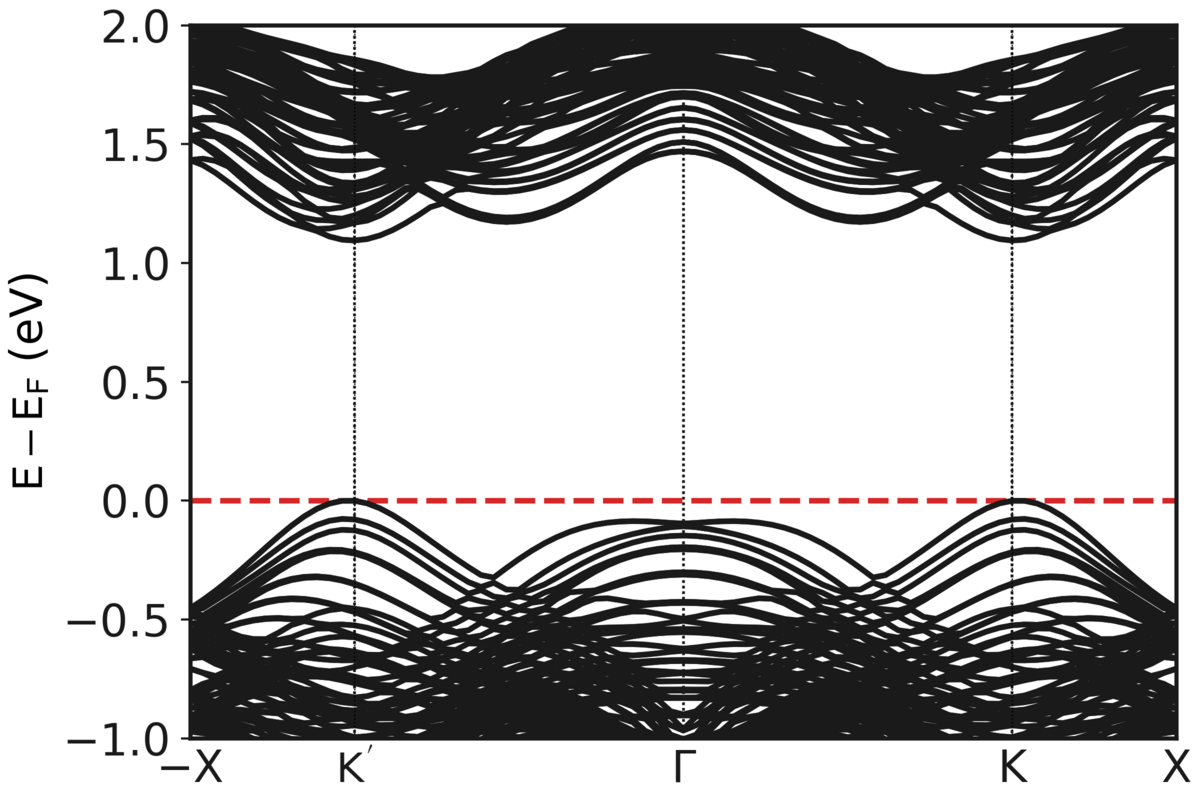}
          \caption{}
    \end{subfigure}
    \begin{subfigure}{0.32\textwidth}
          \includegraphics[width=1\textwidth]{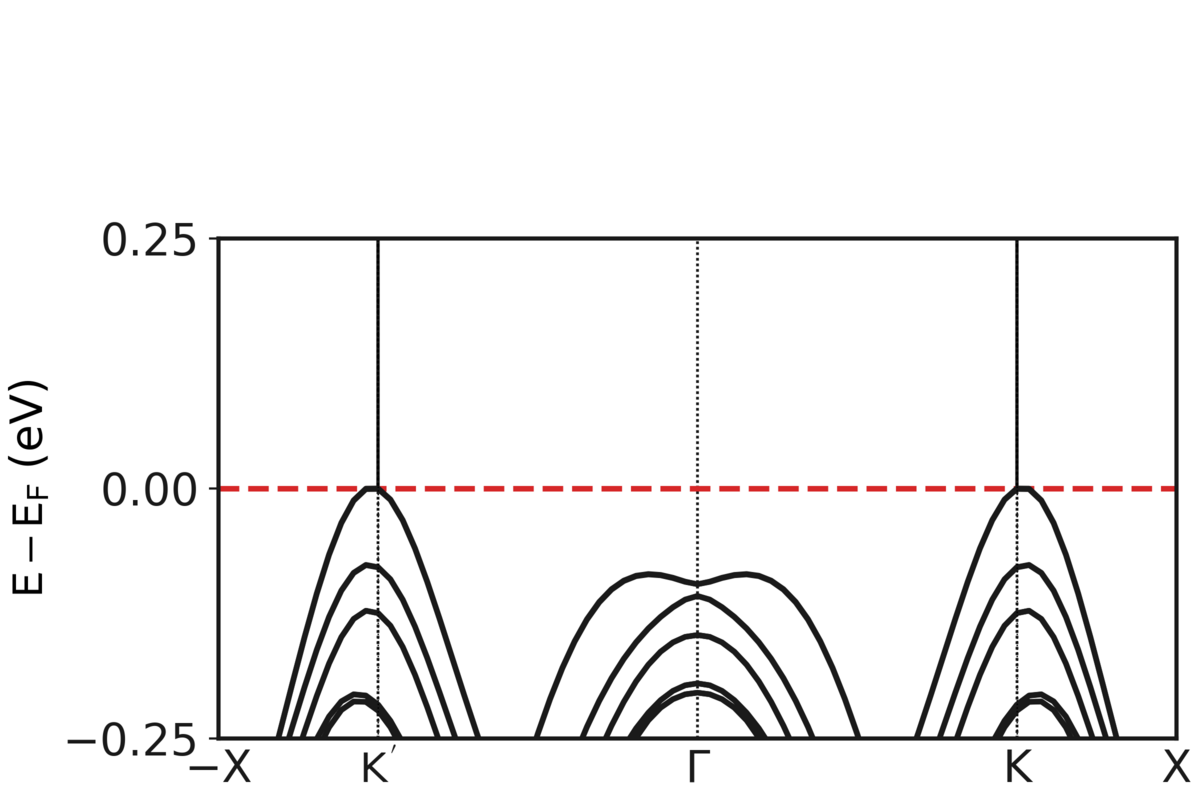}
         \caption{\label{fig_band_structure_homobilayer_20_zoom}}
    \end{subfigure}
    \caption{Band structure changes in homobilayer \ch{WSe2} due to wrinkling at different compressions of a) 0\% b) 2.5\% c) 5\% d) 7.5\% e) 10\% f)12.5\% g) 15\% h)17.5\% i) 20\% j)20\% enlarged area of the band structure of 20\% compressed system. Rashba-like splitting is well apparent. Band gaps of the systems are reduces with strain with the CB at K being more sensitive to strain. An indirect to direct transition occurs also for the higher strained systems. }
    \label{fig_band_structure_homobilayer}
\end{figure}

\begin{figure}
     \includegraphics[width=0.32\textwidth]{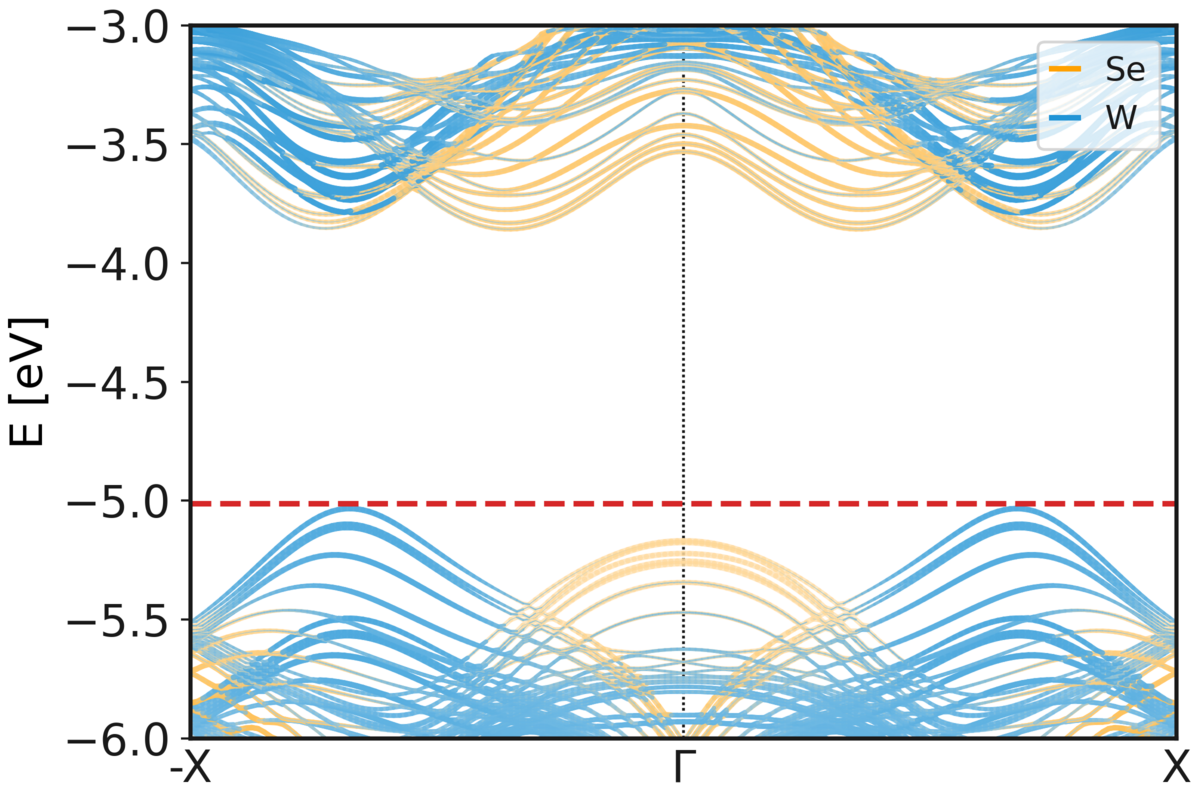}
     \includegraphics[width=0.32\textwidth]{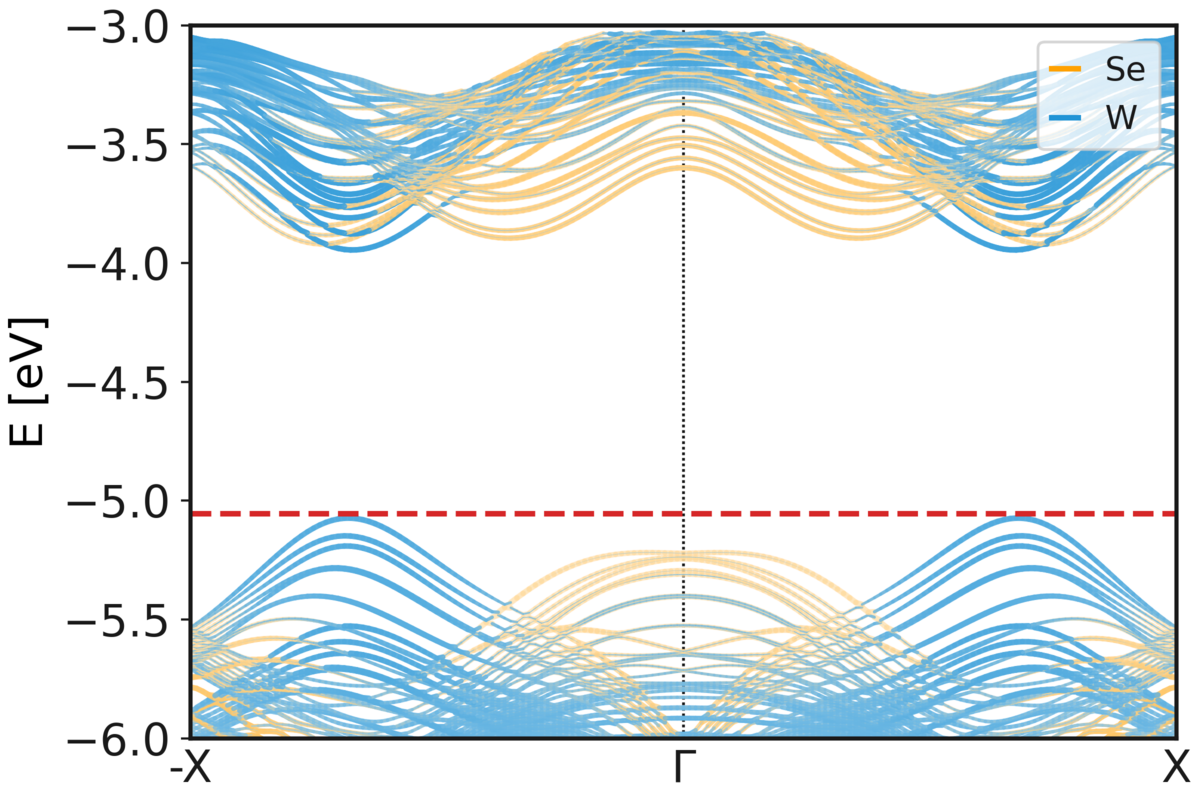}
     \includegraphics[width=0.32\textwidth]{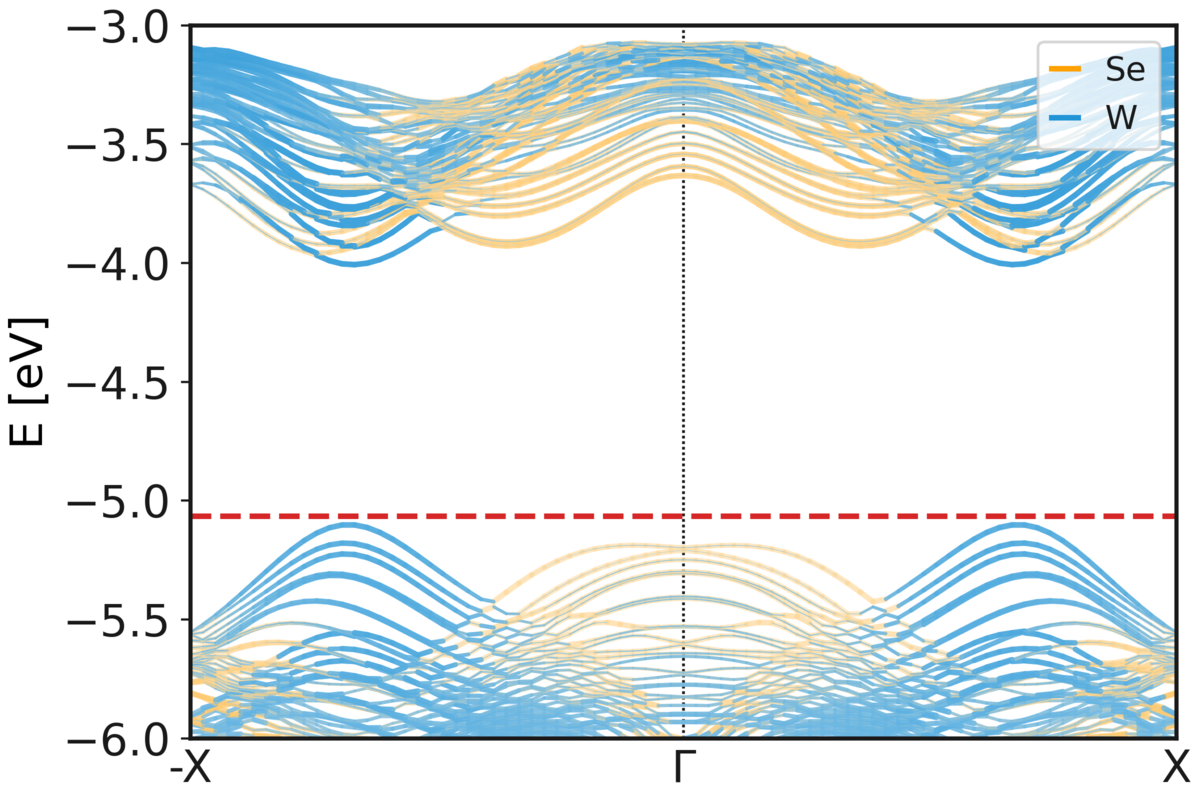}
    
    \caption{Band structure of \ch{WSe2} homobilayer for different compressions (5\%, 10\% and 20\%) without shifting of the energy values. Bands shift downwards, however the shift of the conduction band is larger thus leading to a reduction of the band gap.}
    \label{fig_band_structure_homobilayer_noshift}
\end{figure}

\begin{figure}
    \includegraphics[width=0.32\textwidth]{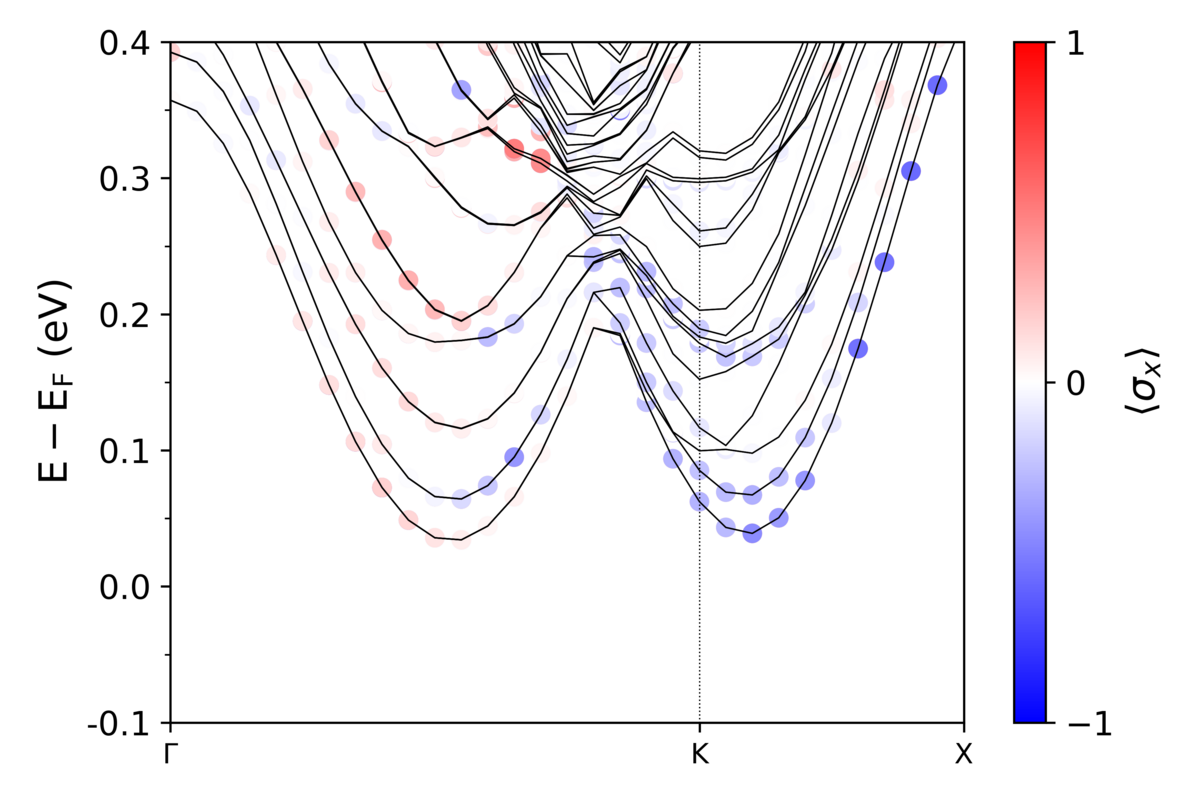}
    \includegraphics[width=0.32\textwidth]{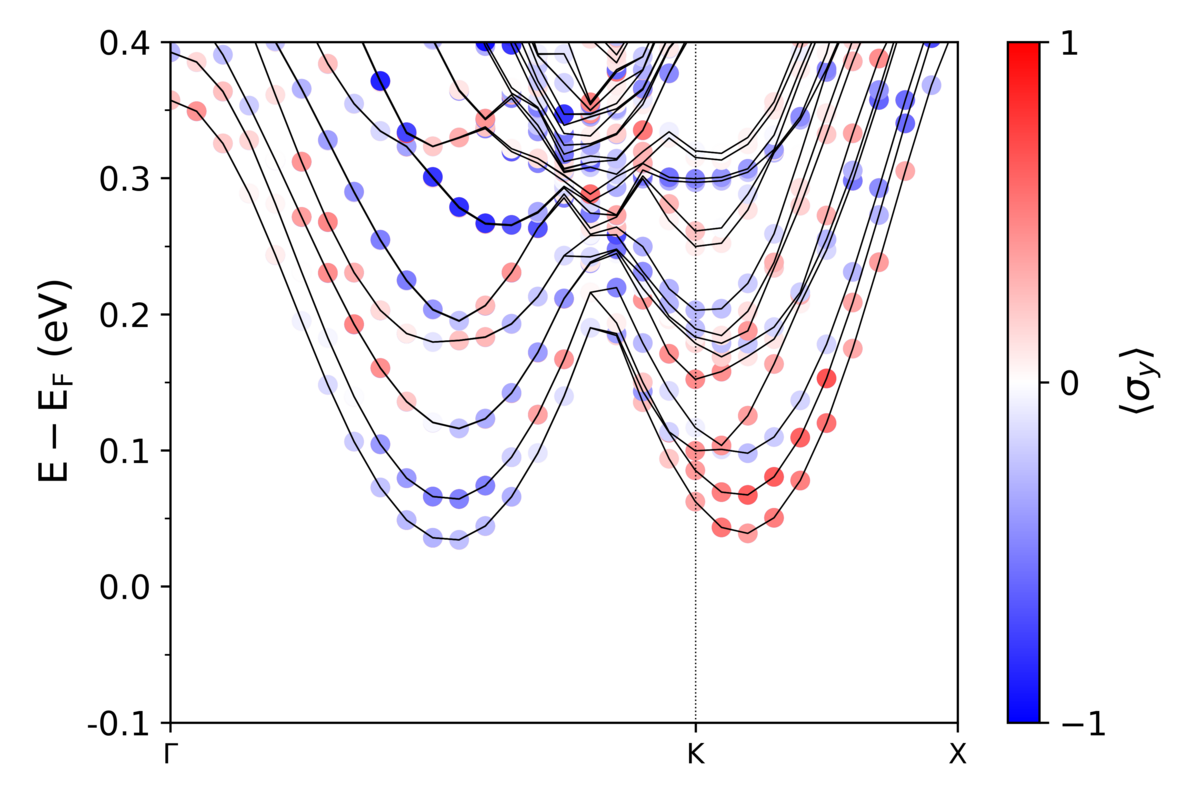}
    \includegraphics[width=0.32\textwidth]{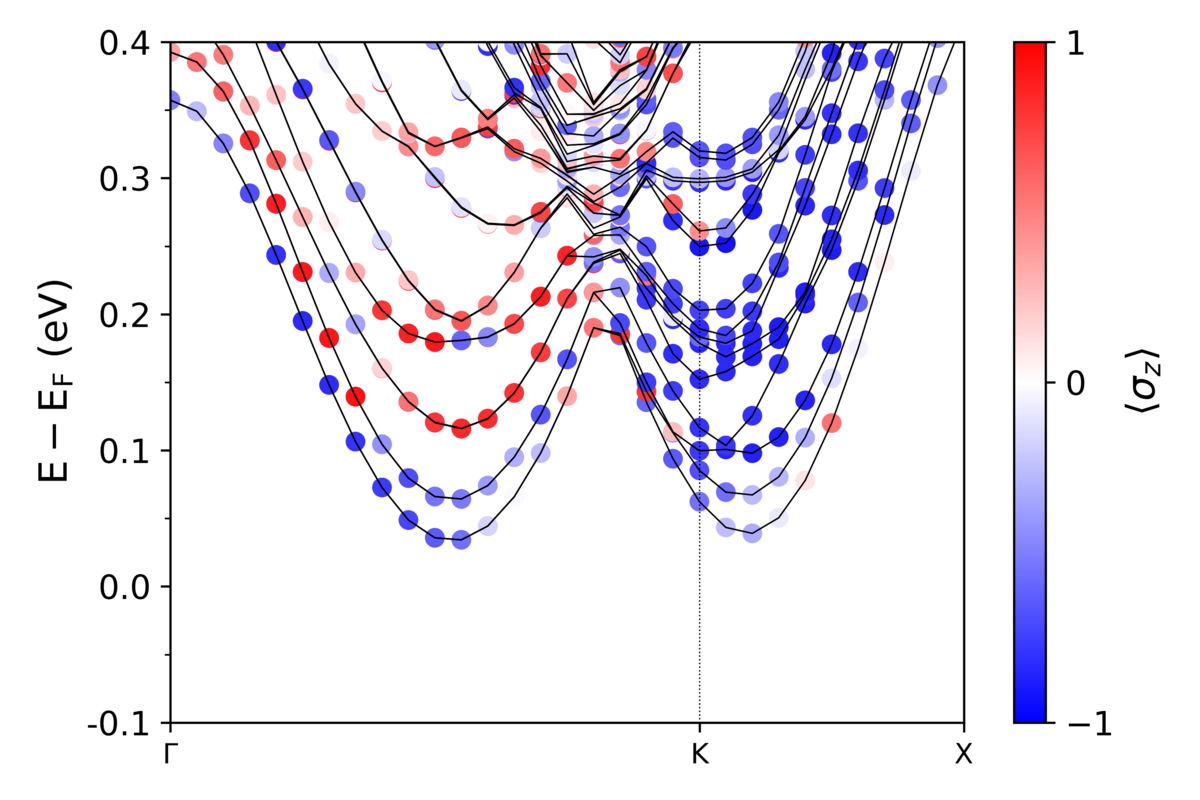}
    \includegraphics[width=0.32\textwidth]{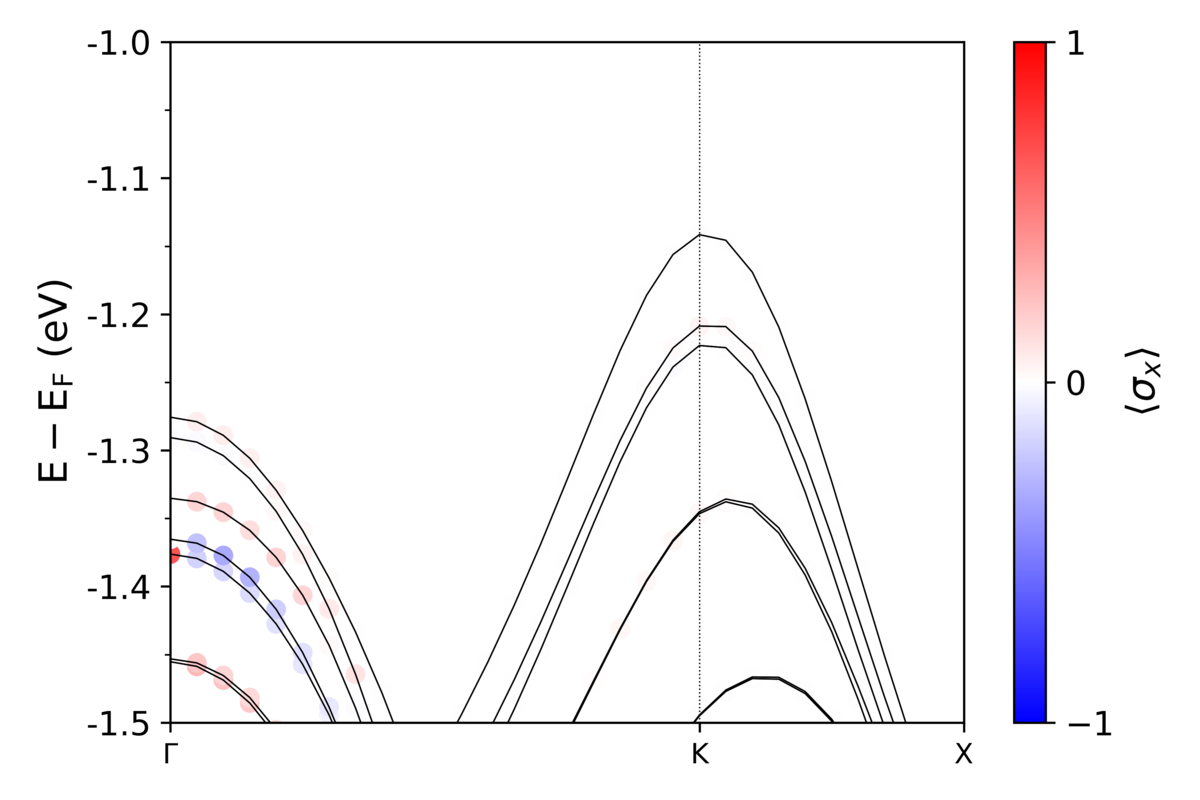}
    \includegraphics[width=0.32\textwidth]{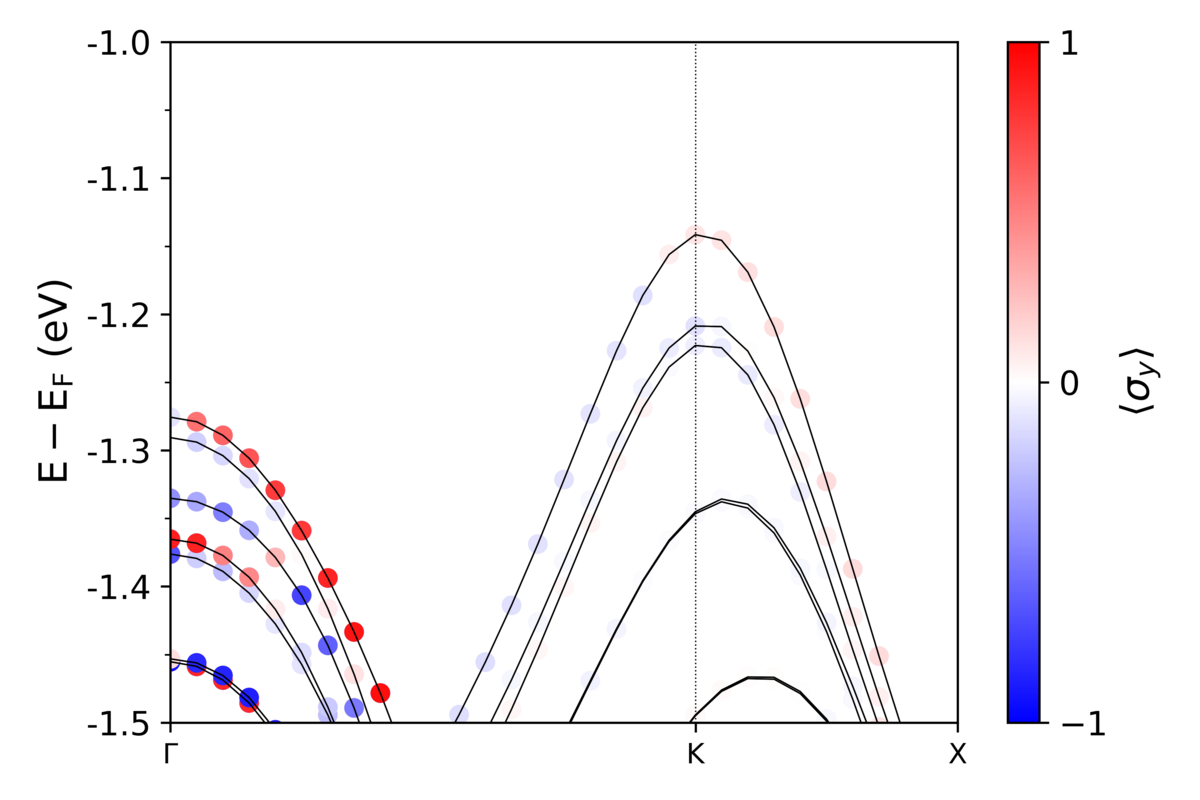}
    \includegraphics[width=0.32\textwidth]{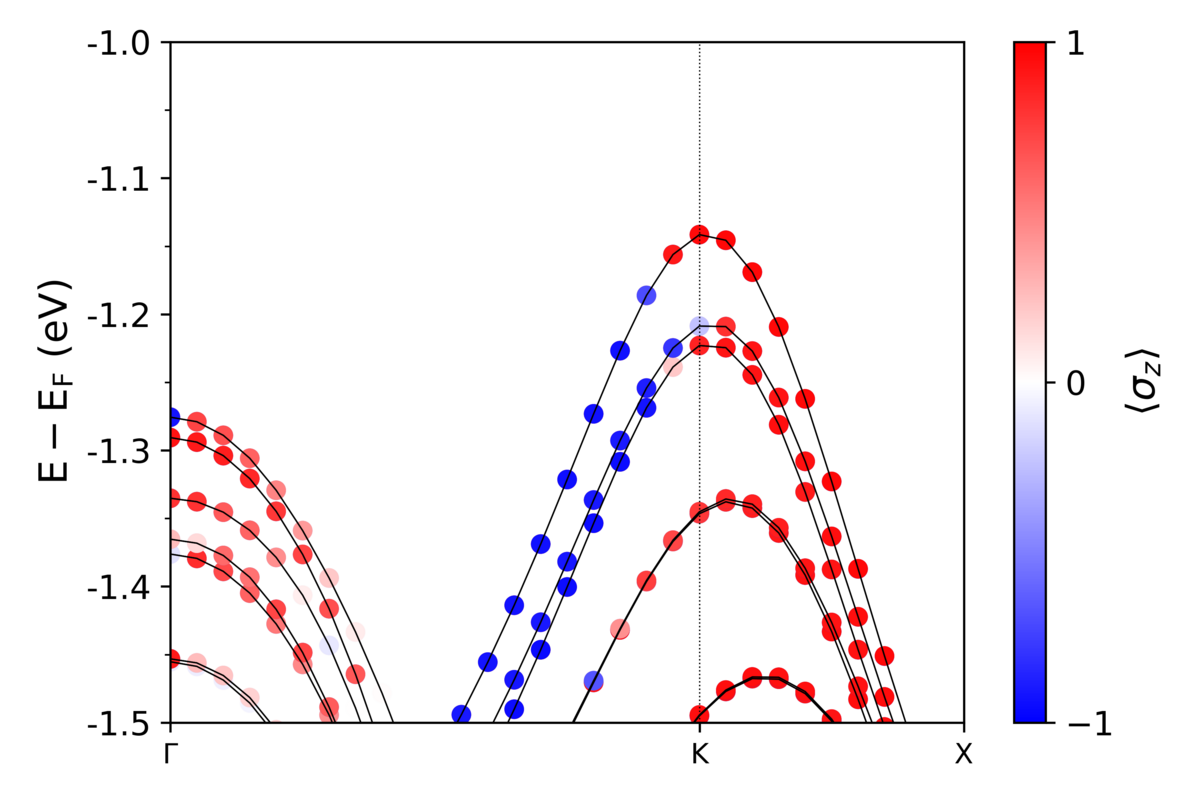}
    \caption{Expectation values of the Pauli matrices $\langle\sigma_i\rangle$ of the wrinkled \ch{WSe2} homobilayer at 2.5\% compression.}
    \label{fig_spin_texture_homo_2.5}
\end{figure}
\begin{figure}
    \includegraphics[width=0.32\textwidth]{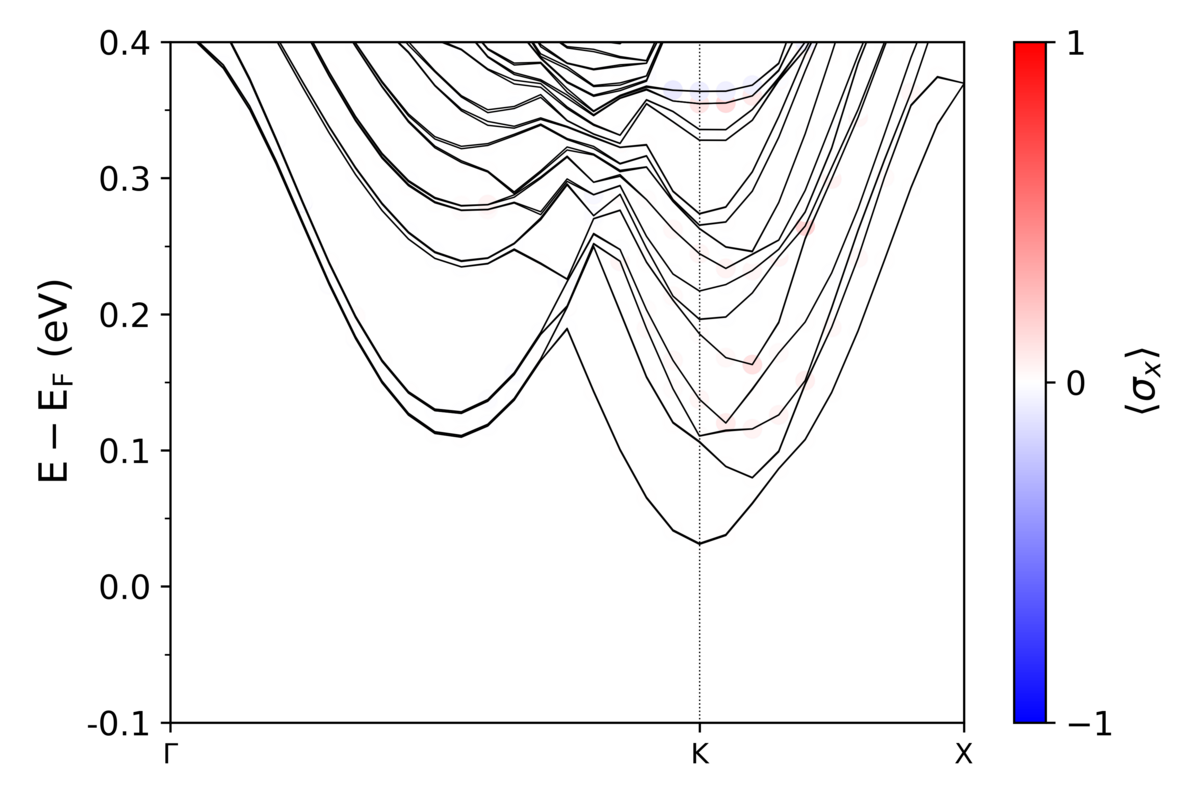}
    \includegraphics[width=0.32\textwidth]{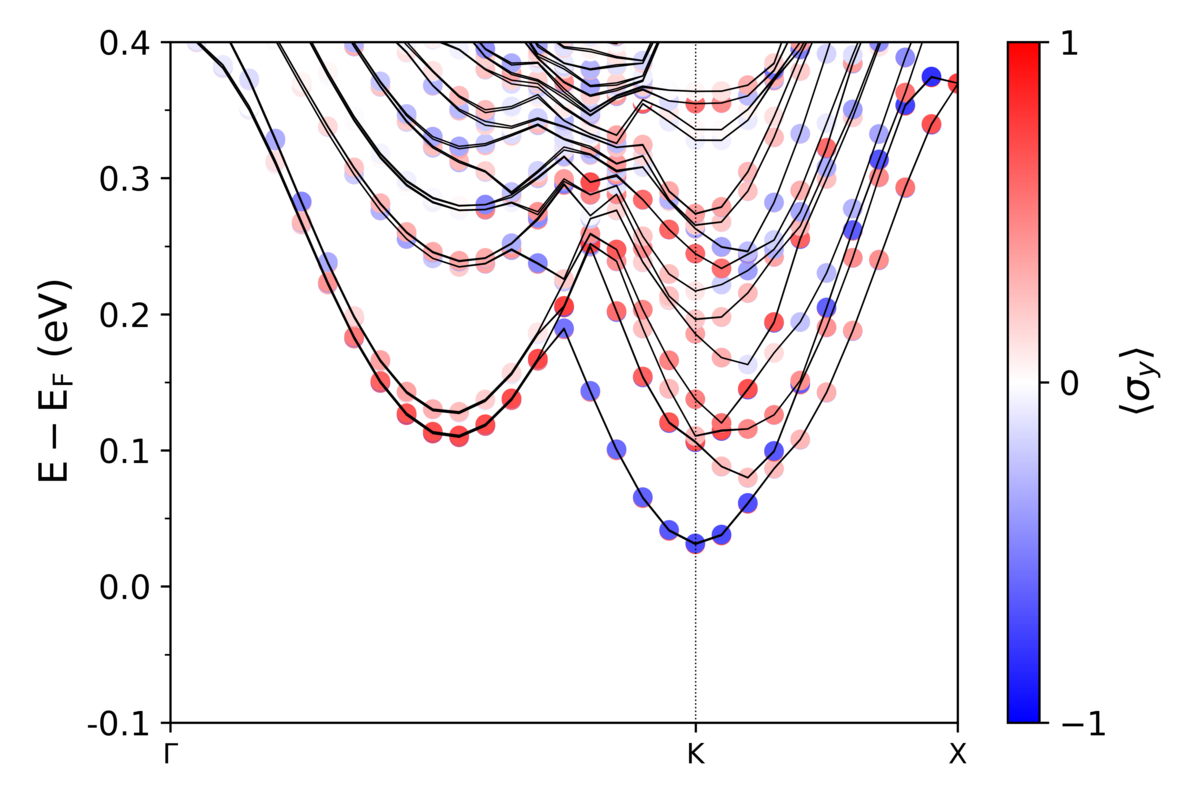}
    \includegraphics[width=0.32\textwidth]{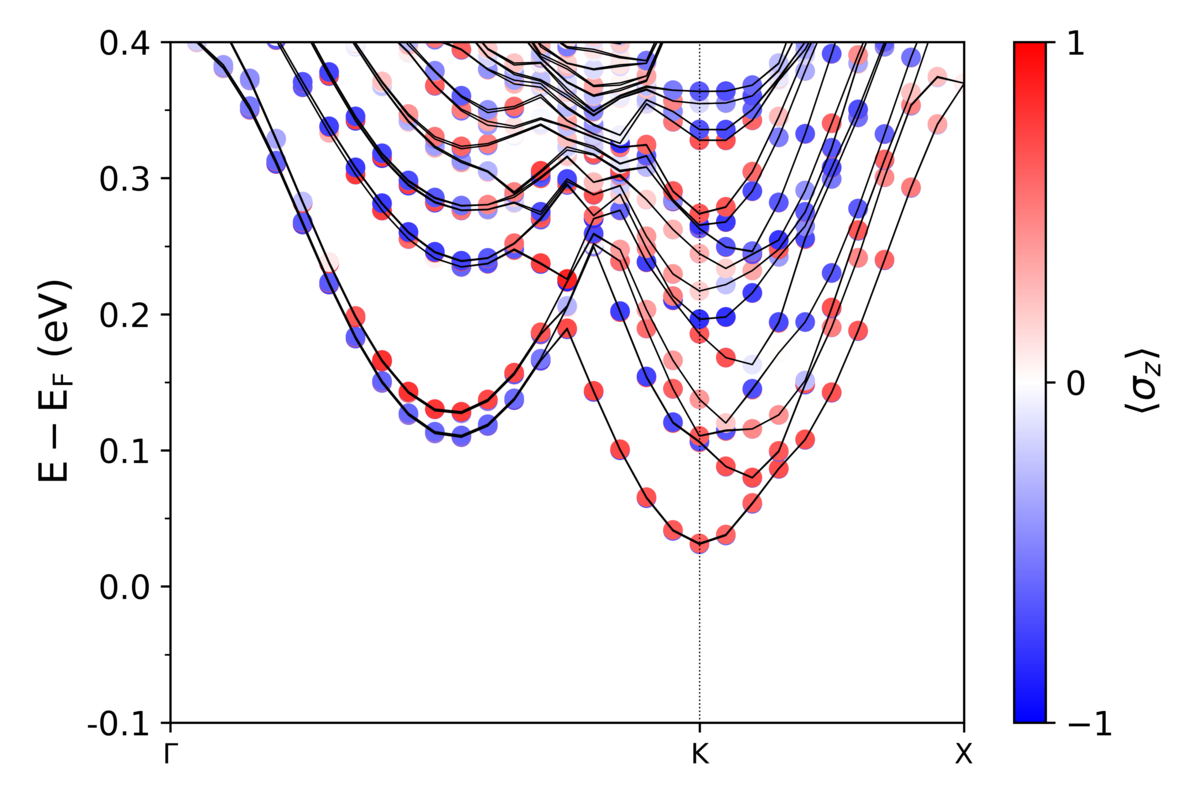}
    \includegraphics[width=0.32\textwidth]{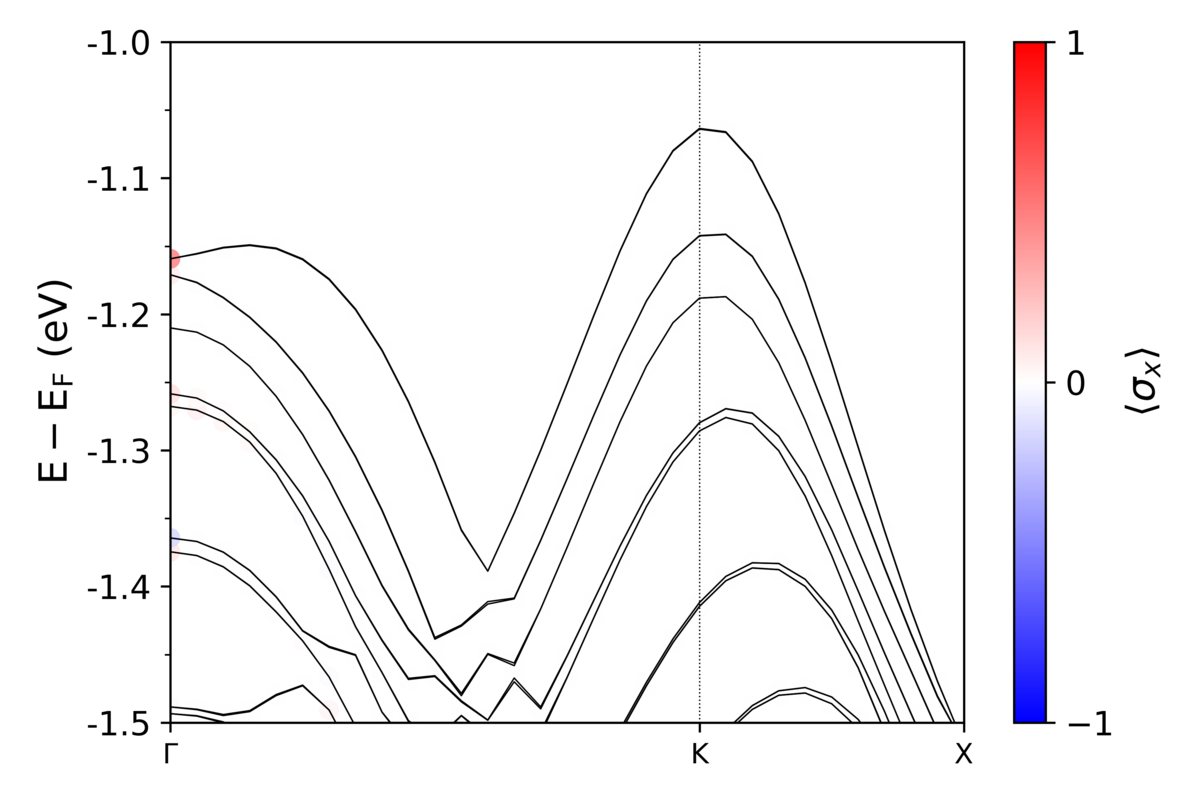}
    \includegraphics[width=0.32\textwidth]{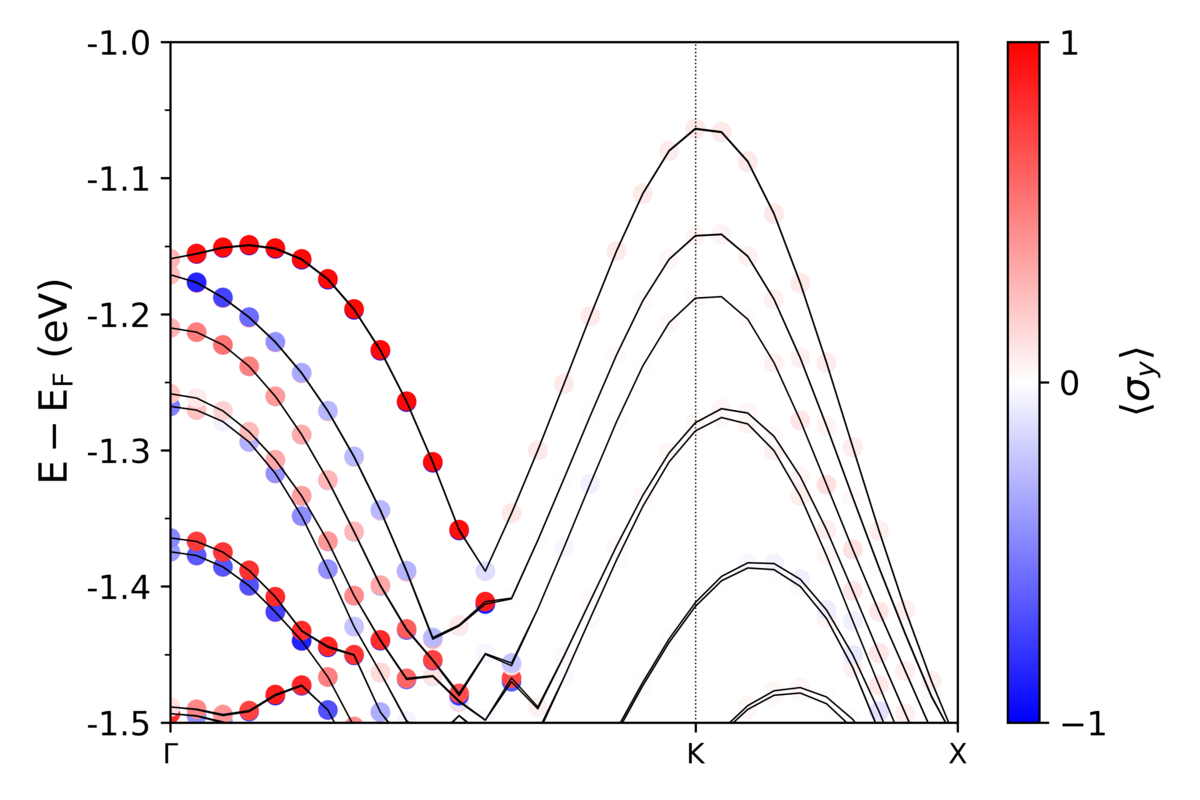}
    \includegraphics[width=0.32\textwidth]{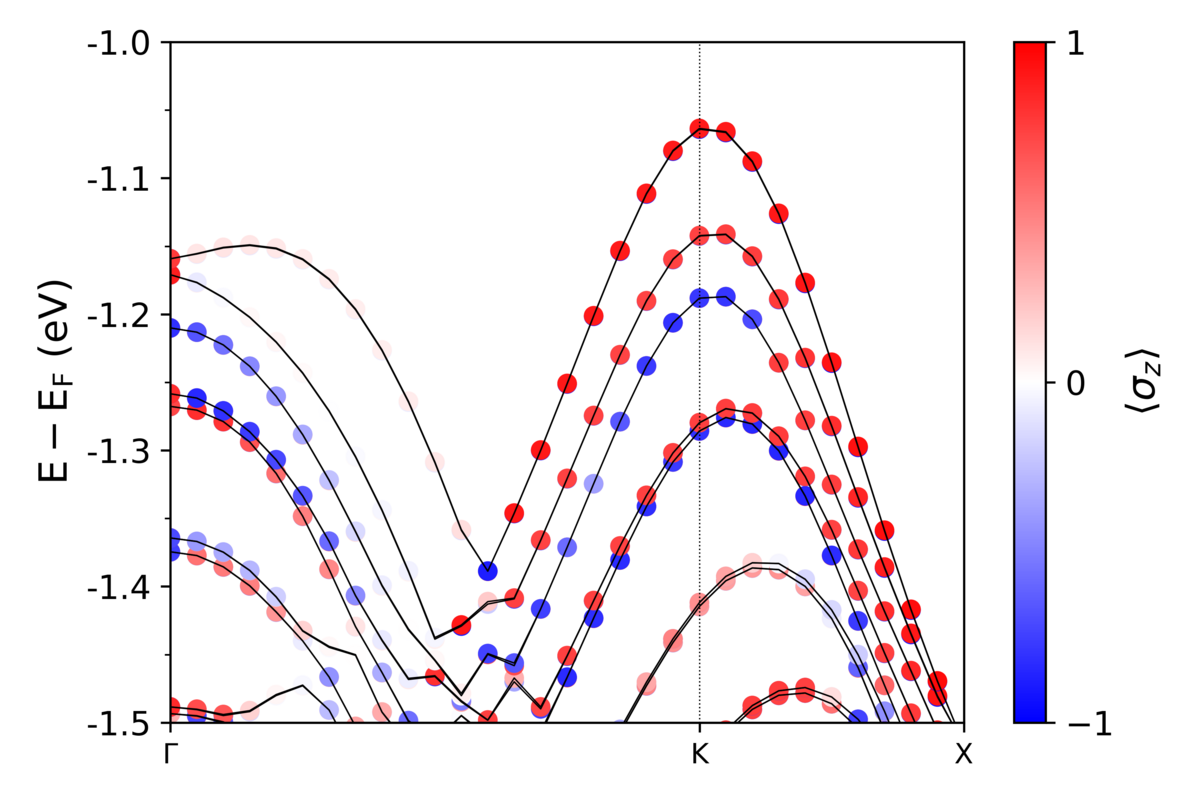}
    \caption{Expectation values of the Pauli matrices $\langle\sigma_i\rangle$ of the wrinkled \ch{WSe2} homobilayer at 20\% compression.}
    \label{fig_spin_texture_homo_20}
\end{figure}

\begin{figure}
\begin{subfigure}{0.32\textwidth}
        \includegraphics[width=1.\textwidth]{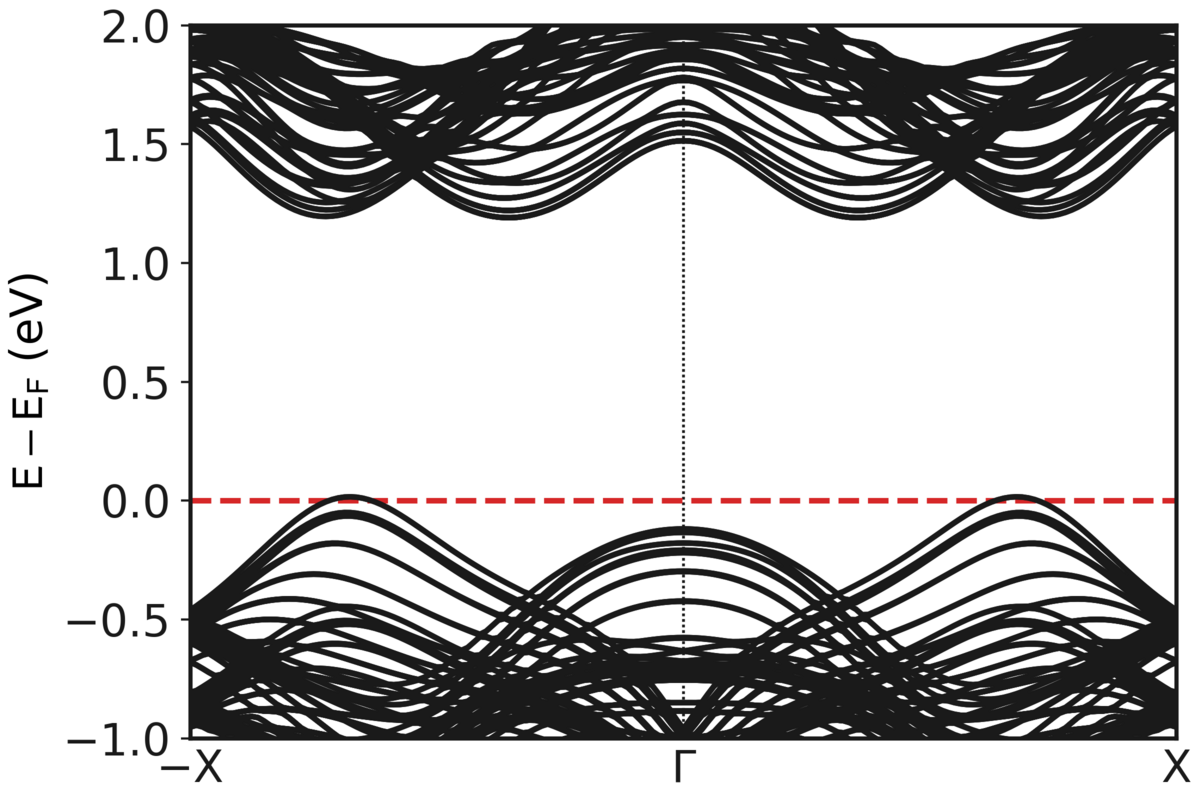}
        \caption{}
\end{subfigure}
\begin{subfigure}{0.32\textwidth}
     \includegraphics[width=1.\textwidth]{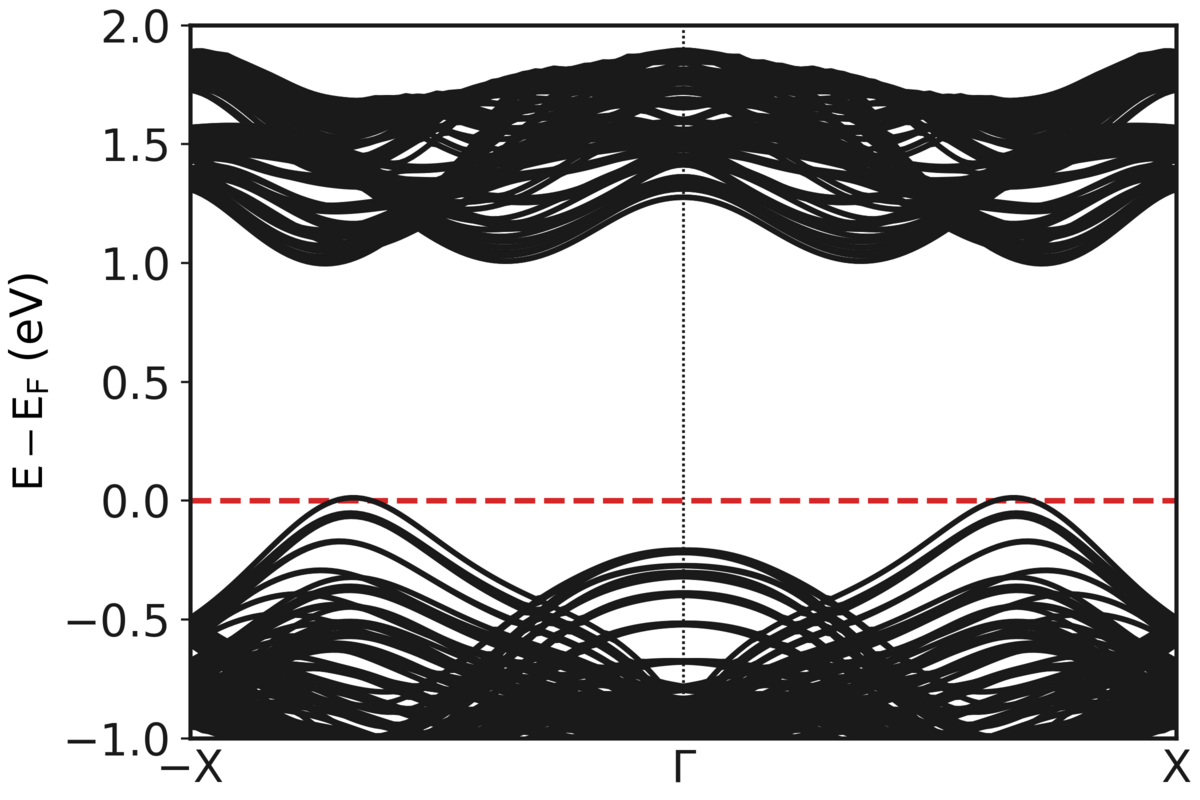}
     \caption{}
\end{subfigure}
\begin{subfigure}{0.32\textwidth}
       \includegraphics[width=1.\textwidth]{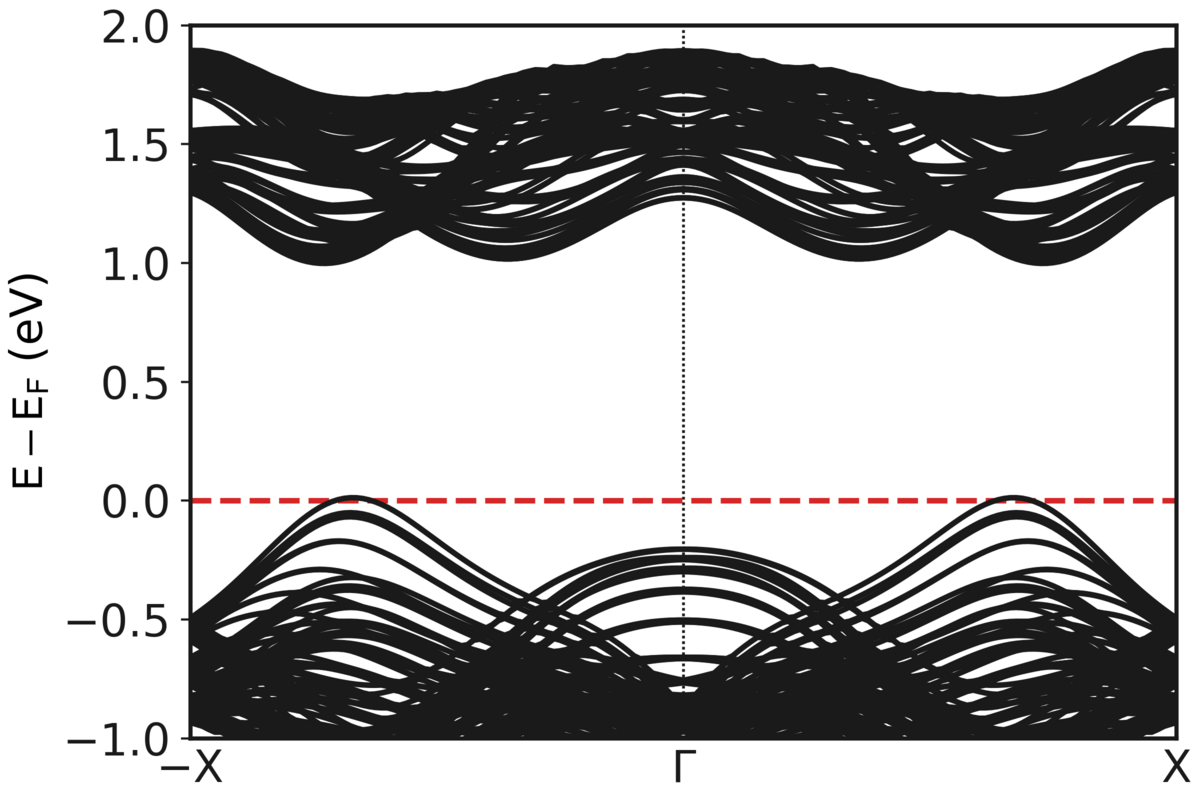}
       \caption{}
\end{subfigure}
\begin{subfigure}{0.32\textwidth}
    \includegraphics[width=1.\textwidth]{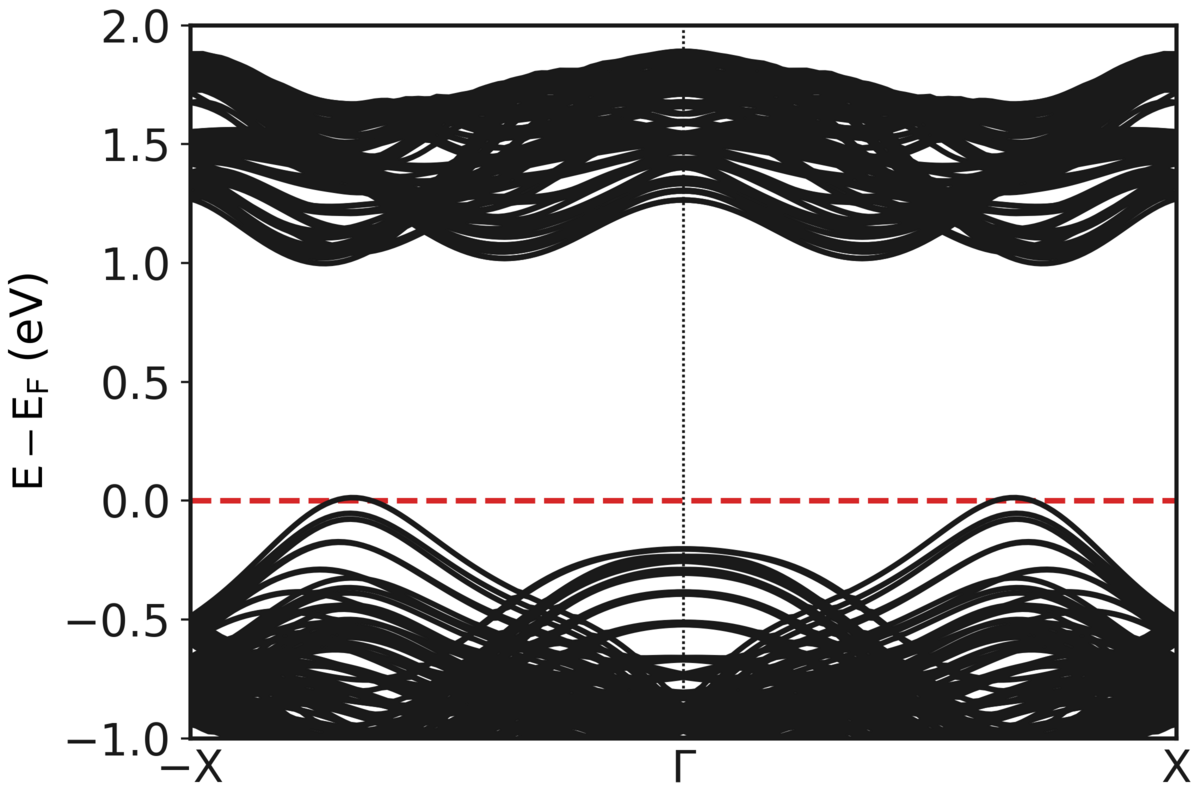}
    \caption{}
\end{subfigure}
\begin{subfigure}{0.32\textwidth}
    \includegraphics[width=1.\textwidth]{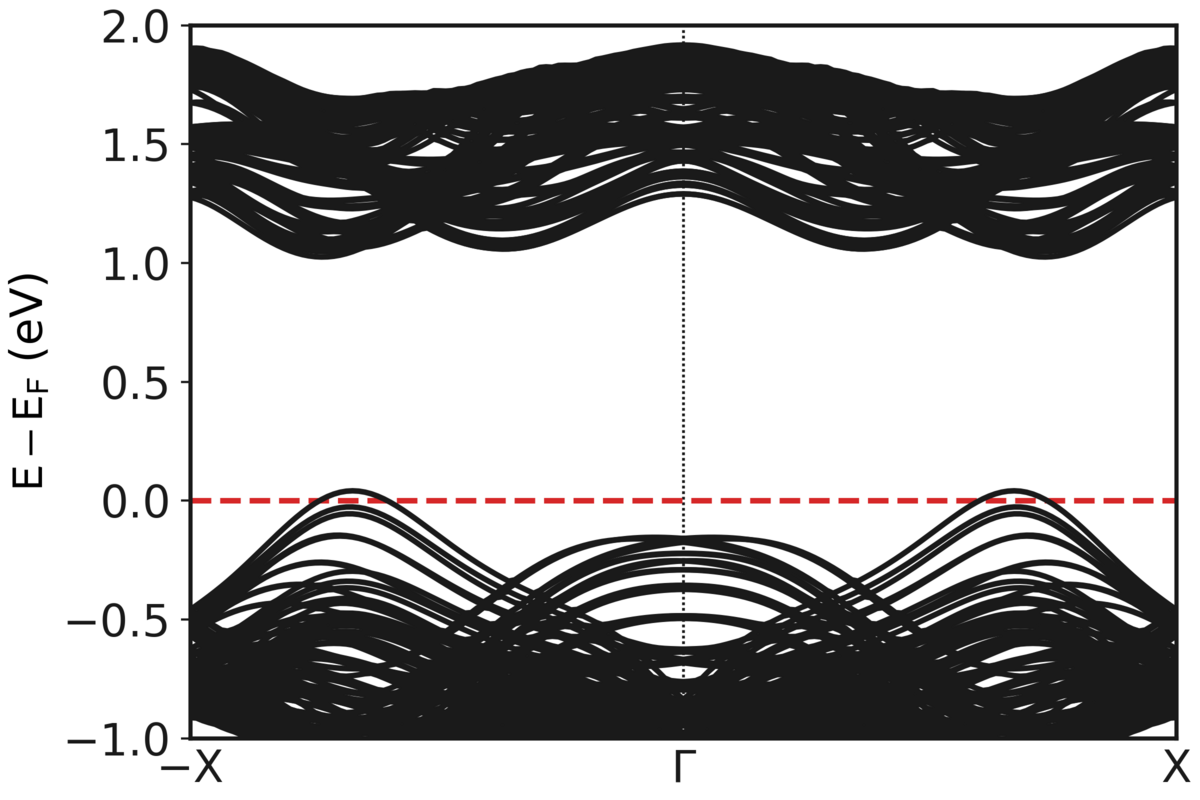}
    \caption{}
\end{subfigure}   
\begin{subfigure}{0.32\textwidth}
    \includegraphics[width=1.0\textwidth]{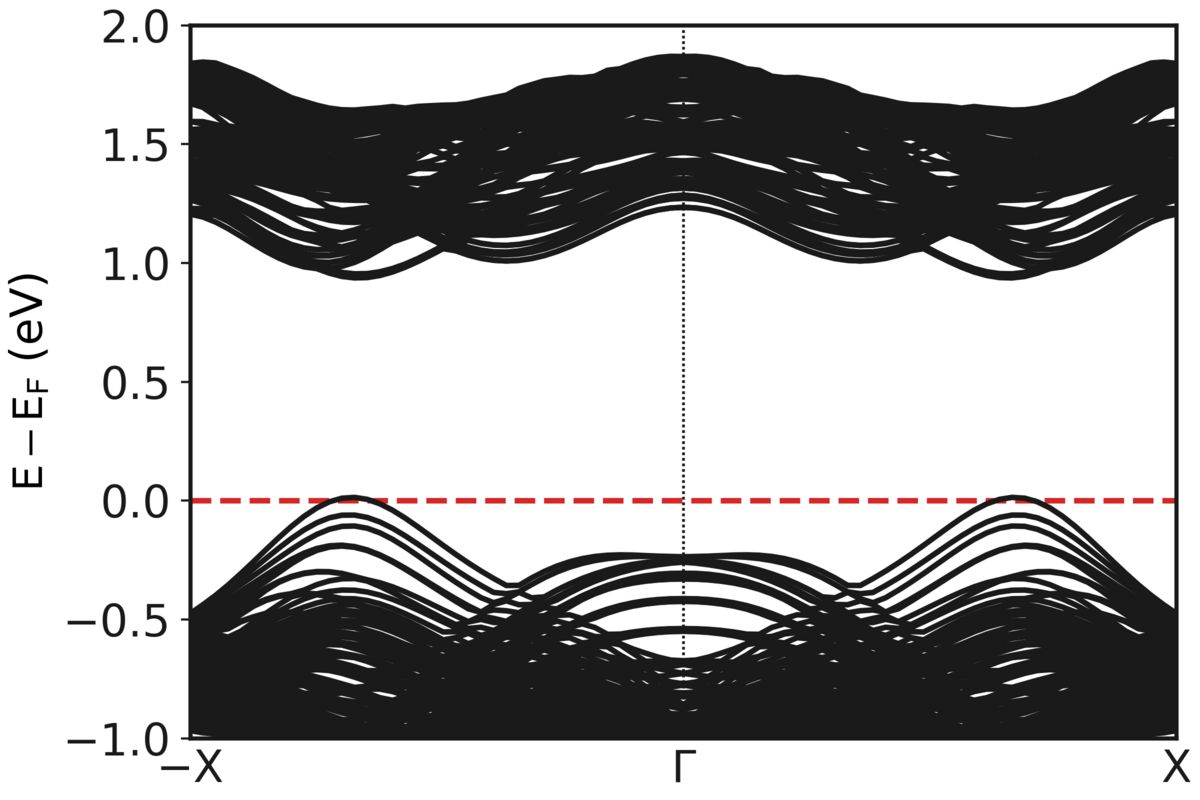}
\caption{}
\end{subfigure}
\begin{subfigure}{0.32\textwidth}
    \includegraphics[width=1.\textwidth]{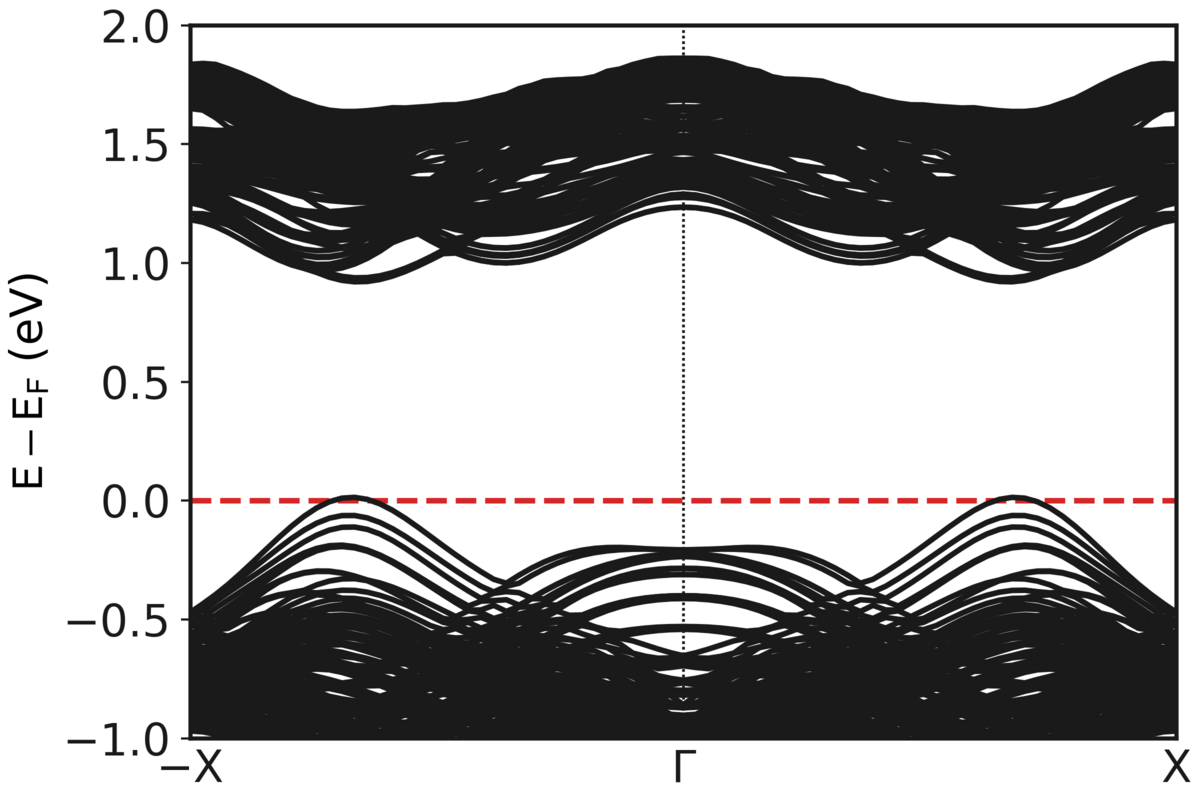}
    \caption{}
\end{subfigure}
\begin{subfigure}{0.32\textwidth}
    \includegraphics[width=1.\textwidth]{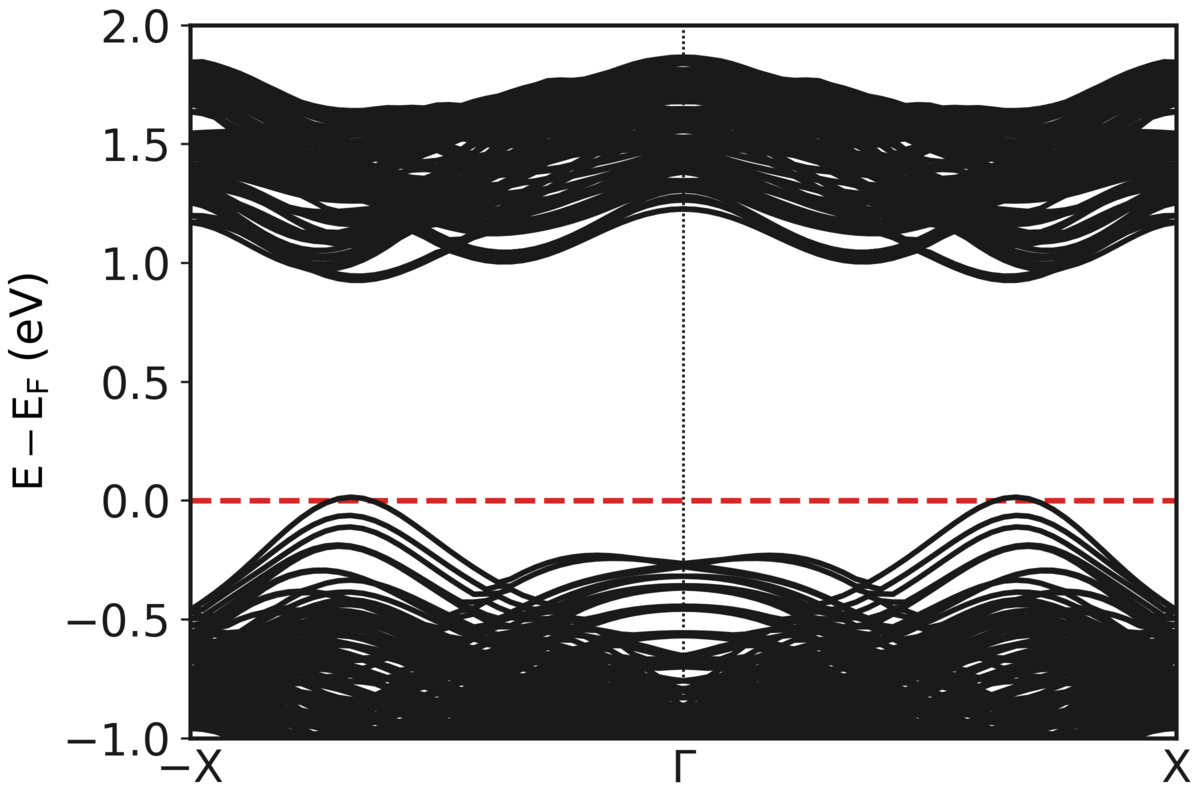}
    \caption{}
\end{subfigure}
    \caption{Band structure of \ch{WSe2$/$MoSe2} heterobilayer for different compressions a) 2.5\% b) 5\% c) 7.5\% d) 10\% e) 12.5 f) 15\% g)17.5\% h) 20\%.}
    \label{fig_band_structure_heterostructure}
\end{figure}

\begin{figure}
    \includegraphics[width=0.49\textwidth]{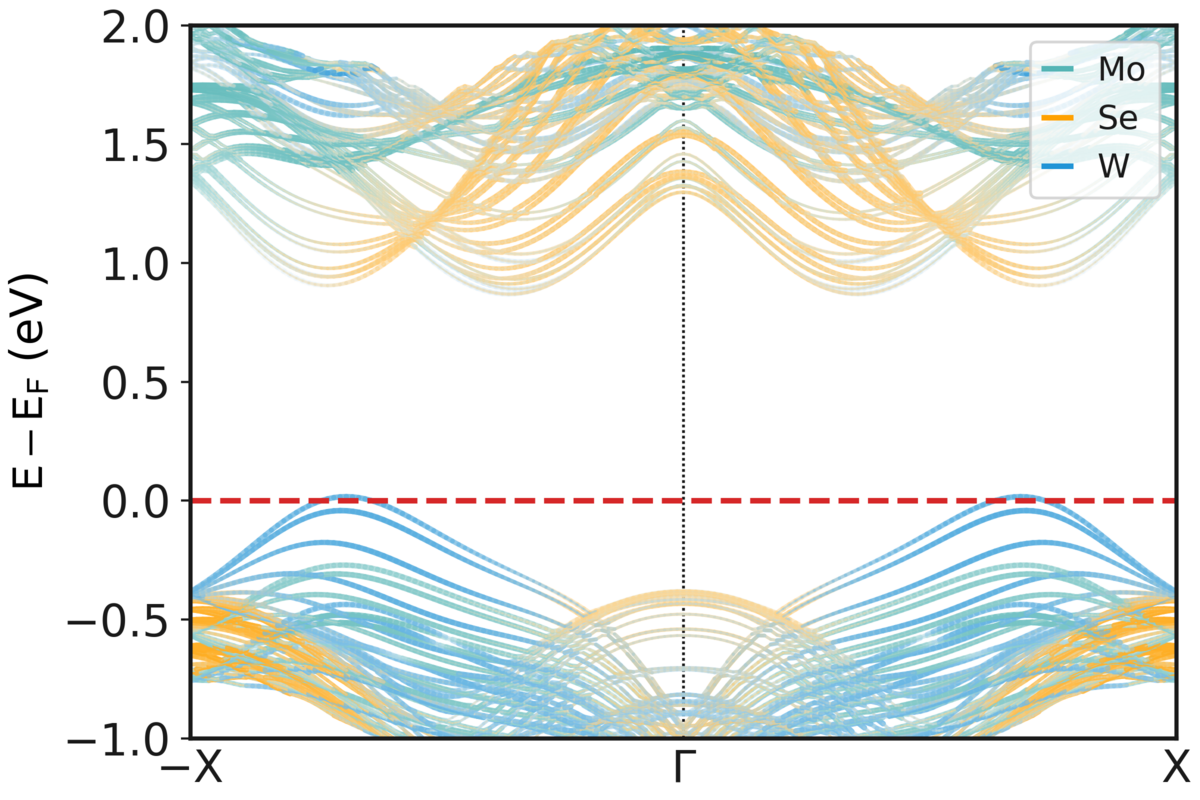}
    \includegraphics[width=0.49\textwidth]{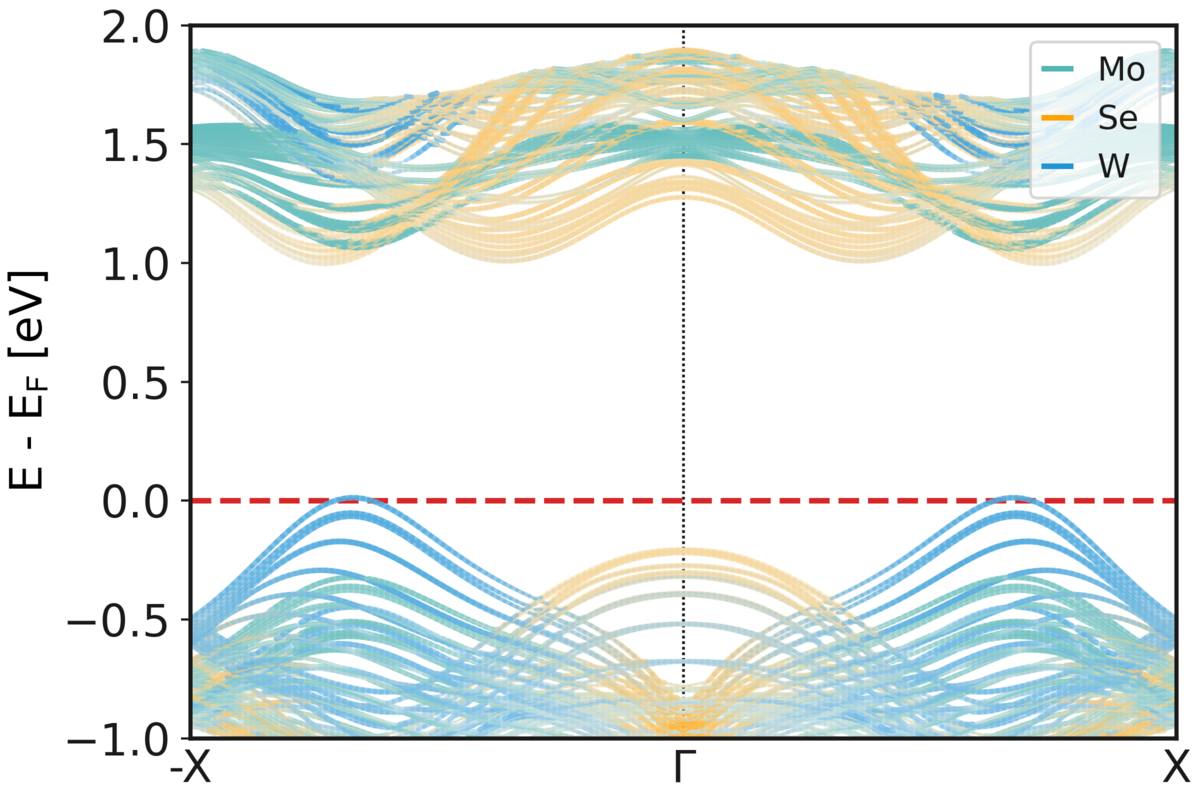}
    \caption{Wrinkling relaxes some of the strain in the system of heterebilayer \ch{WSe2$/$MoSe2}. The band structure of heterobilayer \ch{WSe2$/$MoSe2} at 5\% compression for left) without wrinkling (i.e. uniaxial strained) and right) relaxed also in the out of plane direction and wrinkles are formed}
    \label{fig_band_structure_5percent_w_wo_wrinkle}
\end{figure}

\begin{figure}
    \includegraphics[width=1.0\textwidth]{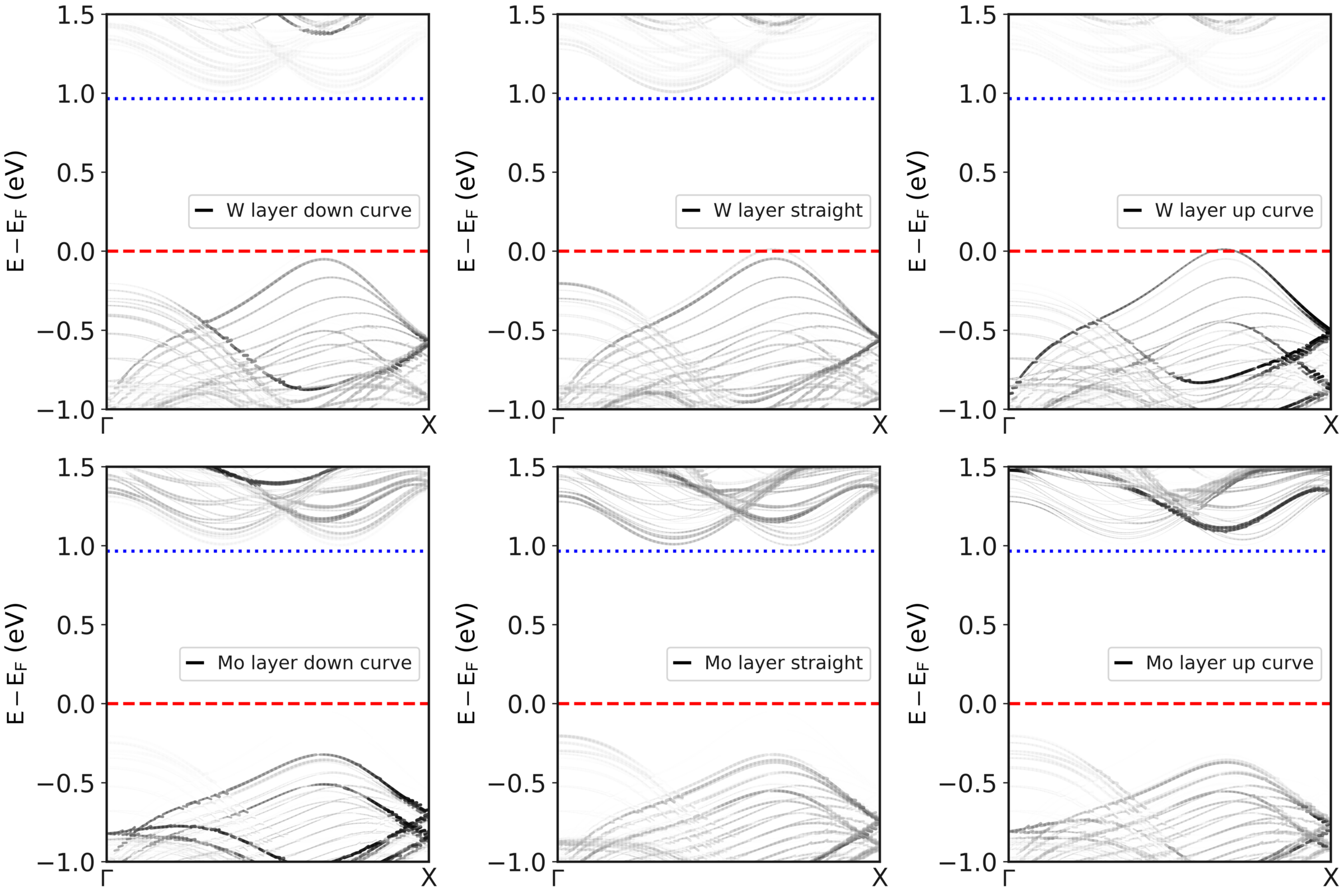}
    \includegraphics[width=0.5\textwidth]{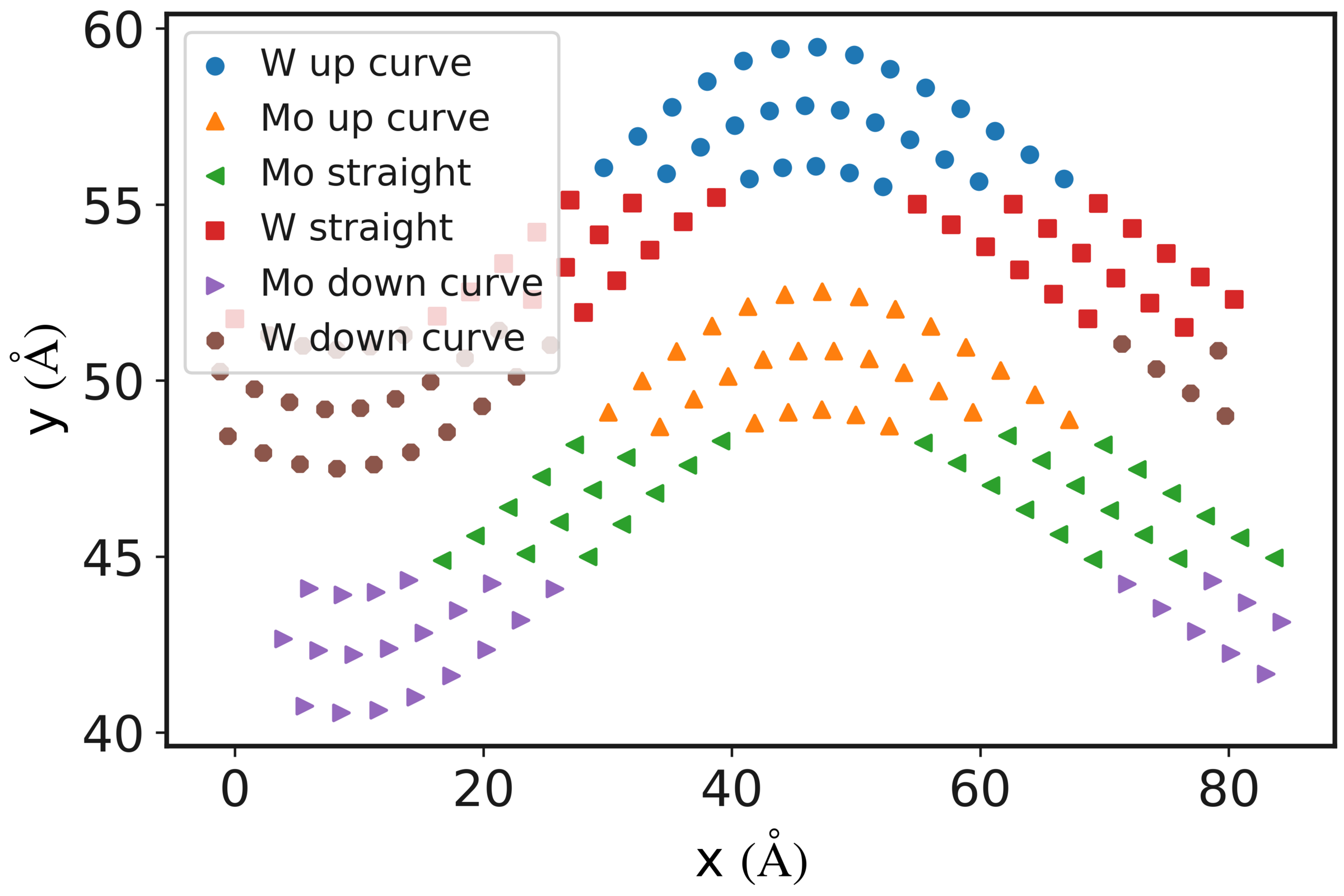}
    \caption{Band structure of \ch{WSe2$/$MoSe2} heterobilayer at 2.5\% compression projected on different sections of the wrinkle, the location of VBM (red) and CBM (blue) are also indicated to help the eye, additionally (below) the position of contributing atoms to the band projection is given.}
    \label{fig_band_heterobilayer_WSe2MoSe2_different_layer_different_sections_2.5}
\end{figure}

\begin{figure}
    \includegraphics[width=0.32\textwidth]{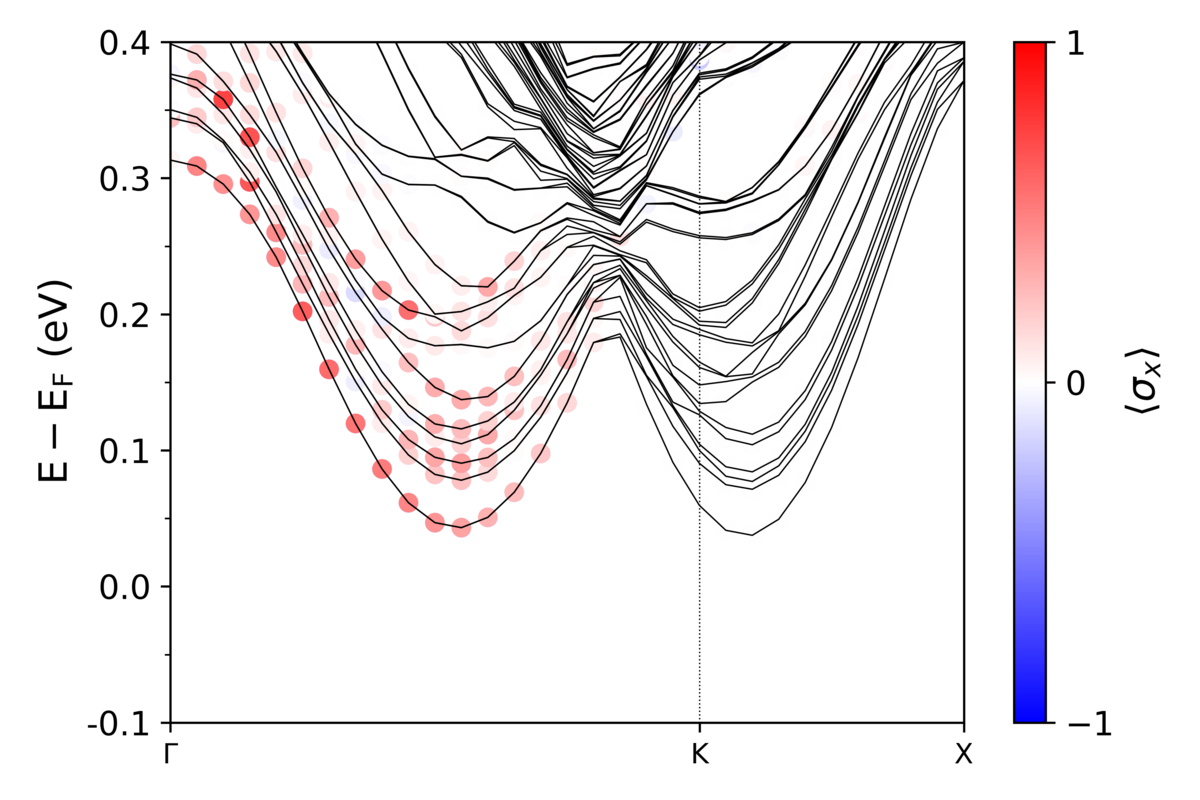}
    \includegraphics[width=0.32\textwidth]{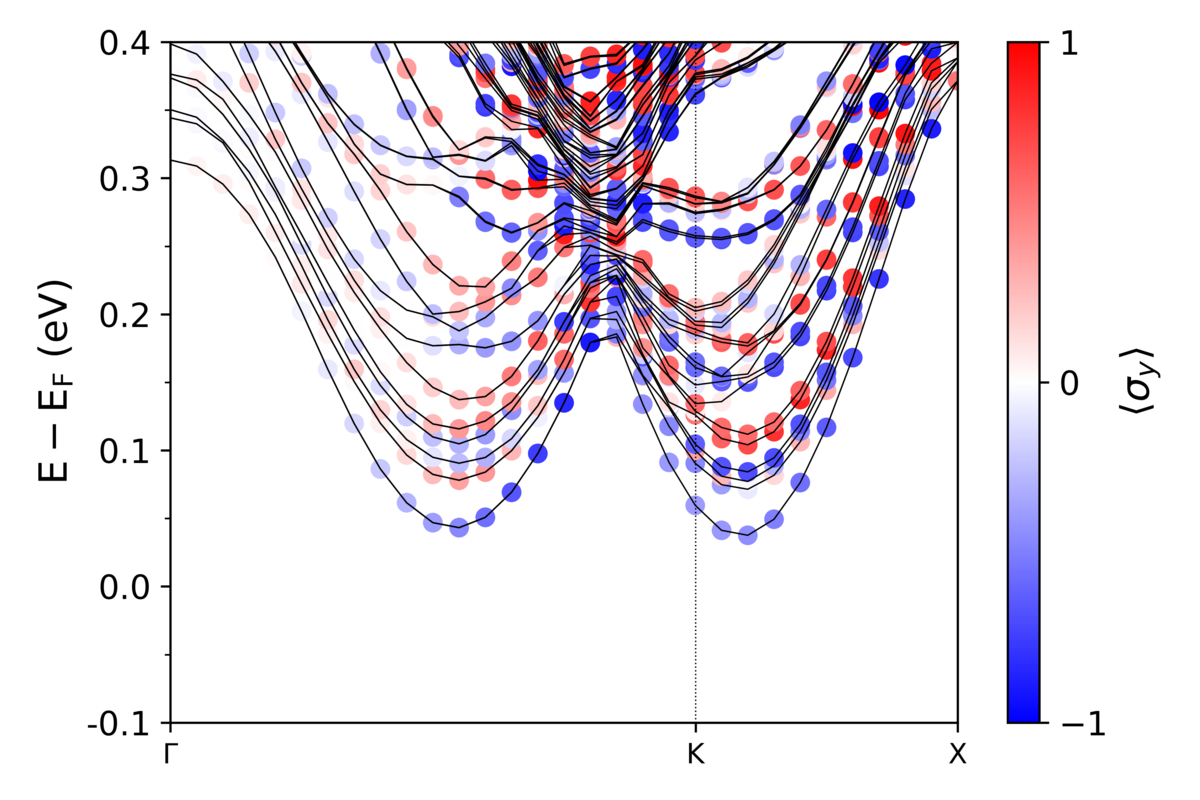}
    \includegraphics[width=0.32\textwidth]{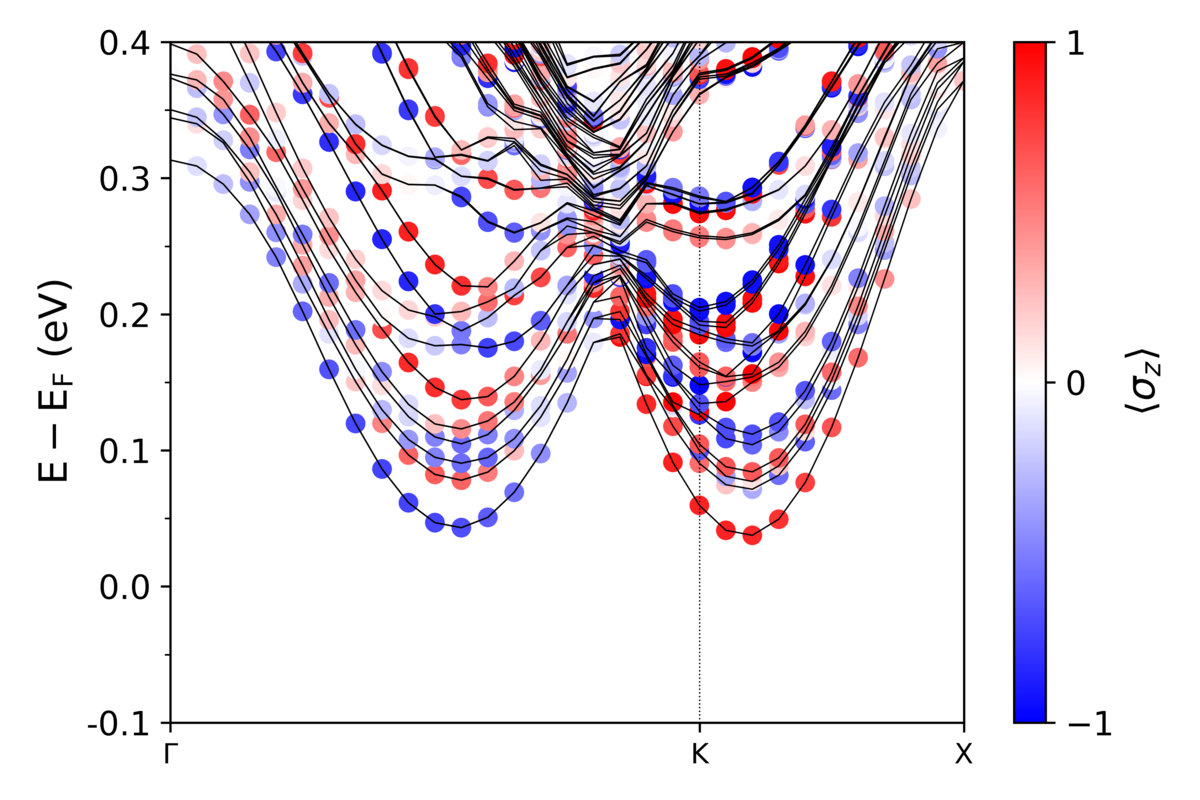}    
    \includegraphics[width=0.32\textwidth]{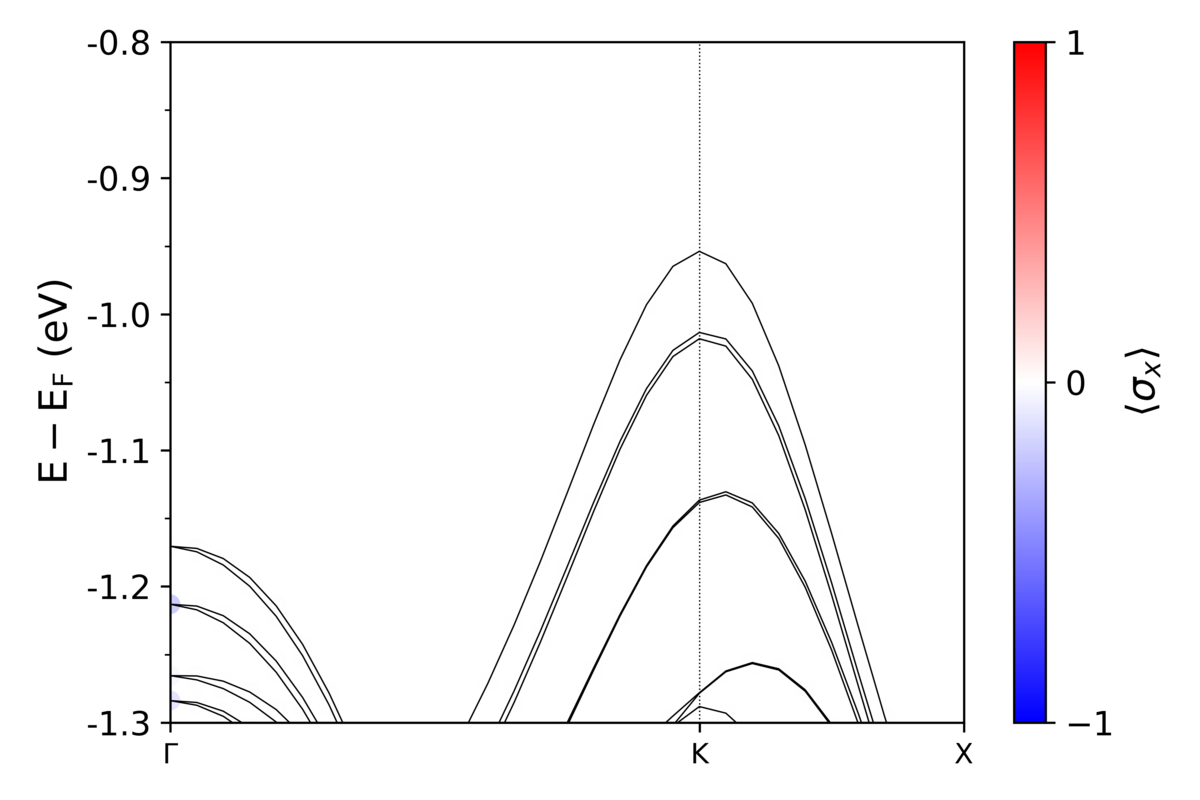}
    \includegraphics[width=0.32\textwidth]{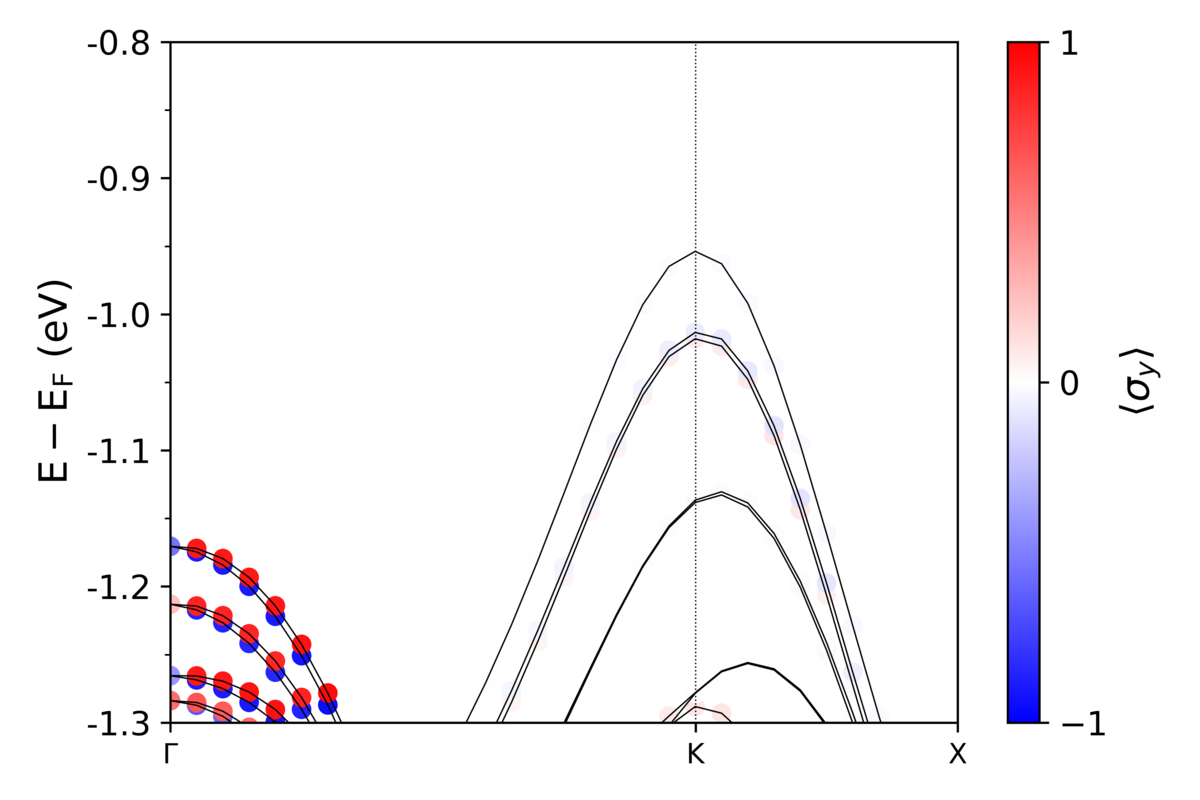}
    \includegraphics[width=0.32\textwidth]{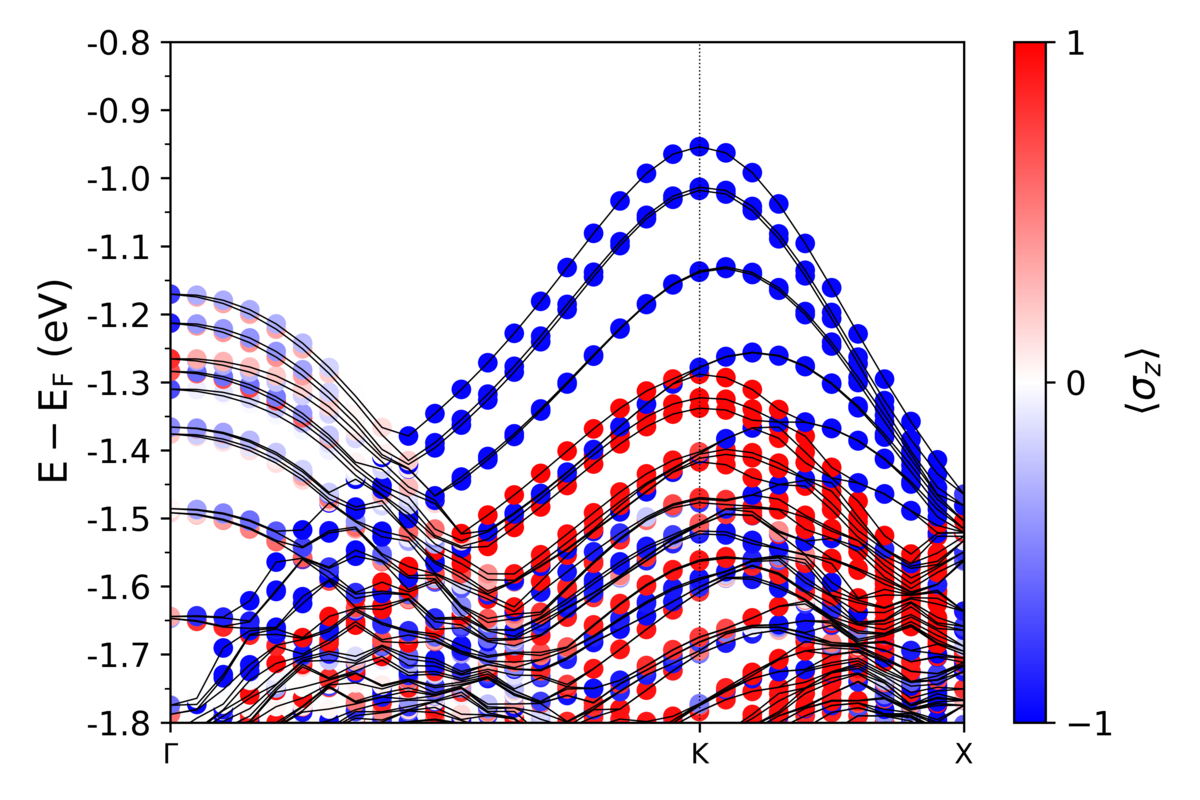}
    \caption{Expectation values of the Pauli matrices $\langle\sigma_i\rangle$ of the wrinkled \ch{WSe2$/$MoSe2} at 2.5\% compression}
    \label{fig_spin_texture_2.5}
\end{figure}
\begin{figure}
    \includegraphics[width=0.32\textwidth]{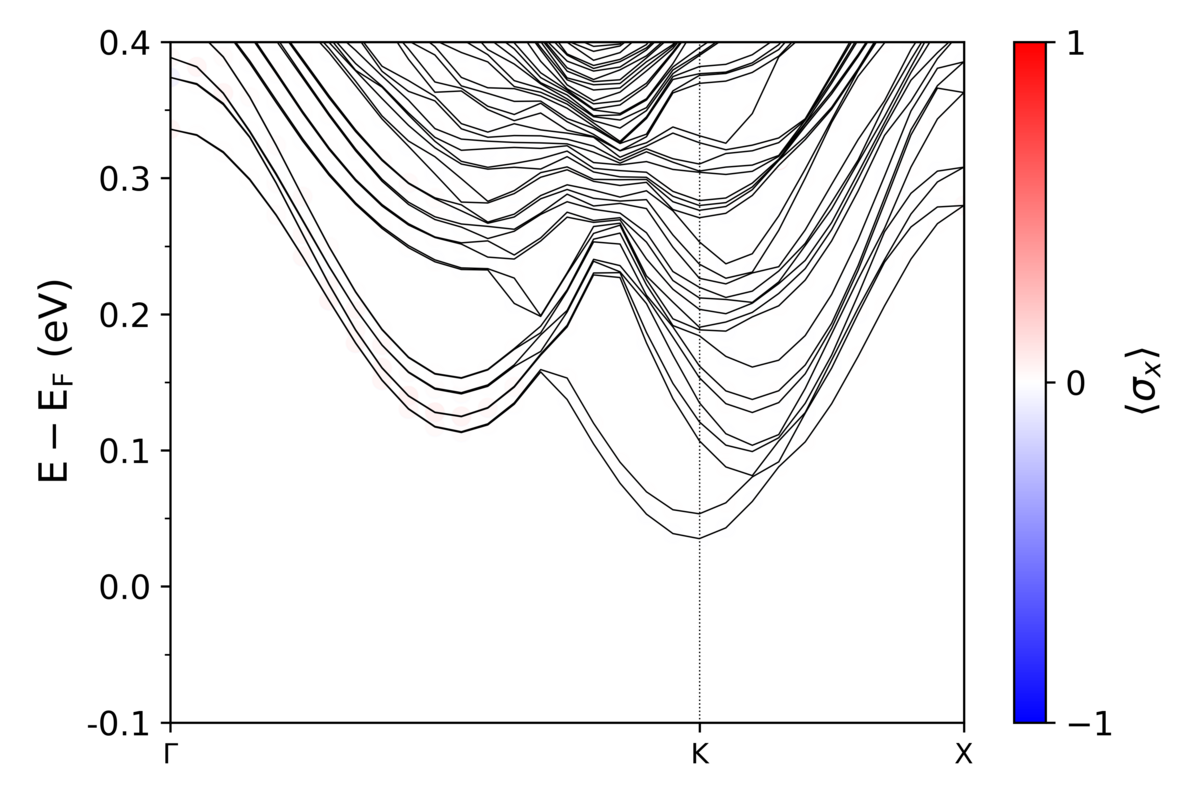}
    \includegraphics[width=0.32\textwidth]{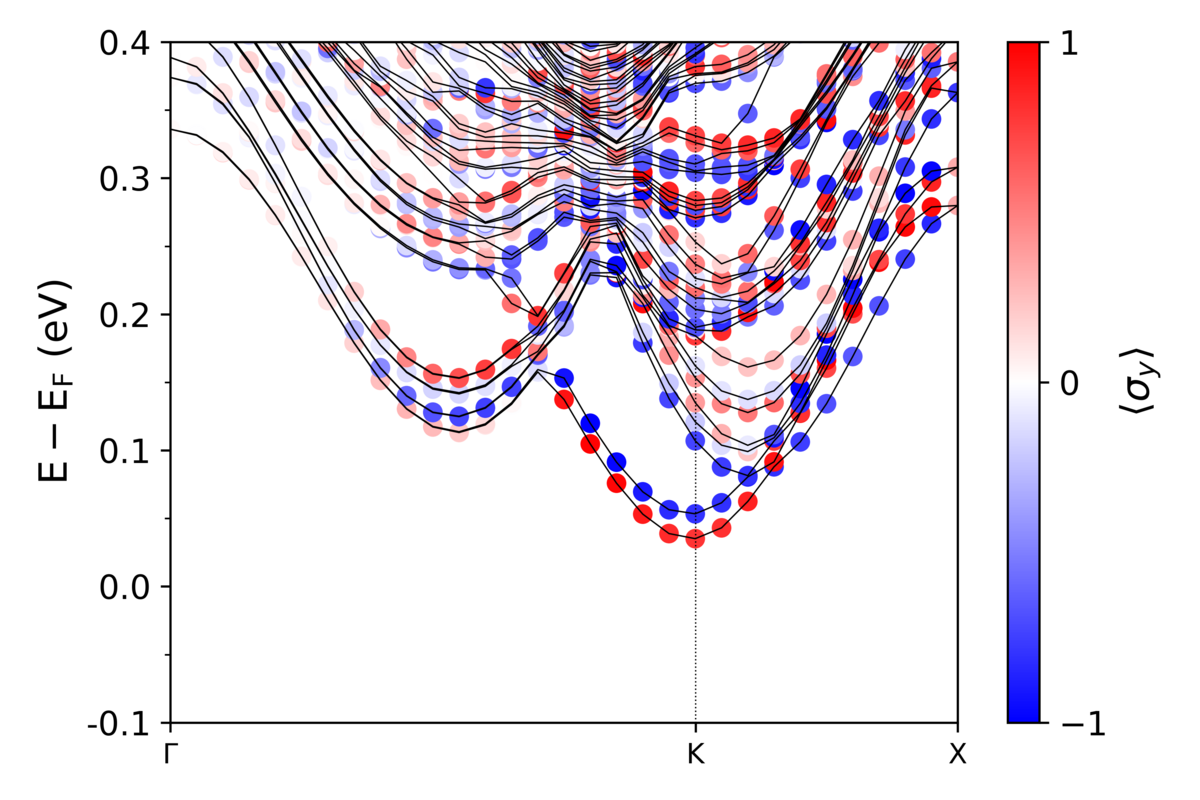}
    \includegraphics[width=0.32\textwidth]{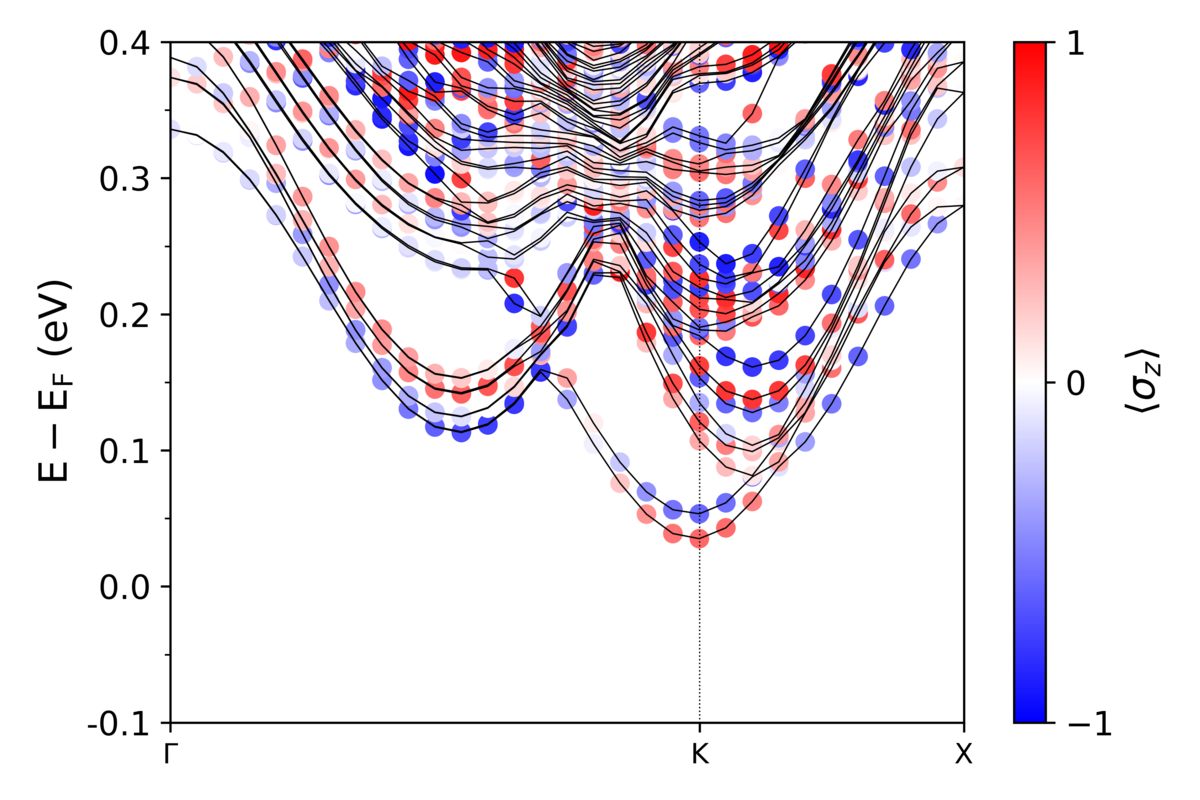}    
    \includegraphics[width=0.32\textwidth]{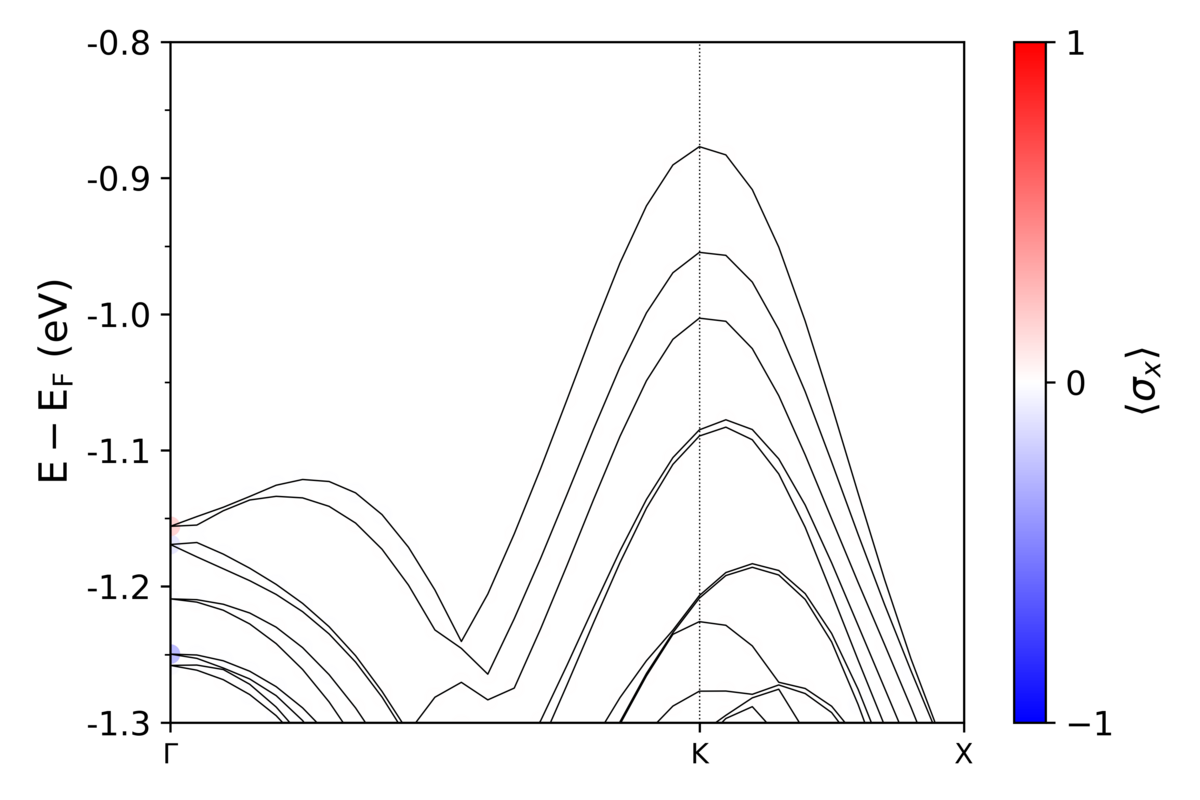}
    \includegraphics[width=0.32\textwidth]{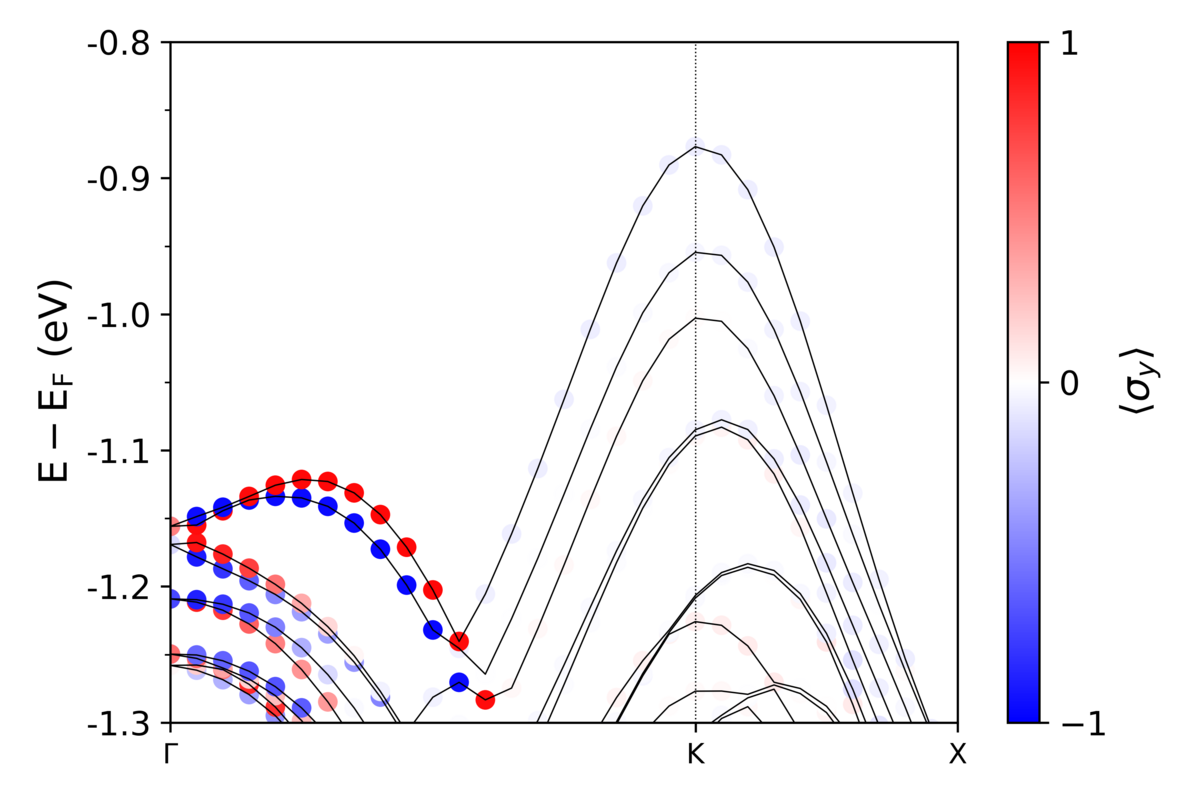}
    \includegraphics[width=0.32\textwidth]{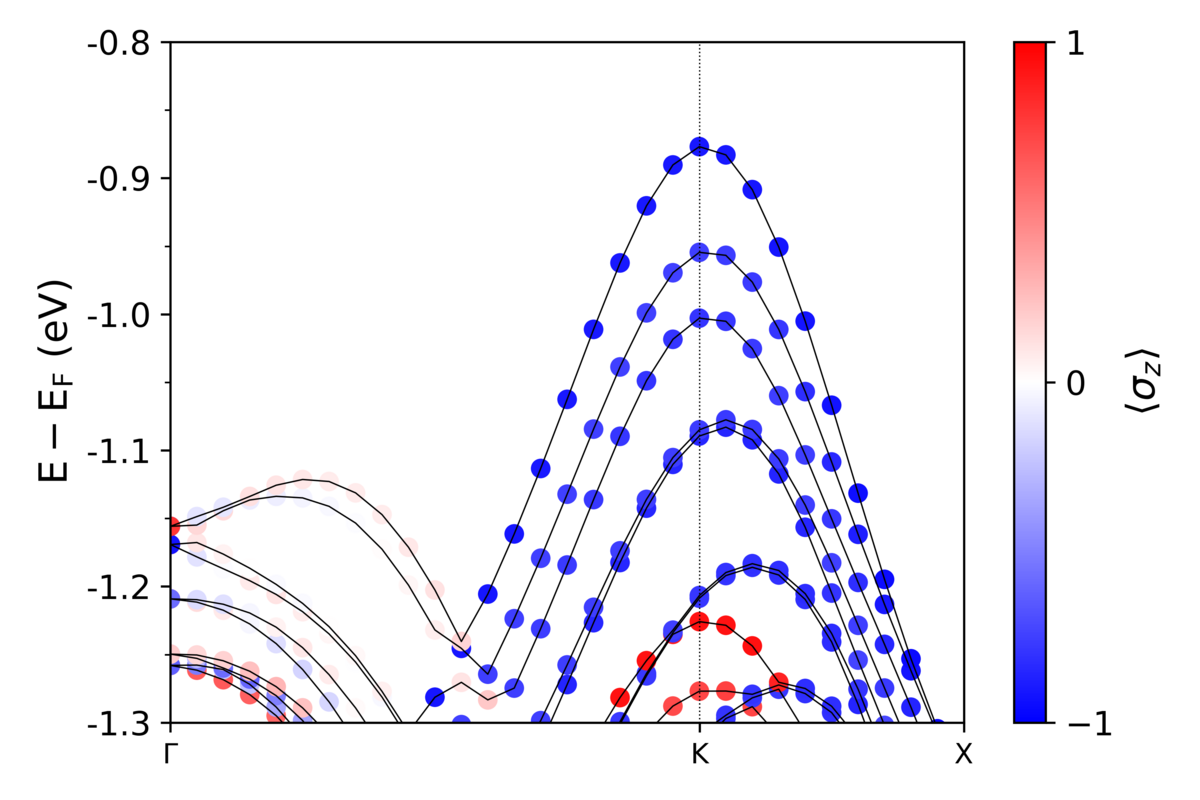}
    \caption{Expectation values of the Pauli matrices $\langle\sigma_i\rangle$ of the wrinkled \ch{WSe2$/$MoSe2} at 20\% compression, red (blue) line indicates VBM (CBM) energies to help the eye}
    \label{fig_spin_texture_20}
\end{figure}

\begin{figure}
\begin{subfigure}{0.32\textwidth}
    \includegraphics[width=1.0\textwidth]{images/spin_texture_heterobilayer/Spin_texture_z_cb_2.5.png}
    \caption{}
\end{subfigure}
\begin{subfigure}{0.32\textwidth}
     \includegraphics[width=1.0\textwidth]{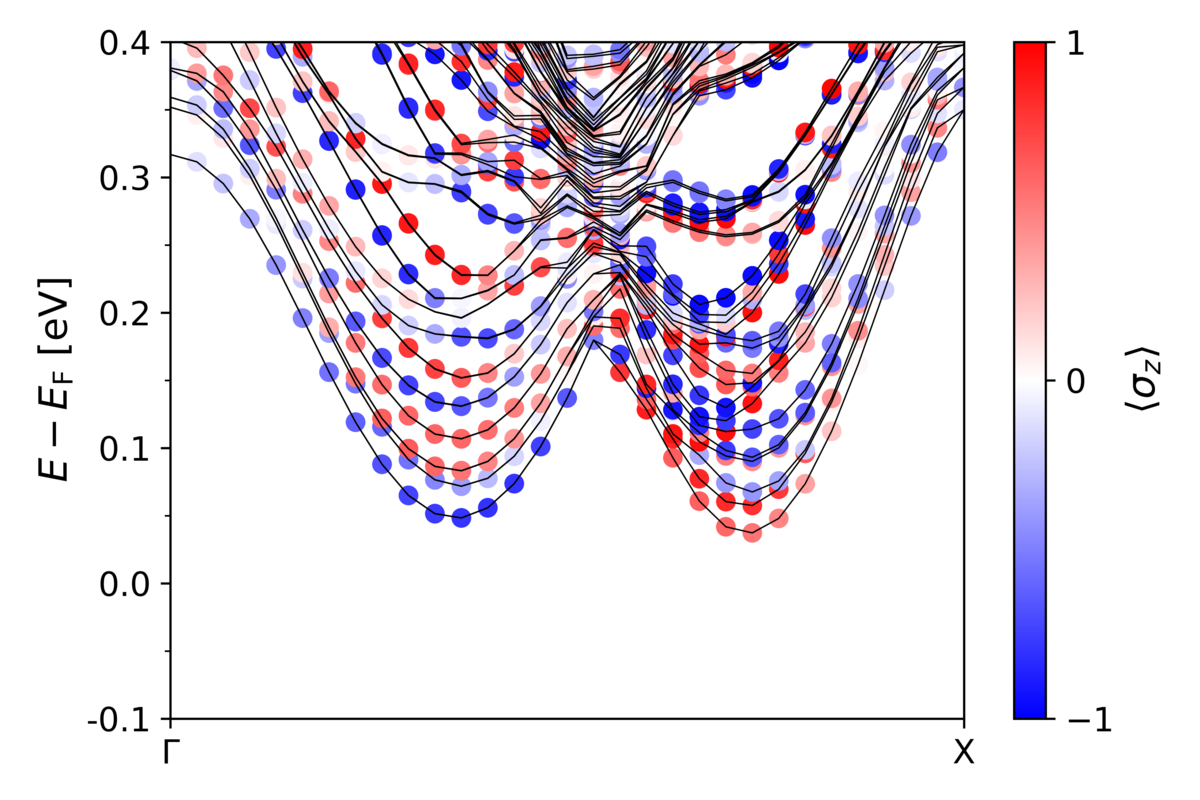}
     \caption{}
\end{subfigure}
   \begin{subfigure}{0.32\textwidth}
       \includegraphics[width=1.0\textwidth]{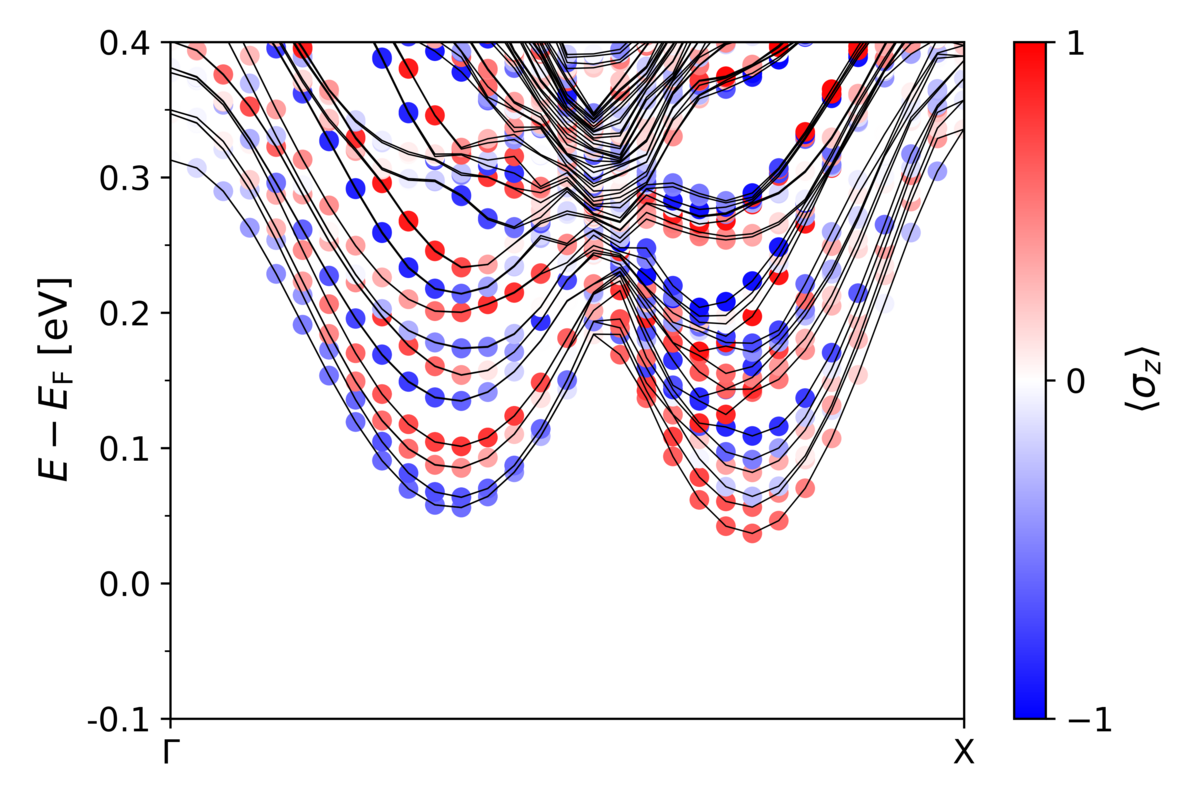}
       \caption{}
\end{subfigure}
\begin{subfigure}{0.32\textwidth}
    \includegraphics[width=1.0\textwidth]{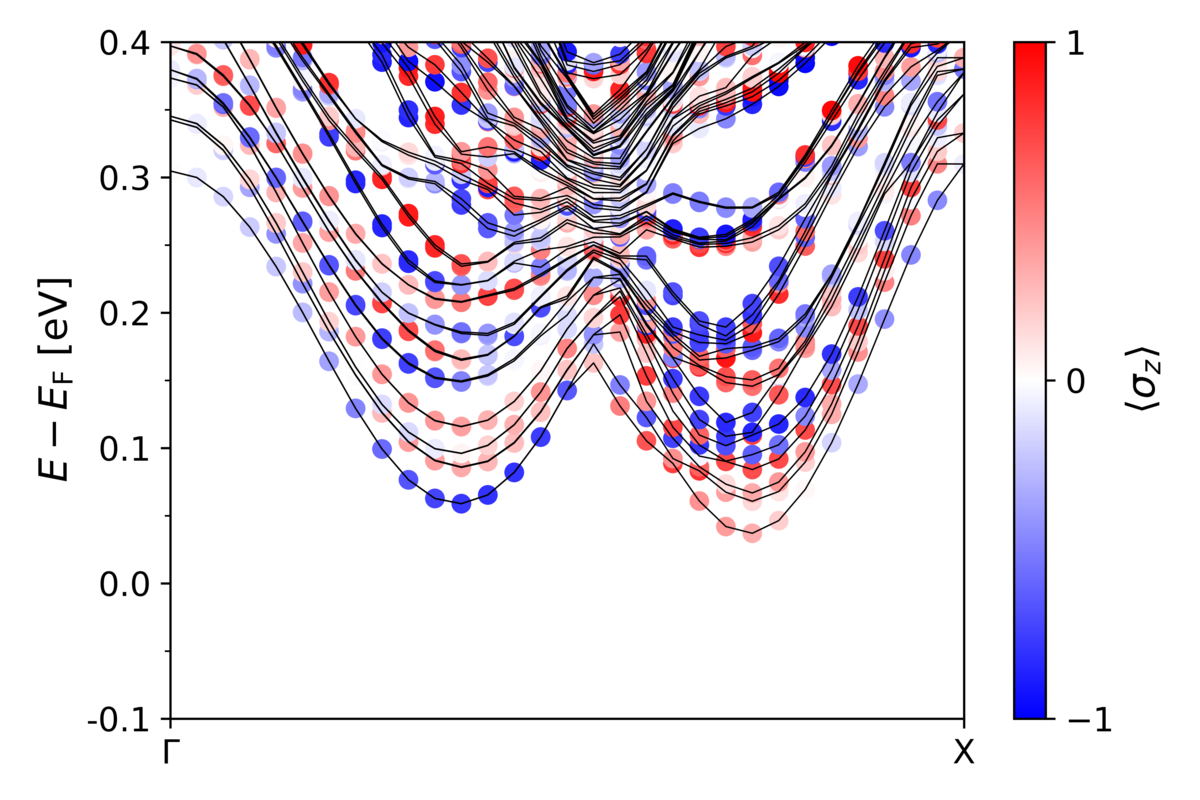}
    \caption{}
\end{subfigure}
 \begin{subfigure}{0.32\textwidth}
    \includegraphics[width=1.0\textwidth]{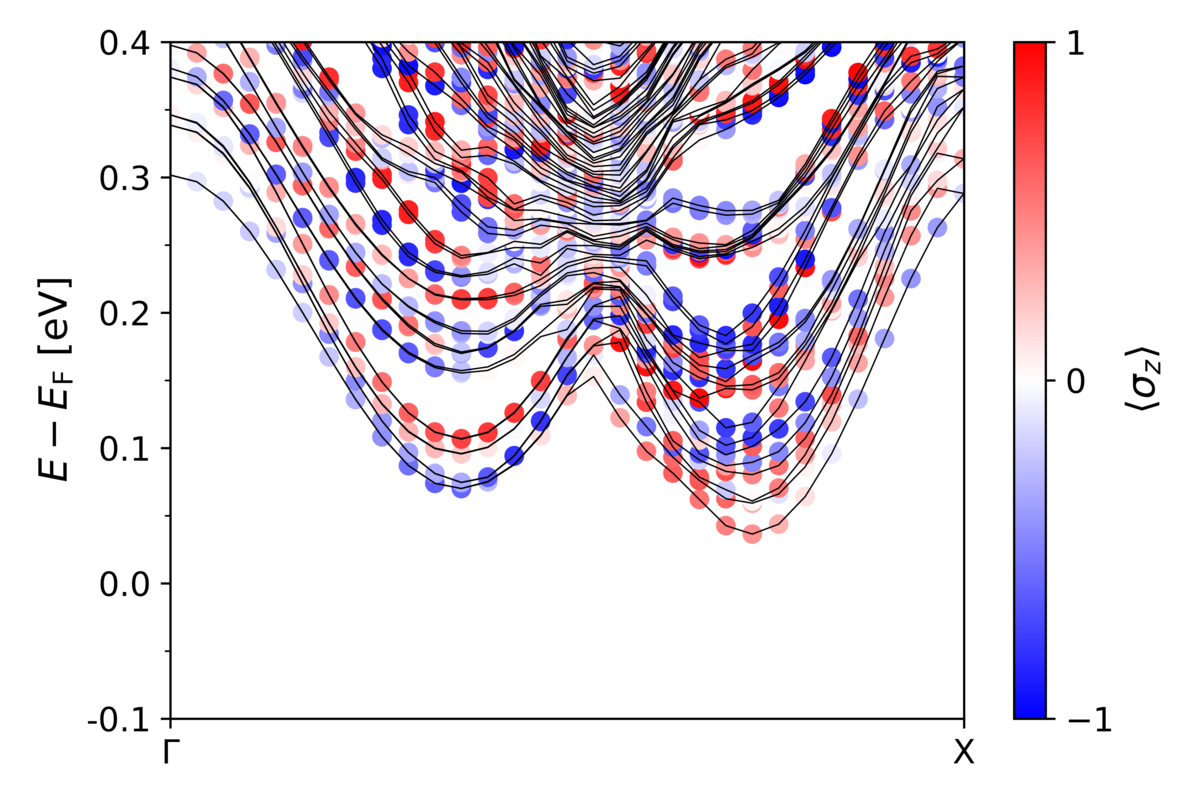}
    \caption{}
 \end{subfigure}
 \begin{subfigure}{0.32\textwidth}
     \includegraphics[width=1.0\textwidth]{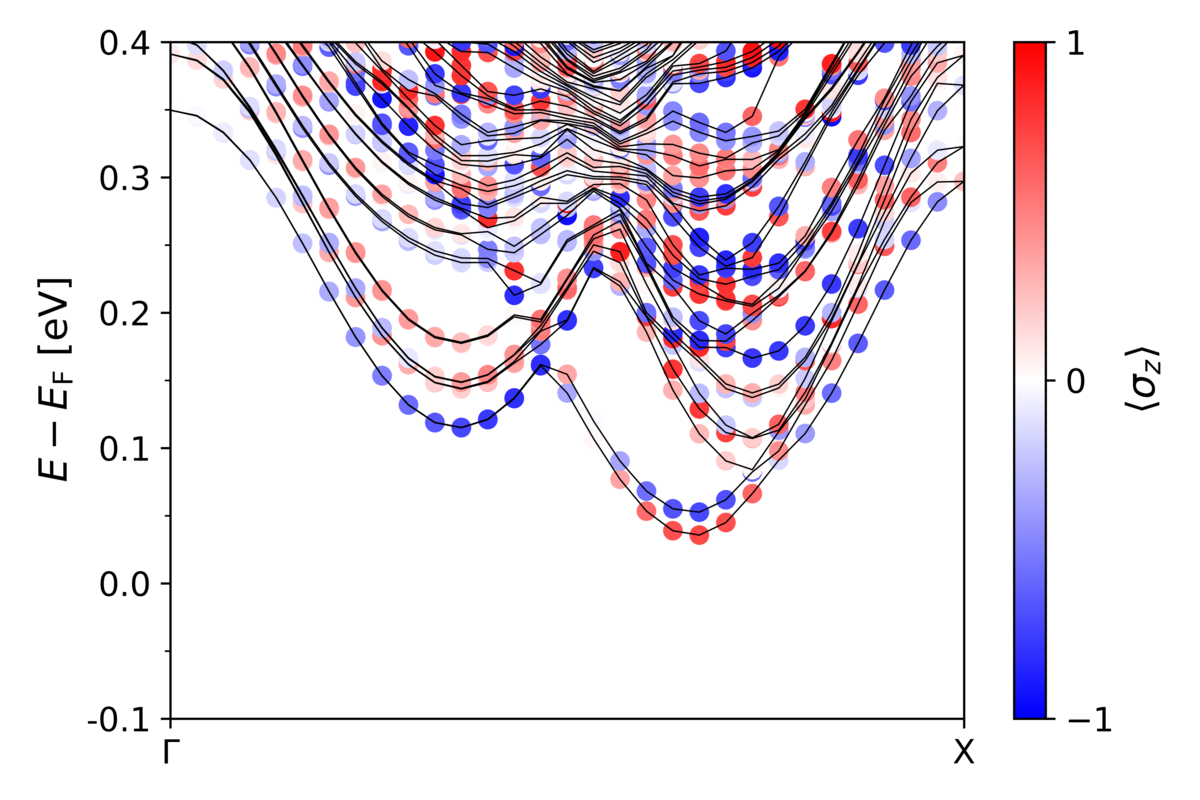}
     \caption{}
 \end{subfigure} 
 \begin{subfigure}{0.32\textwidth}
     \includegraphics[width=1.0\textwidth]{images/spin_texture_heterobilayer/Spin_texture_z_cb_20.png}
     \caption{}
 \end{subfigure}       
    \caption{Z component of the expectation values of the Pauli matrices $\langle\sigma_z\rangle$ for the lowest conduction bands (CB) of the wrinkled \ch{WSe2$/$MoSe2} at different strain a) 2.5 \% b) 5\% c) 7.5 \% d) 10\% e) 12.5 \% 15 \% f) 17.5 \% g) 20 \%}
    \label{fig_spin_texture_cb}
\end{figure}

\begin{figure}
    \begin{subfigure}{0.32\textwidth}
        \includegraphics[width=1\textwidth]{images/spin_texture_heterobilayer/Spin_texture_z_vb_2.5.png}
        \caption{}
    \end{subfigure}
    \begin{subfigure}{0.32\textwidth}
        \includegraphics[width=1\textwidth]{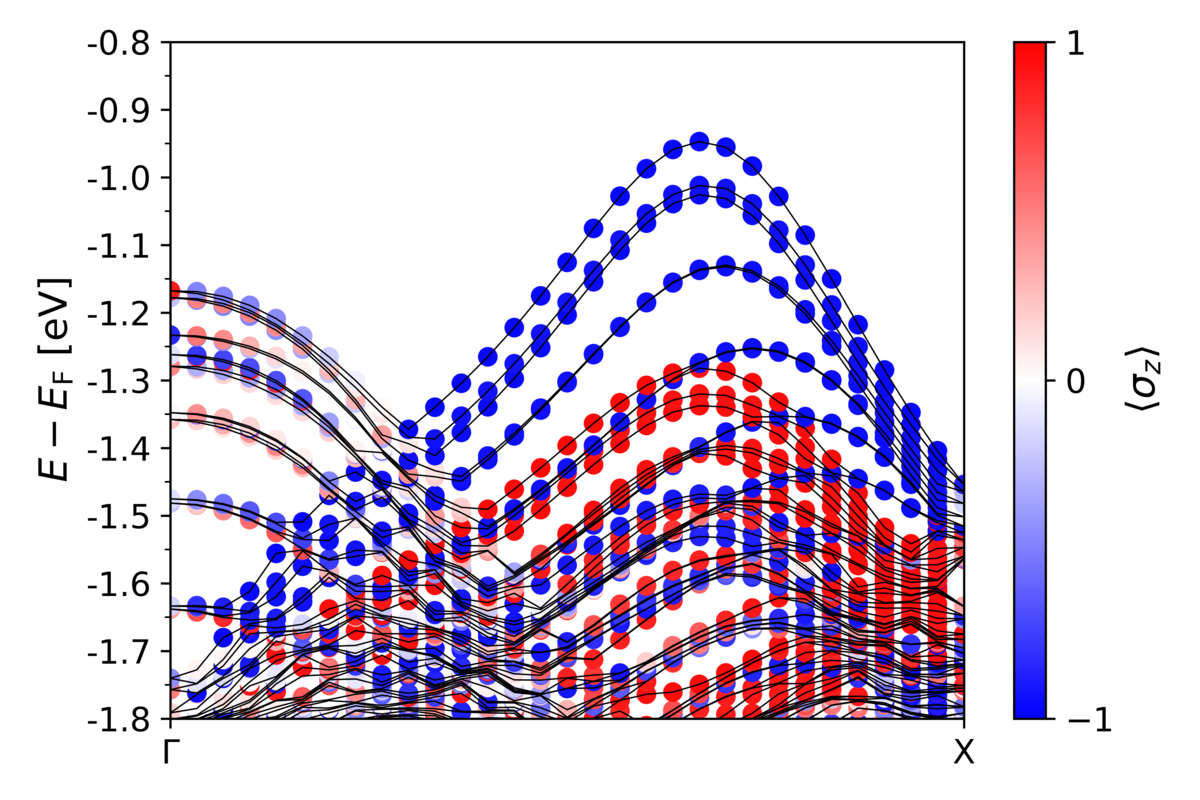}
        \caption{}
    \end{subfigure}
    \begin{subfigure}{0.32\textwidth}
        \includegraphics[width=1\textwidth]{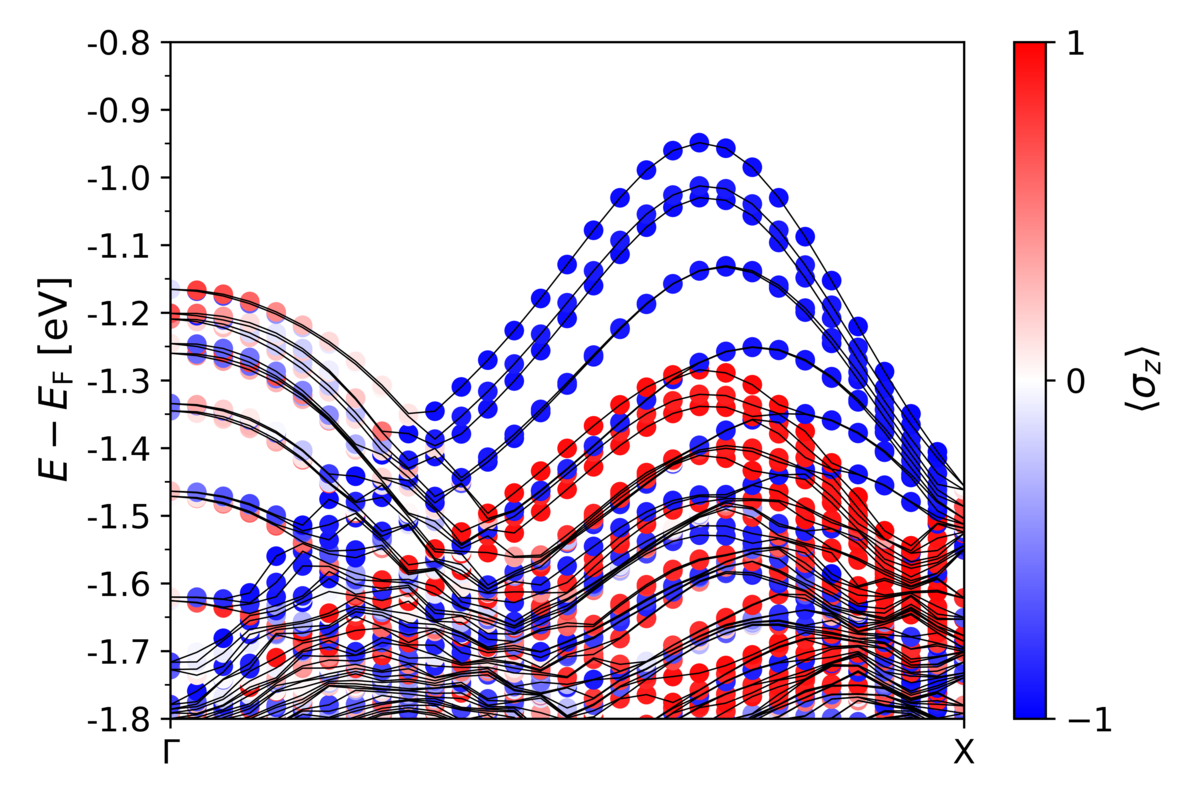} 
        \caption{}
    \end{subfigure}
    \begin{subfigure}{0.32\textwidth}
        \includegraphics[width=1\textwidth]{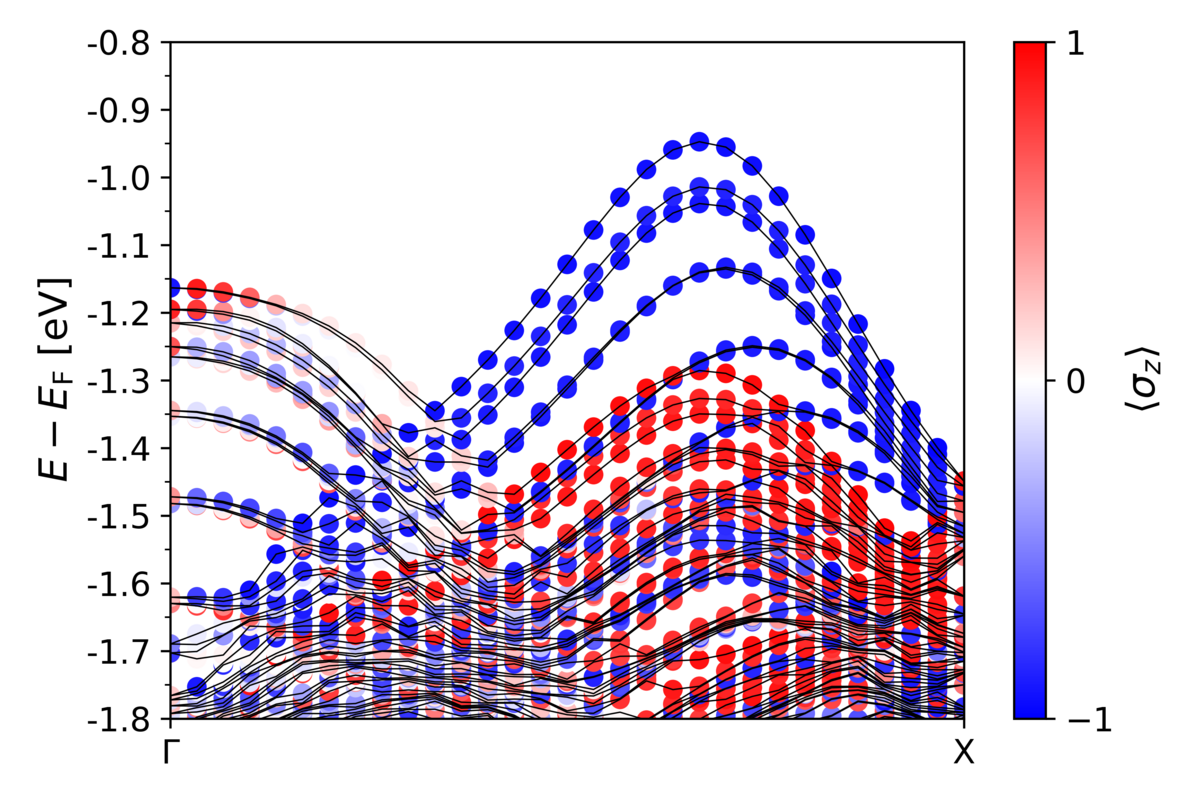}
        \caption{}
    \end{subfigure}
    \begin{subfigure}{0.32\textwidth}
        \includegraphics[width=1.0\textwidth]{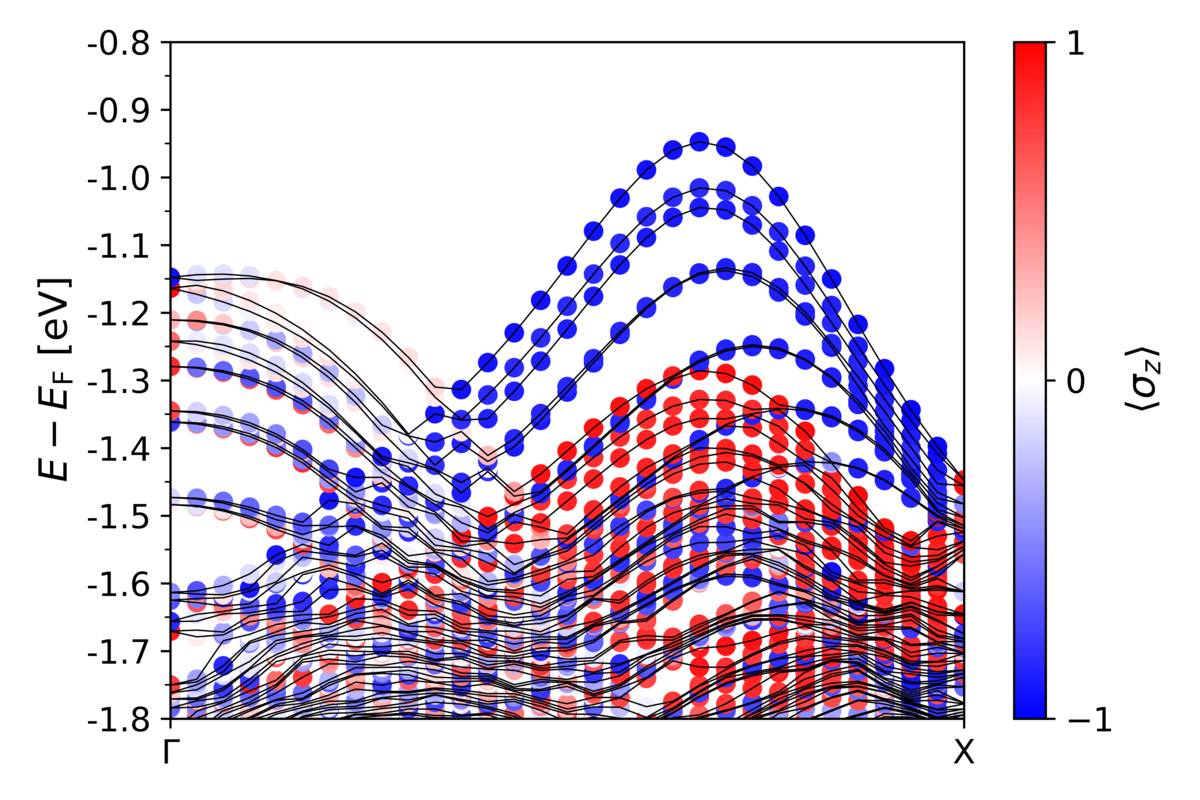}
        \caption{}
    \end{subfigure}
    \begin{subfigure}{0.32\textwidth}
         \includegraphics[width=1.0\textwidth]{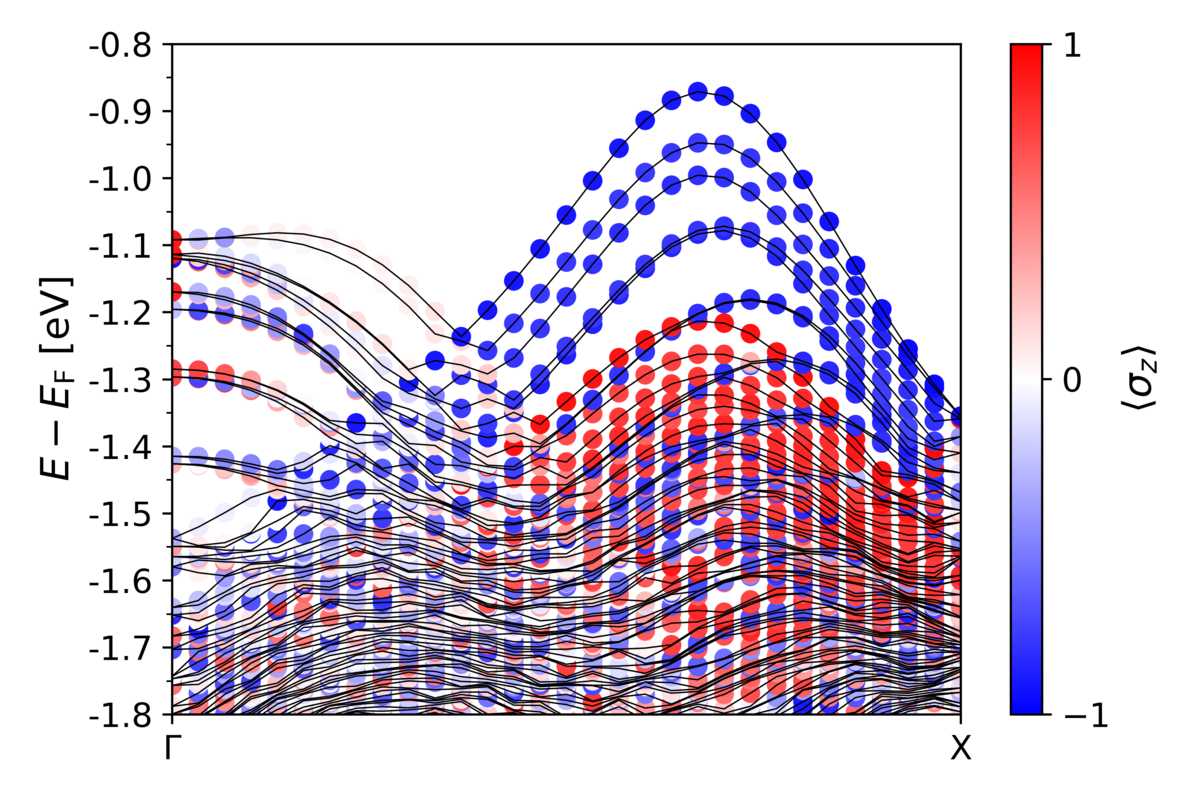}
         \caption{}
    \end{subfigure}
   \begin{subfigure}{0.32\textwidth}
        \includegraphics[width=1.0\textwidth]{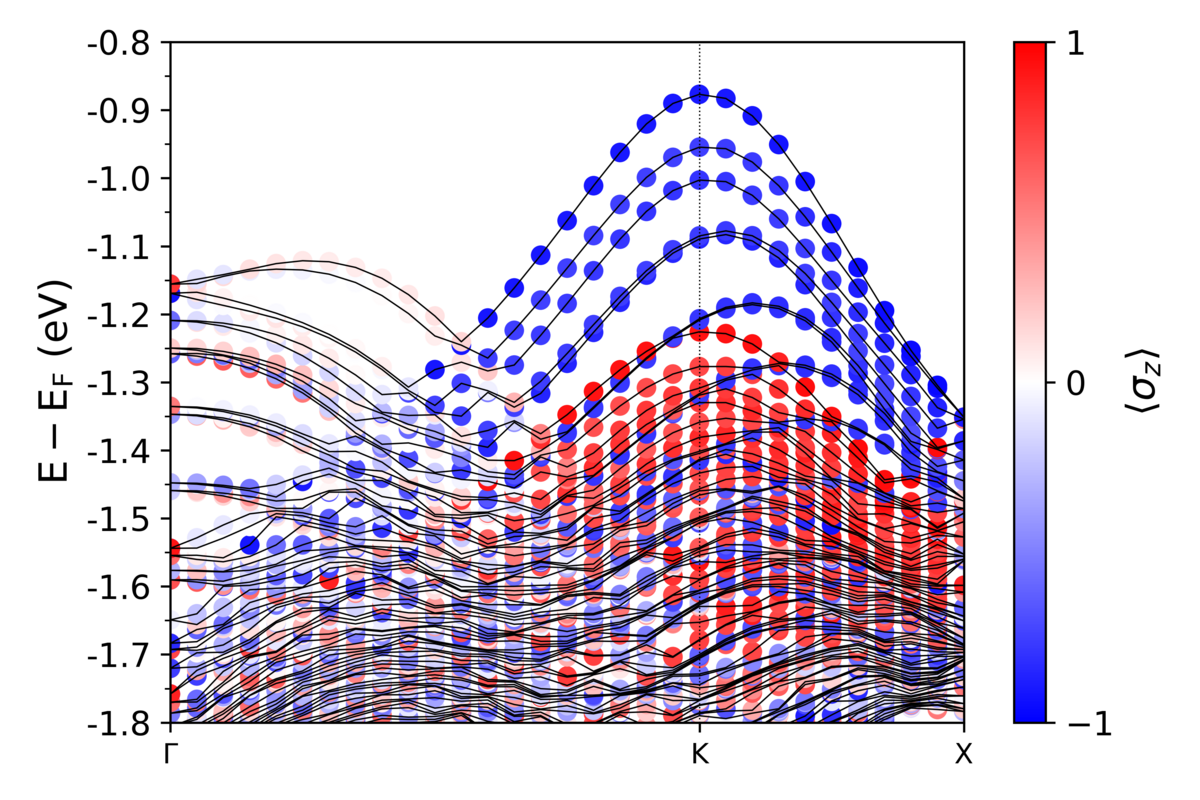}
        \caption{}
   \end{subfigure}  
    \caption{Z component of the expectation values of the Pauli matrices $\langle\sigma_z\rangle$ for the highest valence bands (VB) of the wrinkled \ch{WSe2$/$MoSe2} at different strain a) 2.5\% b) 5\% c) 7.5\% d) 10\% e)12.5 f) 17.5\% g) 20\%}
    \label{fig_spin_texture_vb}
\end{figure}



\begin{table}
\caption{Maximum and minimum strain at each layer of the wrinkled heterobilayer \ch{WSe2$/$MoSe2} structure, defined as $\epsilon=\frac{d-d_{flat}}{d_{flat}}$, where d is the \ch{M-M} distance and M stands for \ch{Mo} or \ch{W} \label{tabel_strain_heterobilayer}}
\begin{adjustbox}{max width=\textwidth}
\begin{tabular}{ c|c|c|c|c} 
  \textbf{Compression}     & \textbf{Maximum strain \ch{Mo} layer}  & \textbf{Minimum strain \ch{Mo} layer} & \textbf{Maximum strain \ch{W} layer} & \textbf{Minimum strain \ch{W} layer} \\ \hline
2.5            & 0.0005  & -0.0064  & 0.0005 & -0.0064 \\
5              & 0.0027  & -0.0049  & 0.0027 & -0.0049 \\
7.5            & 0.0011  & -0.0080  & 0.0010 & -0.0080 \\
10             &0.0038   & -0.0070  & 0.0038 & -0.0070   \\
12.5           & 0.0064  & -0.0088  & 0.0064 & -0.0087\\
15             & 0.0114  & -0.0042  & 0.0114 & -0.0042 \\
17.5           & 0.0143  & -0.0041  & 0.0143 & -0.0041 \\
20             & 0.0135  & -0.0070  & 0.0135 & -0.0070 \\   
\end{tabular}
\end{adjustbox}
\end{table}

\section*{\ch{WSe2/MoSe2} heterobilayer wrinkle}

\begin{table}
    \centering
    \caption{Different band gaps of heterobilayer \ch{WSe2$/$MoSe2} wrinkles at the backfolded K point for different compression (at least 50 \% localization). \label{table_band_gap_hetero_section}} 
    \begin{adjustbox}{max width=\textwidth}
    \begin{tabular}{c|c|c|c|c}
\textbf{Compression}  & \textbf{Interlayer band gap} & \textbf{The smallest band gap} & \textbf{\ch{W} intralayer band gap} & \textbf{\ch{Mo} intralayer band gap} \\
\hline
2.5          & 1.083  & 1.017 & 1.362 & 1.417 \\
5            & 1.053  & 1.002 & 1.330 & 1.387 \\
7.5          & 1.068  & 1.004 & 1.340 & 1.403 \\
10           & 1.031  & 1.002 & 1.365 & 1.369   \\
12.5         & 1.024  & 1.003 & 1.355 & 1.362 \\
15           & 0.923  & 0.923 & 1.190 & 1.264   \\
17.5         & 0.907  & 0.907 & 1.171 & 1.249 \\
20           & 0.912  & 0.912 & 1.183 & 1.261 \\
    \end{tabular}

    \label{table_inter-intralayer_bandgap_heterobilayer}
\end{adjustbox}
\end{table}

\begin{table}
\caption{Structural parameters of the heterobilayer \ch{WSe2$/$MoSe2} in {\AA} for different compressions- A and R values are extracted using the fitted curve explained in the SI \label{table_structural_heterobilayer}}
\begin{tabular}{ c|c|c|c|c|c|c} 
    \hline
     \textbf{Compression}   &  \textbf{$\rm \lambda$} & \textbf{$ \rm  A_{W_{layer}}$}& \textbf{$\rm A_{Mo_{layer}}$} & \textbf{$\rm A_{both}$} & \textbf{ $\rm R_{{min}_{W_{layer}}} $} & \textbf{$\rm R_{{min}_{Mo_{layer}}} $} \\ \hline
      2.5 & 83.137 &  4.309&  4.308&  7.793 & 32.192 & 27.348    \\
      5   & 81.005 &  6.095&  6.103&  9.555 & 28.072 & 33.868    \\
     7.5  & 78.874 &  7.382&  7.394& 10.829 &22.201  & 27.537    \\
     10   & 76.741 &  8.506&  8.52 & 11.945 & 18.248 & 16.35     \\
     12.5 & 74.610 &  9.473&  9.487& 12.886 & 15.733 & 13.707    \\
     15   & 72.479 & 10.158& 10.14 & 13.541 & 8.201  & 11.219    \\
     17.5 & 70.347 & 10.944& 10.923& 14.326 & 11.576 &  9.765    \\     
     20   & 68.215 & 11.57 & 11.526& 15.047 &10.273  &  8.625    \\   
\end{tabular}
\end{table}

\begin{figure}
    \centering
    \includegraphics[width=0.49\textwidth]{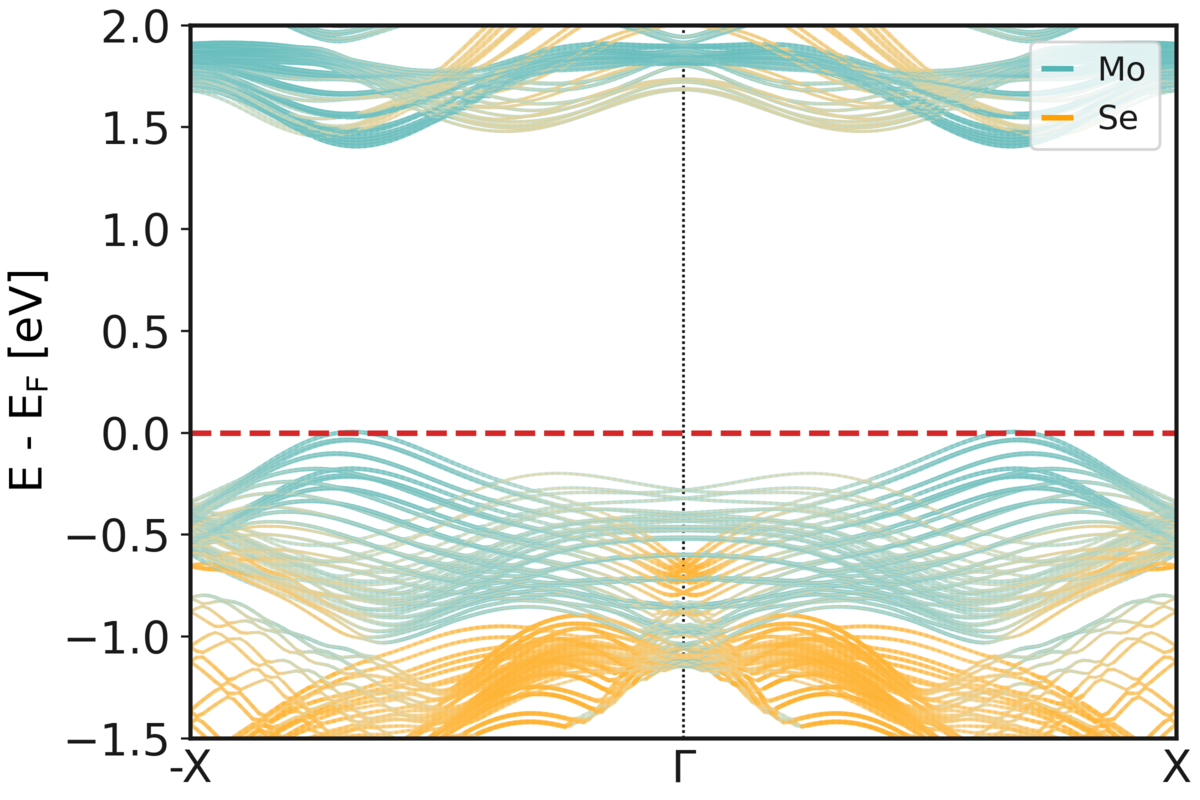}
    \includegraphics[width=0.49\textwidth]{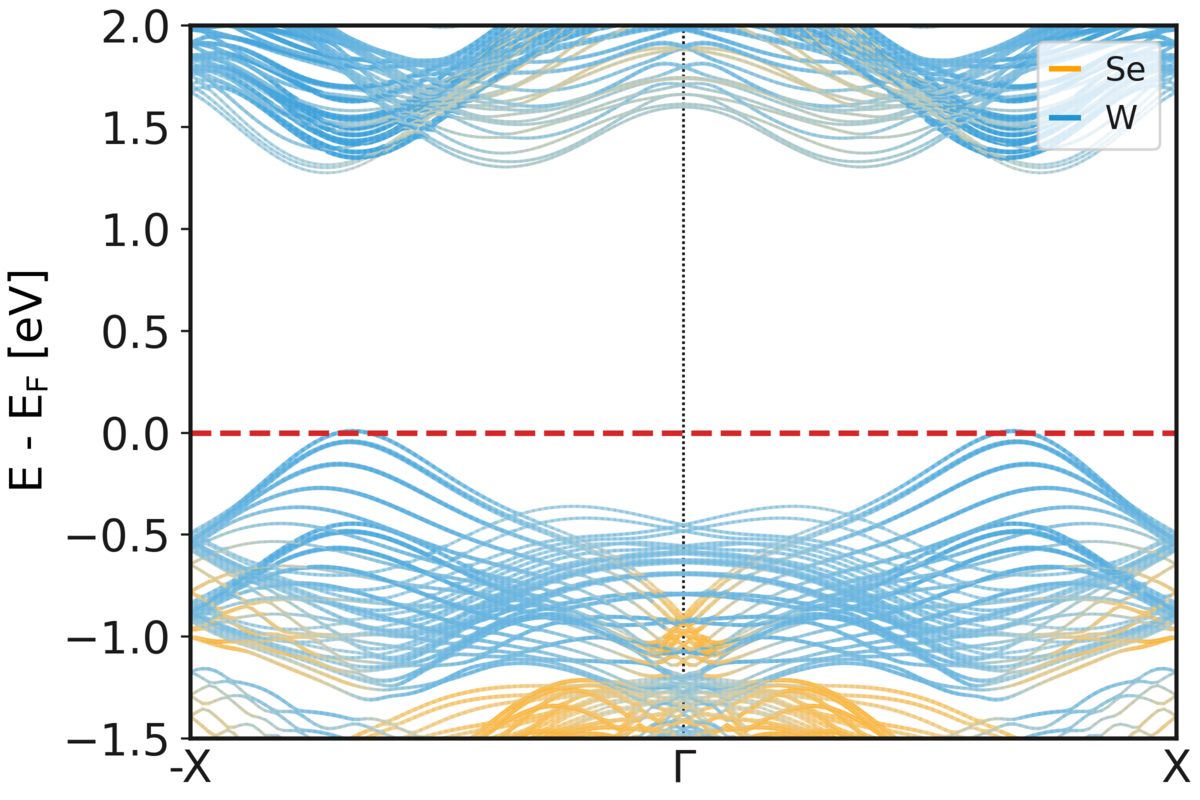}
    \caption{The heterobilayer reduces the Rashba-like splitting in the wrinkle heterobilayer \ch{WSe2$/$MoSe2}. The band structure of the separated layer geometry of the 12.5\% compressed heterobilayer \ch{WSe2$/$MoSe2}. Left) \ch{MoSe2} and right) \ch{WSe2} }
    \label{fig_layer_sperated_heterobilayer12.5}
\end{figure}

\end{document}